\documentclass[preprint]{aastex631}
\usepackage{CJK}
\usepackage{amsfonts}
\usepackage{amsmath}
\usepackage{mathrsfs}
\usepackage{epsfig}
\usepackage{subfigure}

\newcommand\diff{\,\mathrm{d}}

\def\m{\,\mathrm{m}}

\def\s{\,\mathrm{s}}
\def\p{\,\mathrm{p}}

\def\C{\,\mathrm{C}}
\def\N{\,\mathrm{N}}
\def\O{\,\mathrm{O}}

\def\B{\,\mathrm{B}}
\def\A{\,\mathrm{A}}

\def\c{\,\mathrm{c}}

\def\g{\,\mathrm{g}}

\def\P{\,\mathrm{P}}
\def\Teff{\,T_{\mathrm{eff}}}

\def\yr{\,\mathrm{yr}}
\def\cd{\,\mathrm{c}\ \mathrm{d}^{-1}}
\def\logg{\,\log{g}}

\def\raa{Res.\ Astron.\ Astrophys.\ }

\shorttitle{An HADS crossing the Hertzsprung Gap}
\shortauthors{J.-S. Niu \& H.-F. Xue}

\begin{document}
\begin{CJK*}{UTF8}{gbsn}

\title{A rapidly evolving high-amplitude $\delta$ Scuti star crossing the Hertzsprung Gap}

\correspondingauthor{Jia-Shu Niu}
\email{jsniu@sxu.edu.cn}

\author[0000-0001-5232-9500]{Jia-Shu Niu (牛家树)}
\affil{Institute of Theoretical Physics, Shanxi University, Taiyuan 030006, China}
\affil{State Key Laboratory of Quantum Optics and Quantum Optics Devices, Shanxi University, Taiyuan 030006, China}
\affil{Collaborative Innovation Center of Extreme Optics, Shanxi University, Taiyuan, Shanxi 030006, China}

\author[0000-0001-6027-4562]{Hui-Fang Xue (薛会芳)}
\affil{Department of Physics, Taiyuan Normal University, Jinzhong 030619, China}
\affil{Institute of Computational and Applied Physics, Taiyuan Normal University, Jinzhong 030619, China}

\begin{abstract}
  In this work, we report the discovery of the rapidly evolving high-amplitude $\delta$ Scuti star KIC 6382916 (J19480292+4146558) which is crossing the Hertzsprung gap. According to the analysis of the archival data, we find three independent pulsation modes, whose amplitudes and frequencies vary distinctly in 4 years. The linear period variation rates of the first two modes are about 3-4 times larger than the best seismic model constructed by the standard evolution theory, while that of the third one is about 8 times larger than the first two modes. What is more interesting is that, almost all the combinations of the third mode have frequency peaks $0.0815\ \cd$ away from them in the frequency domain. A framework is proposed to interpret the markedly large frequency and amplitude variation rates of the third mode, in which we employ a new pulsation mode (resonating integration mode) generated by the resonance between a radial p-mode and a nonradial mixed mode. Moreover, a global analysis of the interactions between the three independent pulsation modes and their harmonics/combinations is performed based on the interaction diagrams of their amplitudes and phases, which would be a useful tool for the future asteroseismology research. 
\end{abstract}

\section{Introduction}      

People cannot clearly witness the stellar evolution process of a single star obviously in most cases because of its extremely secular time-scale, except for some special time nodes in it (such as the supernova explosion \citep{SN1987A}). However, in some specific evolutionary phases, we have the chance to witness such processes gradually on human timescales. When a star is leaving from the main sequence, the hydrogen nuclei fusion in its core is gradually transferring into the shell. In the Hertzsprung-Russell diagram, its evolutionary phase falls into the Hertzsprung gap, which is one of the most rapidly evolving phases in the life of a star \citep{Kippenhahn2012}.

$\delta$ Scuti stars are a class of short-period pulsating variable stars with periods between 15 minutes and 8 hours and spectral classes of A-F, which locate them on the main-sequence (MS) or post-MS evolutionary phases at the bottom of the classical Cepheid instability strip and excited by the $\kappa$ mechanism \citep{Breger2000,Handler2009,Uytterhoeven2011,Holdsworth2014}. High-amplitude $\delta$ Scuti stars (hereafter HADS) are a subclass of $\delta$ Scuti stars that usually have larger amplitudes ($\Delta V \geq 0_{\cdot}^{m}3$) and slower rotations \citep{Breger2000}.\footnote{ However, with the accumulation of the samples, these criteria become unclear now. For example, the HADS KIC 9408694  has an amplitude of less than 0.1 Kmag at the highest-amplitude modes and a fast rotation ($v \sin i \approx 100\ \mathrm{km\ s}^{-1}$) \citep{Balona2012}.  } Most of the HADS show single or double radial pulsation modes, and some of them have three radial pulsation modes or even some nonradial pulsation modes \citep{Niu2013,Niu2017,Xue2018,Wils2008,Xue2020}.
Theoretically, the period variation rates ($(1/P)(\diff P/ \diff t)$) of HADS caused by the evolution should roughly have a value from $10^{-10}\ \mathrm{yr}^{-1}$ on the MS to $10^{-7}\ \mathrm{yr}^{-1}$ on the post-MS phase \citep{Breger1998}. According to the seismic models, the observed period variation rates of some HADS (such as AE UMa and VX Hya \citep{Niu2017,Xue2018,Xue2022}) can be interpreted by the evolutionary effect self-consistently, and each of these stars should be located in the Hertzsprung gap on Hertzsprung-Russell (H-R) diagram with a helium core and a hydrogen-burning shell. In these works, the linear period variation rates are obtained from the ground-based time-series photometric data accumulated over several decades, but only those of the fundamental pulsation modes have a sufficient precision to be confirmed and tested because of the quality of the data.
The high-precision photometric data from the {\it Kepler space telescope} lasted for about 4 years and could provide an excellent opportunity to determine the linear period and amplitude variation rates not only of the fundamental pulsation mode, but also of other pulsation modes. These observed quantities would tell us more secrets about the stellar evolution and pulsation in this special rapidly evolving phase.

KIC 6382916 (J19480292+4146558, also known as GSC 03144-595) is an HADS that has been monitored extensively by the {\it Kepler space telescope} for its pulsation properties, which shows three independent frequencies (pulsation modes) in its light curves \citep{Ulusoy2013}. These three pulsation modes were at first identified as the nonradial $l=1$ pulsation modes in \citet{Ulusoy2013}, and then identified as the fundamental, first overtone, and second overtone radial p-modes \citep{Mow2016}. Using the long-cadence (LC) photometric data of  KIC 6382916 lasted from BJD 2454953 to 2456424 (Quarter 0-17), we extract the first 23 frequencies with the largest amplitudes (which have a cutoff of 1.3 mmag, see Appendix \ref{app:1} and Table \ref{tab:freq_solution} for more details) to study their variations and interactions. These frequencies are also identified as the labels of the corresponding pulsation modes in this work, from F0 to F22. Theses 23 frequencies are composed of three independent frequencies, F0, F1, and F5 (also labeled as $f_0$, $f_1$, and $f_2$) together with their harmonics/combinations.

\section{Methods and Analysis}
In order to extract the variation information of the amplitudes and frequencies over time, we use the short-time Fourier transformation to deal with the LC photometric data. The prewhitening process in a time window of 120 days is performed when the window is moving from the start to the end time of the LC data, with a step of 20 days. In each step, the frequencies are fixed as that have been extracted in the complete LC data (see Table \ref{tab:freq_solution}), while the amplitudes and phases are obtained by the nonlinear least-square fitting. Then, we get the amplitude and phase variations in the 23 frequencies. Using the Fourier-phase diagram method \citep{Paparo1998,Bowman2016,Xue2018}, the variations in the frequencies can be derived from the variations in the phases.

When the dominating quasi-periodic signals in the variations in amplitudes and phases have been fully handled, we obtain their nonperiodic linear and quadratic variations of them (see Appendix \ref{app:2} for more details).\footnote{Hereafter, we do not consider the periodic signal when we talk about the variations in the amplitudes and phases, unless it is specifically mentioned.} The amplitude of $f_0$ ($A_0$) is relatively stable, that of $f_1$ ($A_1$) slightly decreases while that of $f_2$ ($A_5$) distinctly increases by about 28\% in 4 years. On the other hand, the phases (subtracted by its average) of $f_0$, $f_1$, and $f_2$ vary with the same trend but different levels, which corresponds to the increasing periods of them with different linear period variation rates: $ (1/P_0)(\diff P_0/ \diff t) = (3.0 \pm 1.2) \times 10^{-7}\ \mathrm{yr}^{-1}$, $ (1/P_1)(\diff P_1/ \diff t) = (3.2 \pm 1.0) \times 10^{-7}\ \mathrm{yr}^{-1}$, and $ (1/P_2)(\diff P_2/ \diff t) = (2.4 \pm 0.6) \times 10^{-6}\ \mathrm{yr}^{-1}$. The harmonics/combinations of $f_0$, $f_1$, and $f_2$ also show complex amplitude and frequency variations, which are summarized in the appendix, Table \ref{tab:freq_solution}.

\begin{figure*}[htp]
  \centering
  \includegraphics[width=0.99\textwidth]{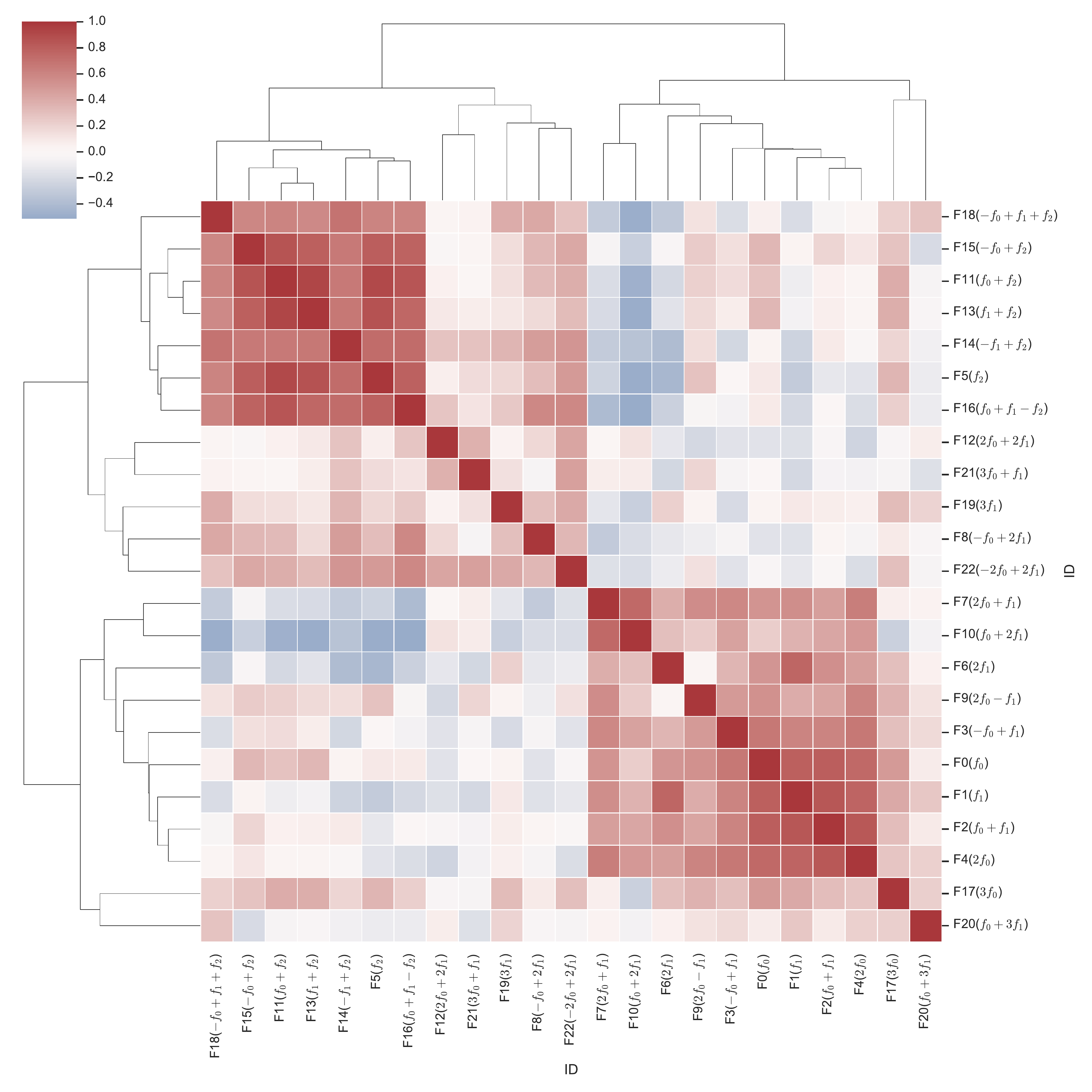}
  \caption{Interaction diagram of amplitudes of the 23 pulsation modes.}
  \label{fig:corr_amp}
\end{figure*}

Noting the complex amplitude variations in the F0-F22 pulsation modes, it is quite interesting to study the amplitude interactions between them, which also indicate the energy transfer between them. Here, we introduce the interaction diagram (see Figure \ref{fig:corr_amp}) to show the amplitude interactions between the pulsation modes. In Figure \ref{fig:corr_amp}, the color of a small square represents the correlation coefficient\footnote{Here we use the Spearman's rank correlation coefficient, which is a measure of how well the relationship between two variables can be described by a monotonic function.} between the amplitudes of the labeled pulsation modes whose columns and rows intersect at the square. The dendrograms on the upper and left represent the agglomerative hierarchical clustering (AHC) process \citep{AHC} in the amplitude interaction space, which is spanned by the vectors (corresponding to each of the pulsation modes) consisting of the correlation coefficients between a specific pulsation mode to all of them. In general, the AHC process reorders the pulsation modes in order to cluster those who have similar behaviors in the interaction space together. More details can be found in Appendix \ref{app:3}.

At first glance of Figure \ref{fig:corr_amp}, the pulsation modes are clustered into two groups (separated between F22 and F7), which correspond to the main features of increase in and periodicity of the amplitudes. In the upper group, all the $f_2$ related modes are closely clustered together, which have obvious increasing amplitudes. On the other hand, it is obvious that F1, F6, F7, and F10 are $f_1$ related and anticorrelated with the $f_2$ related modes, which indicates that they have decreasing amplitudes. What is more interesting is that F12, F21, F19, F8, and F22 are not $f_2$ related, but clustered together with the $f_2$ related modes. After a closer look on these pulsation modes in the frequency domain, we find an interesting discovery: almost all these modes have partners $\Delta \omega = 0.0815\ \cd$ away from them.\footnote{For F21, we cannot find any significant peaks (larger than 0.4 mmag) in both sides.} Inspired by this phenomenon, we carefully review all the 23 pulsation modes in the frequency domain carefully, and find that all the $f_2$ related modes have partners $\Delta \omega = 0.0815\ \cd$ away from them.\footnote{The zoom-in of all the 23 pulsation modes in the frequency domain is shown in Figure \ref{fig:spec01}, \ref{fig:spec02}, \ref{fig:spec03}, and \ref{fig:spec04}.} It indicates that $f_2$ could be a nonradial modes with $l=1$\footnote{Here, all the $1 \le l \le 3$ modes are possible. However, if we consider the relative large amplitude and all the doublets in the frequency domain, $l=1$ is the most likely scenario.} and all these $f_2$ related pulsation modes and their partners are locked by the star's rotation (with a rotation frequency of $\Delta \omega = 0.0815\ \cd$), which are interacting with each other. Additionally, all the $f_2$ related modes (F5, F11, F13, F14, F15, and F18) have partners $\Delta \omega = 0.0815\ \cd$ smaller than them, and the $-f_2$ related modes (F16) have partners $\Delta \omega = 0.0815\ \cd$ larger than them, which implies that $f_2$ has an azimuthal order of $m=1$. If this is true, the linear period variation rates of all $f_2$ (including $-f_2$) related modes with significant nonzero values should be dominated by the slowed-down rotation of the star, which gives rise to the linear period variation rates that are about an order of magnitude greater than those of the purely $f_0$ and $f_1$ related modes without partners.

Based on the above inferences, it is easy to understand the significant nonzero $(1/P)(\diff P/ \diff t)$ values reaching up to an order of magnitude from $10^{-6}$ to $10^{-5}$, which are seriously influenced by the slowed-down rotation and the interactions with their partners. In Figure \ref{fig:spec01}, \ref{fig:spec02}, \ref{fig:spec03}, and \ref{fig:spec04}, we show the pulsation modes without significant partners; if their $(1/P)(\diff P/ \diff t)$ have significant nonzero values (F0, F1, F2, F4, and F7), these values fall into a range of about $2.4 \times 10^{-7}\ \yr^{-1}$ to $4.0 \times 10^{-7}\ \yr^{-1}$, which should be the values of global linear period variations that are unaffected (or minimally affected) by the rotation and interactions of the modes.
Considering that $f_0$ and $f_1$ have almost the same period variation and have a period ratio  ($P_1/P_0 = f_0/f_1$) of 0.763 \citep{Petersen1996,Poretti2005}, they could be considered as the fundamental and first overtone radial pulsation modes, respectively.

\begin{figure*}[htp]
  \centering
  \includegraphics[width=0.495\textwidth]{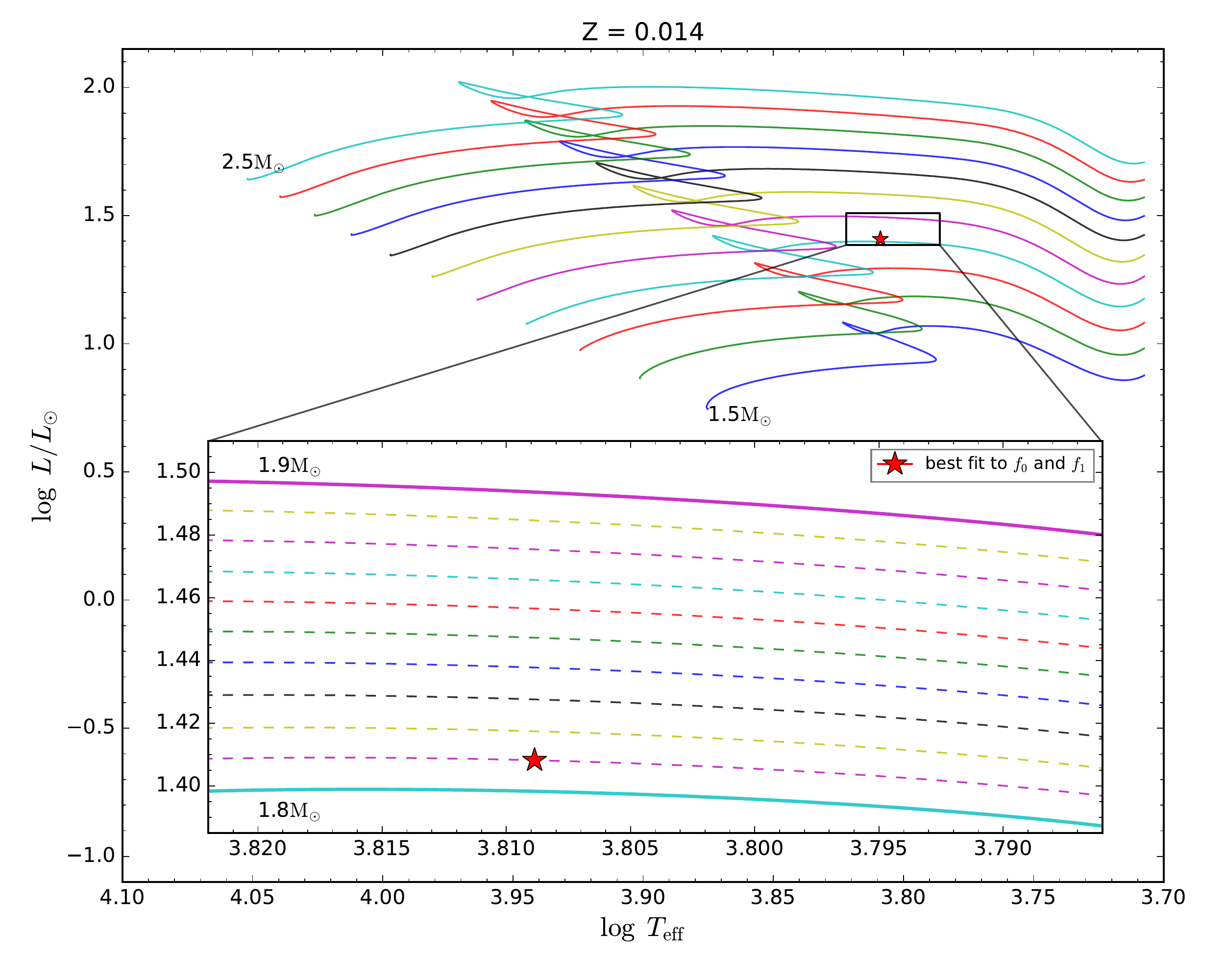}
  \includegraphics[width=0.495\textwidth]{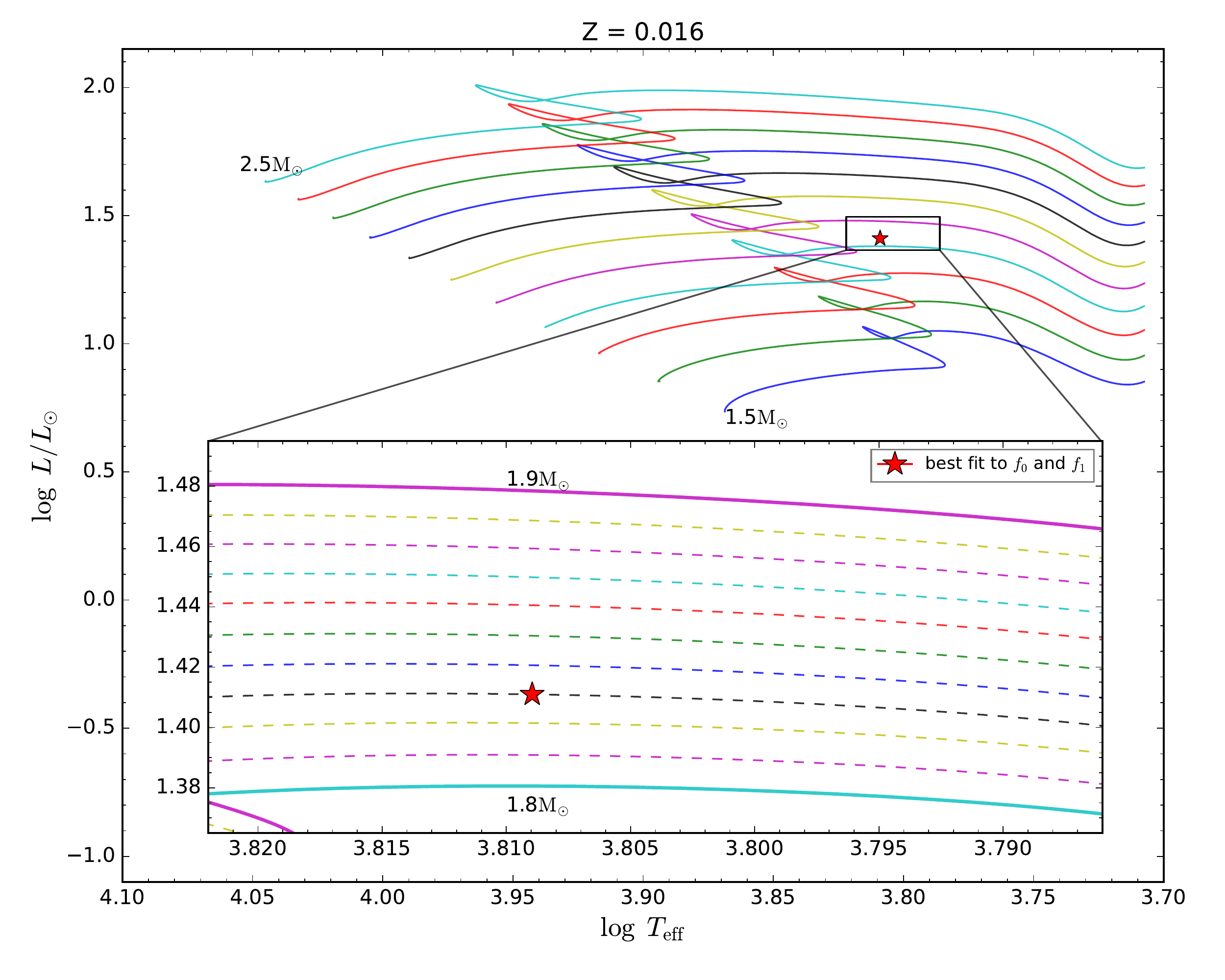}
  \includegraphics[width=0.495\textwidth]{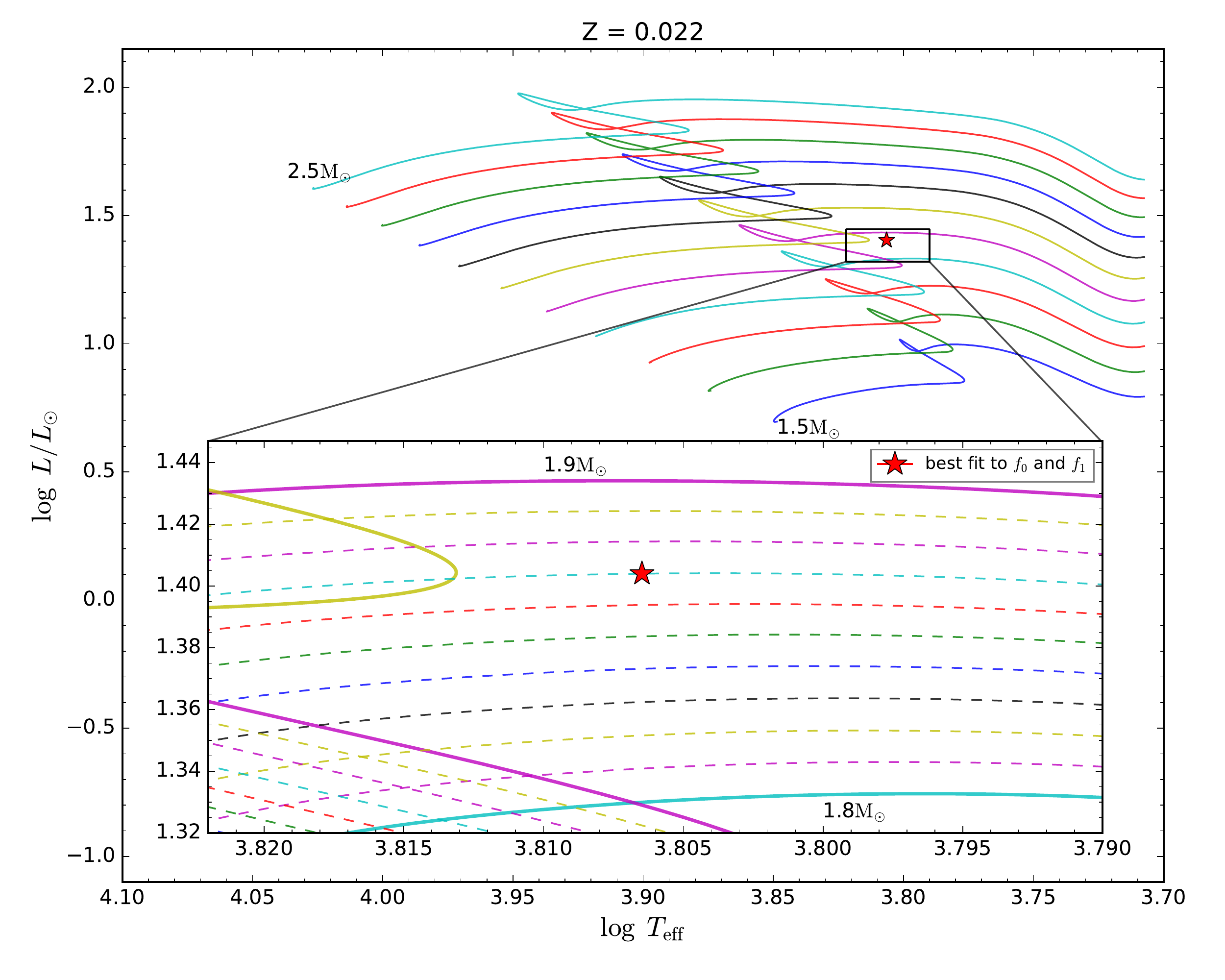}
  \caption{Hertzsprung-Russell diagram with all the evolutionary tracks and the best-fit seismic models of KIC 6382916 with different $Z$ values. The colored solid evolutionary tracks show from the zero-age MS to the post-MS evolutionary stage, with initial masses from 1.50 $M_{\odot}$ to 2.50 $M_{\odot}$. The regions surrounded by the black rectangular boxes are selected to zoom in the best-fit seismic models, which are presented by the red stars. The colored dashed lines present the evolutionary tracks with initial mass steps of 0.01 $M_{\odot}$. }
  \label{fig:best_model}
\end{figure*}

Single-star nonrotating evolutionary models are constructed by using different initial masses with three groups of [Fe/H] ($0.040$ \citep{Mathur2017}, $0.111$ \citep{Xiang2019}, and $0.266$ \citep{Luo2019}) from observations (which are labeled by $Z = 0.014$, $Z = 0.016$, and $Z = 0.022$; see more details in Appendix \ref{app:4}), from the pre-main sequence to the start of the red giant branch. At each of the steps on the evolutionary tracks, the pulsation frequencies are calculated based on the corresponding stellar structure. Figure \ref{fig:best_model} shows the best-fit seismic models of the observed independent frequencies ($f_0$ and $f_1$) together with the corresponding evolutionary tracks for different $Z$ values. The detailed information of the best-fit seismic models and the observations are collected in Table \ref{tab:best-fit}.

The results of the seismic models show that KIC 6382916 is located in the post-MS phase with the mass from 1.81-1.87 $M_{\odot}$. The most incredible thing is that the observed linear period variation rates of $f_0$ and $f_1$ are 3-4 times larger than the theoretical predicted ones. If we ascribe these observed linear period variation rates to the stellar evolution, we find that the star is evolving more rapidly than theoretical prediction and the standard stellar evolution theory cannot precisely describe the star's evolution process in this rapidly evolving phase. Another possible scenario is that the large linear period variations might be caused by mass transfer, which could be induced by the stellar wind or an undiscovered binary companion of the star. These possibilities should be studied and confirmed in the future.

\begin{table*}[htp]
  \centering
  \caption{Information of the best-fit seismic models and the observations.}
  \label{tab:best-fit}
  \begin{tabular}{c|ccc|c}
    \hline
    \hline
      & $Z=0.014$ & $Z=0.016$ & $Z=0.022$ & Observed \\
      \hline
$f_0\ (\cd)$ $(l=0, n_p=0)$ &  4.909573  &  4.909774 &  4.910073 &  $4.909842 \pm 0.000001 $ \\ 
$f_1\ (\cd)$ $(l=0, n_p=1)$ &  6.432067  &  6.431822 &  6.431710 &  $6.431890 \pm 0.000001$ \\ 
$(1/P_0)(\diff P_0/\diff t)\ (\yr^{-1})$ & $7.78 \times 10^{-8}$ & $8.83 \times 10^{-8}$ & $ 1.197 \times 10^{-7}$ & $ (3.0 \pm 1.2) \times 10^{-7}$\\ 
$(1/P_1)(\diff P_1/\diff t)\ (\yr^{-1})$ & $ 7.49 \times 10^{-8}$ & $ 8.50 \times 10^{-8}$ & $ 1.150 \times 10^{-7}$ & $ (3.2 \pm 1.0) \times 10^{-7}$         \\ 
\hline
$f_2^+\ (\cd)$ $(l=0, n_p=2)$&  8.061727 &  8.067854 &  8.085706 &  $8.035410 \pm 0.000002 $ \\ 
$f_2^-\ (\cd)$ $(l=1, n_p=2)$&  7.759899 &  7.993722 &  7.670764 &  $ 7.953927 \pm 0.000002 $ \\ 
$(1/P_2^+)(\diff P_2^+/\diff t)\ (\yr^{-1})$ & $ 7.39 \times 10^{-8}$ & $ 8.39 \times 10^{-8}$ & $ 1.123 \times 10^{-7}$ & $ (2.4 \pm 0.6) \times 10^{-6}$ \\ 
$(1/P_2^-)(\diff P_2^-/\diff t)\ (\yr^{-1})$ & $ -7.98 \times 10^{-8}$ & $ -8.41 \times 10^{-8}$ & $ -1.349 \times 10^{-7}$ & $ (2.4 \pm 0.6) \times 10^{-6}$ \\ 
\hline
$\log \Teff$ &  3.809 &  3.809 &  3.806 & [3.796, 3.867]\footnote{The observed $\log \Teff$ and $\log g$ values are collected from \citet{Bai2019,Ulusoy2013,Luo2019,Xiang2019,Mathur2017,Berger2018,Gaia2019,Gaia2022}.}\\ 
$\logg$ &  3.478 &  3.480 &  3.487 & [3.46, 4.31]\\ 
$\log L/L_{\odot}$ &  1.408 &  1.411 &  1.404 & ---  \\ 
\hline
Mass ($M_{\odot}$) &  1.81 &  1.83 &  1.87 & --- \\ 
Age (Gyr) &  1.39 &  1.37 &  1.36 & --- \\ 
$\log R/R_{\odot}$ &  0.609 &  0.610 &  0.612 & ---\\ 
\hline
  \end{tabular}
\end{table*}

Based on the above best-fit seismic models from the constraint of $f_0$ and $f_1$, we calculate the frequencies of the second overtone mode, which are systematically greater than the observed $f_2$ and labeled as $f_2^+$ (see in Table \ref{tab:best-fit}). The best-fit results of the nonradial ($l=1$, $m=0$) modes to the observed $f_2 - \Delta \omega$ have $n_p = 2, n_g =  23$ (for $Z=0.014$), $n_p = 2, n_g =  21$ (for $Z=0.016$), and $n_p = 2, n_g =  19$ (for $Z=0.022$), which are mixed modes of the p-mode and g-mode types (see in Table \ref{tab:best-fit}) and labeled as $f_2^-$. 
Although in the $Z=0.016$ case the observed $f_2$ can be roughly fitted by $f_2^- + \Delta \omega$ and its large linear period variation rate can be ascribed to the slowed-down rotation of the star, the distinctly increasing amplitude of the $f_2$ mode remains an unsolved problem in this framework.

\begin{figure*}
  \centering
  \includegraphics[width=0.495\textwidth]{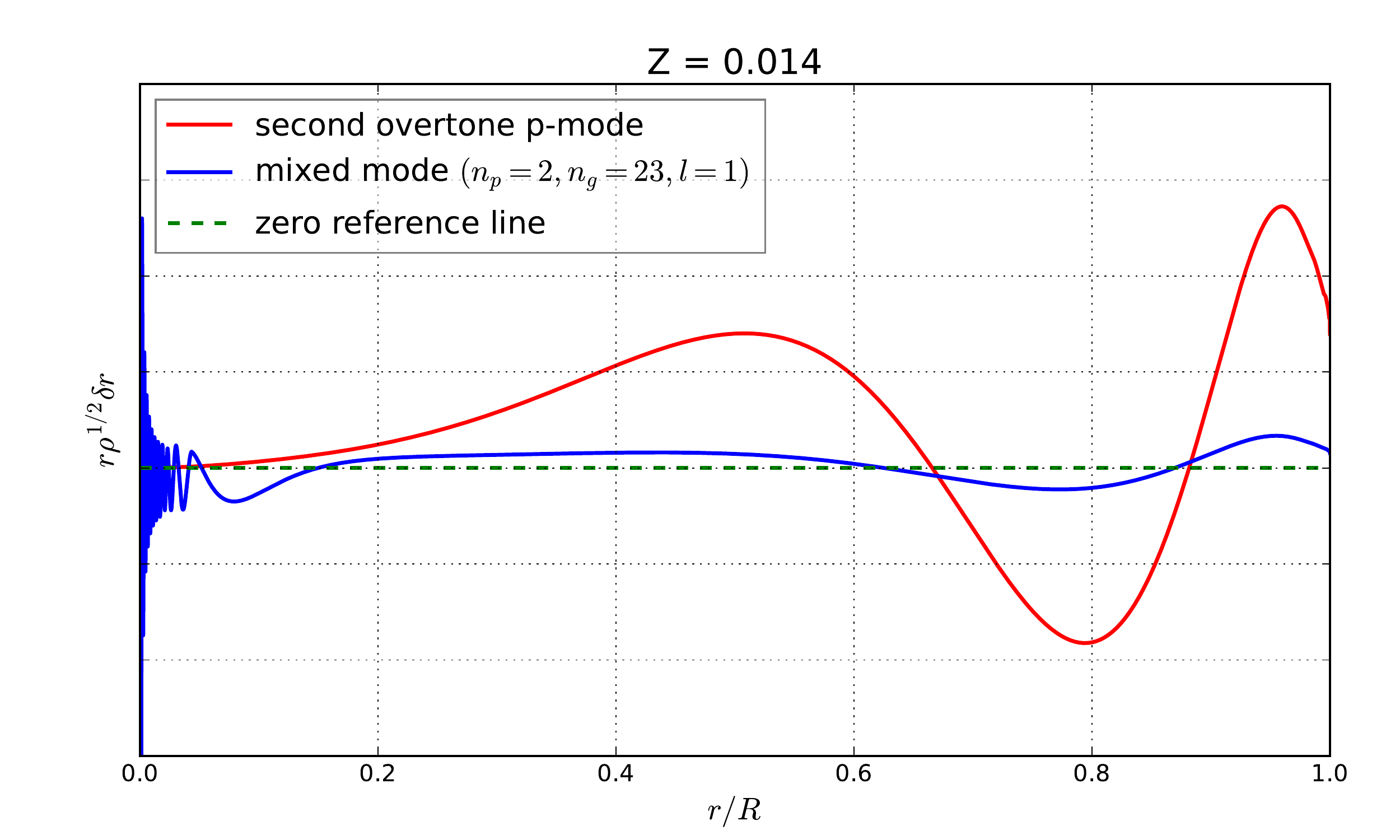}
  \includegraphics[width=0.495\textwidth]{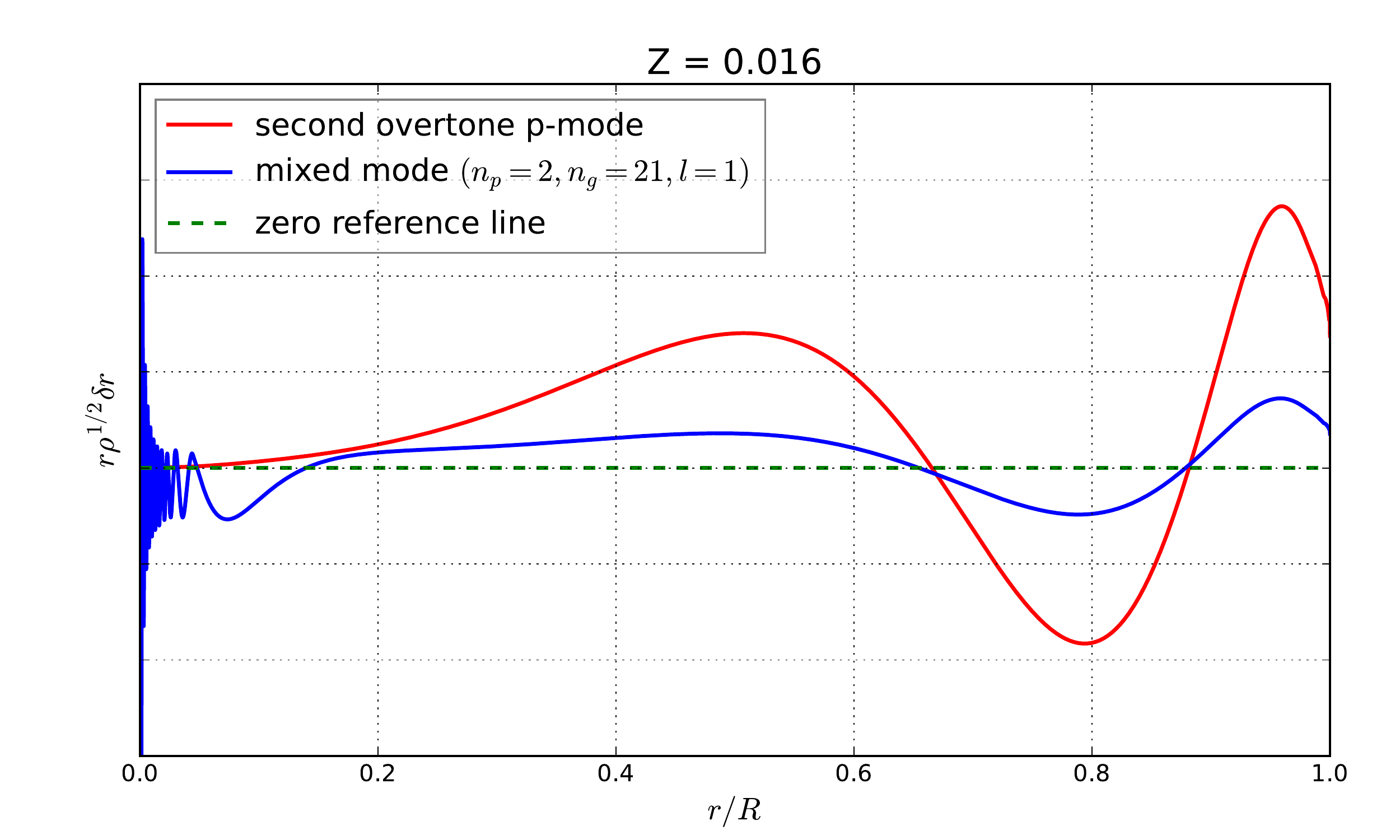}
  \includegraphics[width=0.495\textwidth]{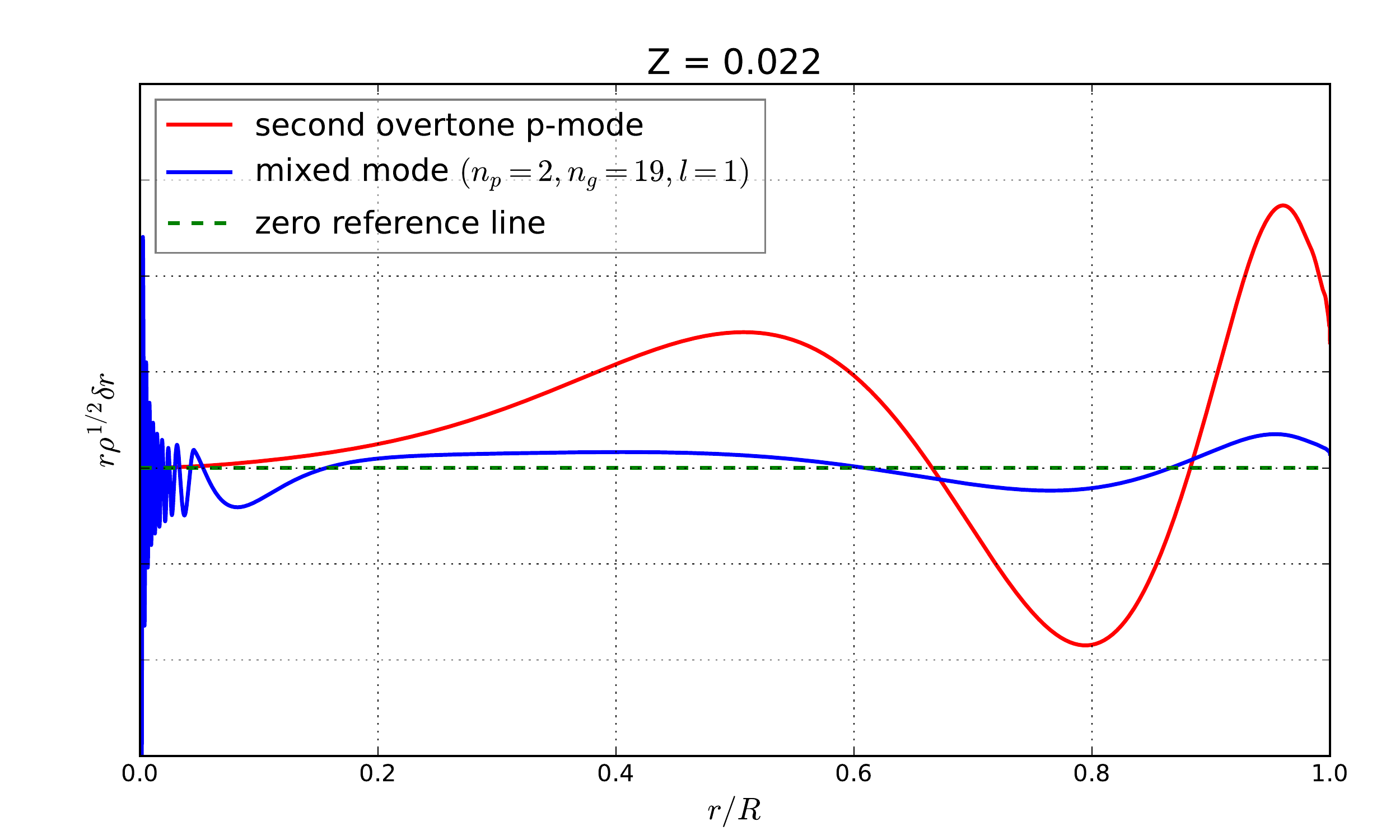}
  \caption{Radial displacement (represented by $r \rho^{1/2} \delta r$) against the fractional radius ($r/R$) for the best-fit seismic models with different $Z$ values. $r$ is the radial distance to the center; $\rho$ is the density at $r$; $\delta r$ is the radial displacement eigenfunction at $r$; $R$ is the radius of the star.}
  \label{fig:best_xi}
\end{figure*}

Here, we propose a framework to interpret all the important characteristics of $f_2$. 
Firstly, a resonance happens between a radial p-mode and a nonradial mixed mode, i.e. $f_2^+$ and $f_2^-$, and generates a resonating integration mode (RI mode) $f_2^{\pm}$. 
The requirements of the resonance can be met by (i) $f_2^+$ and $f_2^-$ having close frequency values and (ii) their radial displacements having close nodes and similar phases along with the radius in the outer layers (see in Figure \ref{fig:best_xi}).
Secondly, the RI mode $f_2^{\pm}$ should have the properties of radial and nonradial pulsation modes, which splits into two frequencies ($f_2^{\pm 1}$ and $f_2^{\pm 2}$, $f_2^{\pm 2} - f_2^{\pm 1} = \Delta \omega$) because of the star's rotation. $f_2^{\pm 1}$ resonates and superimposes with the combination of $f_{0}$/$f_{1}$ (i.e. $-f_0 + 2f_1$) to form the observed partner of $f_2$ (i.e. F8), while $f_2^{\pm 2}$ corresponds to the observed $f_2$. 

In the above framework, the frequency of $f_2^{\pm}$ should be modulated between $f_2^+$ and $f_2^-$ (similar to the situation in \citet{Goupil1998}), and then $f_2$ and its partner ($f_2^{\pm 2}$ and $f_2^{\pm 1}$, respectively) should represent the modulations, which are covered by current long-time scale data sets and could be shown out in a short-time scale data set. 
Because the linear period variation of $f_2$ is dominantly affected by the slowed-down rotation of the star, it shows a rate about 8 times larger than the global ones (that of $f_0$ and $f_1$). 
Taking it a step further, we note that the theoretical linear period variations in $f_2^+$ and $f_2^-$ have opposite directions in the stellar evolution. The distance between $f_2^+$ and $f_2^-$ decreases with the evolution of the star, and the resonance between them will be strengthened and the amplitude of $f_2$ will increase.\footnote{Here, we suppose a mechanism that increases the global linear period variation rates of $f_0$ and $f_1$ and also works for $f_2^+$ and $f_2^-$ to some extent, but keeps the signs of them unchanged.}

About the interactions between the $f_2$ related modes and their rotationally locked partners, on one hand, all of them have components inherited from the RI mode, which gives rise to increasing amplitudes. On the other hand, each of the observed partners is the resonance result of the $f_0$/$f_1$ related mode and the radial component of the RI mode, which will also cause energy transfer between them and then amplitude modulation. All these factors together produce the interaction relationships in Figure \ref{fig:corr_amp}. A more detailed quantitative theory about the framework should be constructed in future works.

For KIC 6382916, following the current amplitude variation rates, the amplitude of $f_2$ will exceed that of $f_0$ at about BJD 2460098 (2023 June), and exceed that of $f_1$ at about BJD 2460196 (2023 September). This prediction can be tested in the near future by the following photometric observations, which also could also provide us with an opportunity to witness the stellar evolution process of a single star gradually in this special evolutionary phase.

\section{Discussions and Conclusions}

The period variation rates of HADS have been studied for a long time as their large amplitudes are always dominated by one pulsation mode, and the classical O-C method can be put into practice based on the accumulation of times of maximum light of about several decades \citep{Breger1998}.  In recent years, based on the continuous high-precision time-series photometric data from space telescopes like {\it Kepler}, the linear period variation rates of HADS can be extracted directly from the light curves, not only for the dominating pulsation modes, but also for other nondominating pulsation modes and the harmonics/combinations of these modes (see, e.g., \citet{Breger2014,Bowman2016,Bowman2021}) . These linear period variation rates ($(1/P)(\diff P/\diff t)$) generally have the values from $\pm (10^{-9} - 10^{-6}) \yr^{-1}$, while some of them are not that reliable and would be updated by additional high-quality photometric data.
Theoretically, the period variation of an HADS can be caused by stellar evolution \citep{Niu2017,Xue2018,Xue2022}, mass transfer \citep{Xue2020}, light traveling time effect in a multiple system \citep{Xue2020}, and pulsational nonlinearity \citep{Breger2014,Zong2016,Bowman2016}. In these possible origins, the first three will affect all pulsation modes globally\footnote{In more detail, the stellar evolution will also lead to different pulsation modes (p-mode and g-mode) to linear period variation rates with different signs but similar absolute values.}, while the last one will affect on the pulsation modes involved in the nonlinear interactions individually. In general, the observed period variation of an HADS is the result of a combination of several or all factors above. Especially, as a potential marker of stellar evolution, it is quite interesting and important whether the observed linear period variation of an HADS can be ascribed to the stellar evolution \citep{Niu2017,Xue2018,Xue2022}.

Harmonics/combinations of independent pulsation modes are common amongst pulsating stars, such as $\delta$ Scuti stars \citep{Breger2014}, $\beta$ Cep stars \citep{Degroote2009}, SPB stars \citep{Papics2017}, $\gamma$ Dor stars \citep{Kurtz2015}, and pulsating white dwarfs \citep{Wu2001}. Mathematically, the harmonics/combinations come from the nonsinusoidal properties of light curves, which indicate the nonlinearity of the star's pulsation physically.\footnote{Such as the nonlinear transformation from the temperature variation to the flux variation ($F=\sigma T^{4}$), and the nonlinear response of the stellar medium to the pulsation waves.} 
In general, the harmonics/combinations are not considered as intrinsic stellar pulsation modes \citep{Brickhill1992,Wu2001} and should mimic the behaviors of the parent's pulsation modes. 
As a result, in practice, the harmonics/combinations are always removed in the prewhitening process and not taken as the key information for the asteroseismology. 
Although the period variation rates of HADS for different pulsation modes have been studied by some pioneering works (see, e.g., \citet{Breger2014,Bowman2016}), those of harmonics/combinations and the global interactions between harmonics/combinations have received little attention, which could provide us with abundant information about the nonlinear characteristics of both the harmonics/combinations themselves and the interactions between them.

For KIC 6382916, each of the period variations in the 23 pulsation modes are dominated by a quasi-periodic plus a linear signal (for phase variation, quadratic signals). The quasi-periodic signals are not the focus of in this work (which might have instrumental origin, or due to the light-time effect in a binary system), and will be studied in future works. 
About the linear period variation, unlike in AI Vel \citep{Walraven1992}, the fundamental and first overtone modes ($f_0$ and $f_1$) have similar period variation rates, which are 3-4 times larger than the theoretical predictions. Considering the similar values on both the $f_0$ and $f_1$ modes (and their harmonics/combinations without significant partners), the most possible origins of this result is a faster stellar evolution than that expected in this specific phase for the mass loss of the star, which needs further studies based on observational evidences.  
$f_2$ has a $(1/P)(\diff P/ \diff t)$ value 8 times larger than those of $f_0$ and $f_1$, which could come from the slowed-down rotation of the star and indicates that $f_2$ is a nonradial mode. 
Meanwhile, some of the harmonics/combinations show different period variations from their parent's modes (such as F22), which could be ascribed to the nonlinear interactions to the partners.

Beside the large linear period variation rates, the $f_2$ mode also has a distinctly increasing amplitude (see in Figure \ref{fig:var_amp_phase02}). Moreover, in the interaction diagram of amplitudes (see in Figure \ref{fig:corr_amp}), the $f_2$ related modes are closely clustered together, which also represents the particularity of $f_2$. 
Inspired by the results of theoretical calculations, we propose a framework to explain the strange behaviors of $f_2$ qualitatively. In this framework, the observed $f_2$ is the nonradial component of an RI mode (split by the rotation of the star), while the RI mode comes from a resonance between a radial p-mode and a nonradial mixed mode. In this case, the linear period variation rate of $f_{2}$ is dominated by the slowed-down rotation and the increasing amplitude comes from the gradually enhanced resonance along with the stellar evolution. In this sense, we can say that we witness the stellar evolution process of a single star gradually in this special evolutionary phase.

In the above framework, the variation rate of the rotation ($\diff \Delta \omega/\diff t$) can be obtained either directly by the observed frequency variations in the $f_2$ related modes and their partners (the differences between them), or indirectly by the observed frequency variations in the three independent frequencies ($f_0$, $f_1$, and $f_2$) based on the frequency decomposition relation (see Appendix \ref{app:5} for more details). 
Moreover, if we assume the conservation of angular momentum in this evolutionary phase and a uniform rotation of the star, the variation in the structure along with the stellar evolution will also cause a slowed-down rotation.
The variation rates of the rotation frequency ($\diff \Delta \omega/\diff t$) obtained from these estimations are shown in Figure \ref{fig:domega}, more details can be found in Appendix \ref{app:5}.
In this work, five pairs can be used to calculate $\diff \Delta \omega/\diff t$: F3-F14, F5-F8, F6-F11, F9-F16, and F15-F22. 	
In Figure \ref{fig:domega}, all the five pairs indicate a slowed-down rotation, although pairs F3-F14 and F6-F11 show large uncertainties and have no definite answer. Pairs F5-F8 and F9-F16 have consistent results with that obtained from $f_0$, $f_1$, and $f_2$ (F0-F1-F5). The different results between the pair F15-F22 and F0-F1-F5 might indicate the nonlinear interaction between F15 and F22 or the influence from a hidden g-mode on pair F15-F22 in this low-frequency domain.
On the other hand, it is obvious that the slowed-down rotation frequency ($|\diff \Delta \omega/\diff t|$) rates from theoretical models are significantly smaller than those from observations, which can be ascribed to a more effective angular momentum transfer based on the differential rotation of the star or an extra angular momentum loss induced by the stellar wind or its undiscovered binary companion.
It needs more precise theoretical models and observational data for further study.

\begin{figure*}[htp]
  \centering
  \includegraphics[width=0.9\textwidth]{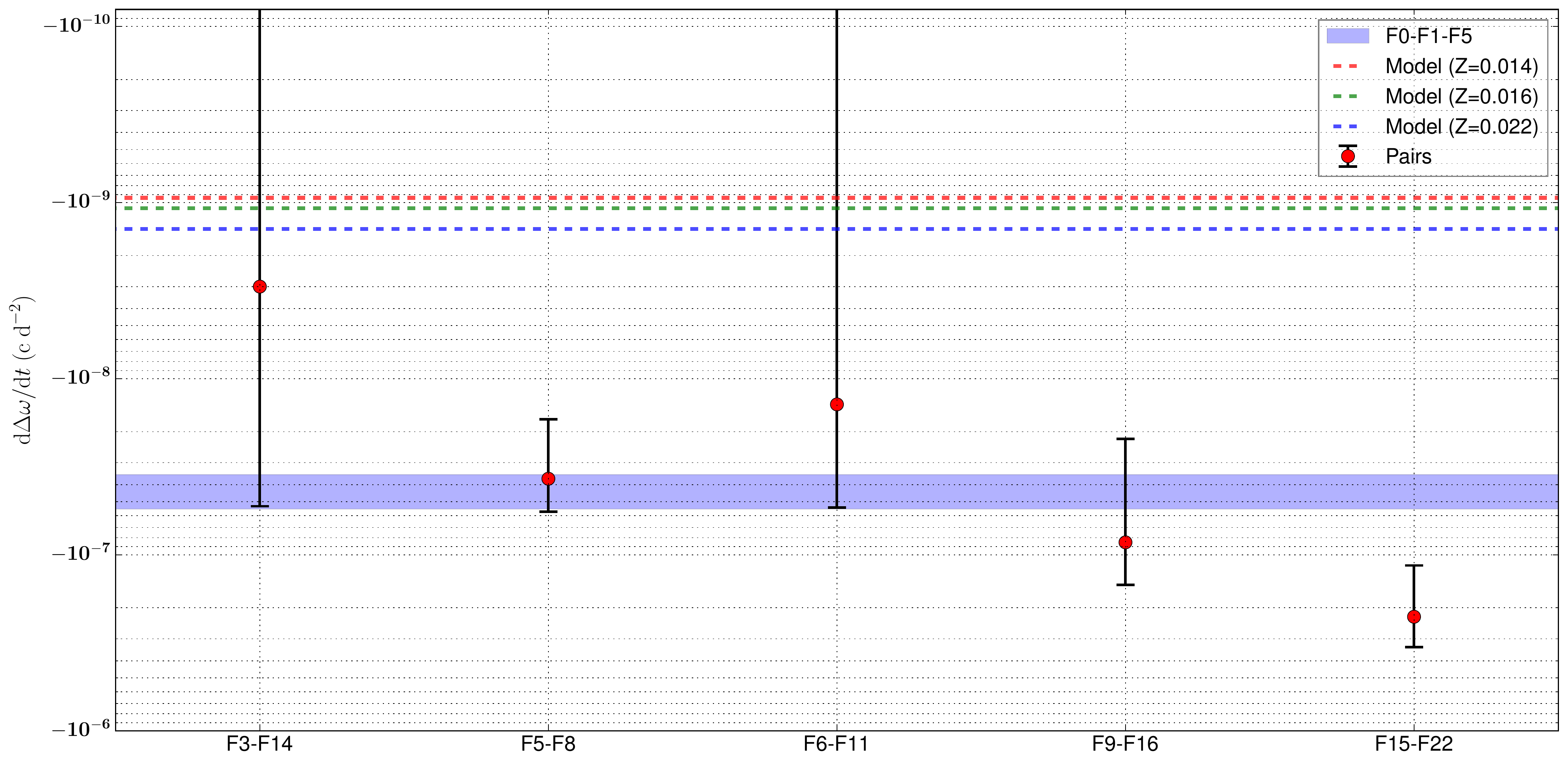}
  \caption{Variation rates of the rotation frequency from different estimates. The red dots with error bars represent those obtained from the five pairs. The blue region represents those obtained from the variation of $f_0$, $f_1$, and $f_2$. The three dashed lines represent those from the best-fit seismic models of different $Z$ values.}
  \label{fig:domega}
\end{figure*}

It is interesting to take a closer look on F22. On the one hand, the frequency variation rate of F22 can be obtained directly from Table \ref{tab:var_p} ($\dot{f}_{22} = (19.12 \pm 10.05) \times 10^{-8}\ \mathrm{c\ d}^{-2}$).\footnote{Here, the method of uncertainty estimation is the same as that in Appendix \ref{app:5}.} On the other hand, we have identified $f_{22} = -2f_{0} + 2f_{1}$, and then we should have $\dot{f}_{22} = -2\dot{f}_{0} + 2\dot{f}_{1}$. Based on $\dot{f}_{0}=(-4.0 \pm 1.5) \times 10^{-9}\ \mathrm{c\ d}^{-2}$ and $\dot{f}_{1}=(-5.7 \pm 1.8) \times 10^{-9}\ \mathrm{c\ d}^{-2}$, we can get $-2\dot{f}_{0} + 2\dot{f}_{1} = (-3.4 \pm 4.6) \times 10^{-9}\ \mathrm{c\ d}^{-2}$. Even taking the large uncertainties into account, it shows that $\dot{f}_{22} \neq -2\dot{f}_{0} + 2\dot{f}_{1}$, which indicates that the observed F22 mode is not a simple superposition of $f_{0}$ and $f_{1}$. The nonlinear interactions from its partner or a hidden g-mode might be the origin of its significantly large frequency variation.

This is not the only case. For $\delta$ Scuti stars KIC 9700322 and KIC 5950759, one can also find partners similar to the case in this work (equally spaced frequency multiplets caused by the rotation of the star) beside the dominant radial pulsation modes \citep{Breger2011,Yang2018}. We believe that this phenomenon is common in slowly rotating HADS, which evolve into a special phase and meet the resonant conditions, and the above framework of the RI mode will be refined in future research.

About KIC 6382916, there are still many interesting questions to be addressed, such as the origin of the quasi-periodic signals in the amplitudes and phases; the origin of the large linear period variation rates of $f_0$ and $f_1$ compared with the theoretical prediction; more quantitative details of the large amplitude increases in $f_{2}$ (28\% in 4 years); whether the amplitude of $f_2$ will exceed those of $f_0$ and $f_1$ in the future; the origin of the observed large slowed-down rotation rate compared with the theoretical predictions; the inconsistency of the frequency variation rates of the harmonics/combinations and their parent's pulsation modes (such as the case in F22). More detailed research and observations are needed.

\section*{Acknowledgments}
We would like to thank the anonymous reviewer for his/her professional and detailed suggestions and Jue-Ran Niu for providing us with an efficient working environment.
J.S.N. acknowledges support from the National Natural Science Foundation of China (NSFC) (No. 12005124 and No. 12147215). H.F.X. acknowledges support from the Scientific and Technological Innovation Programs of Higher Education Institutions in Shanxi (STIP) (No. 2020L0528) and the Applied Basic Research Programs of Natural Science Foundation of Shanxi Province (No. 202103021223320).

\software{{MESA} \citep{Paxton2011,Paxton2013,Paxton2015,Paxton2018,Paxton2019}, {GYRE} \citep{Townsend2013,Townsend2018,Goldstein2020}, {\tt emcee} \citep{emcee}, {seaborn} \citep{seaborn}}

The MESA {\tt work} folder for reproducing the best-fit models with different $Z$ values is available on Zenodo under an open-source Creative Commons Attribution license: \dataset[doi:10.5281/zenodo.7133342]{https://doi.org/10.5281/zenodo.7133342}.

\clearpage


\clearpage
\appendix
\restartappendixnumbering

\section{Photometric Data Reduction}
\label{app:1}
\begin{figure*}[htp]
  \centering
  \includegraphics[width=0.9\textwidth]{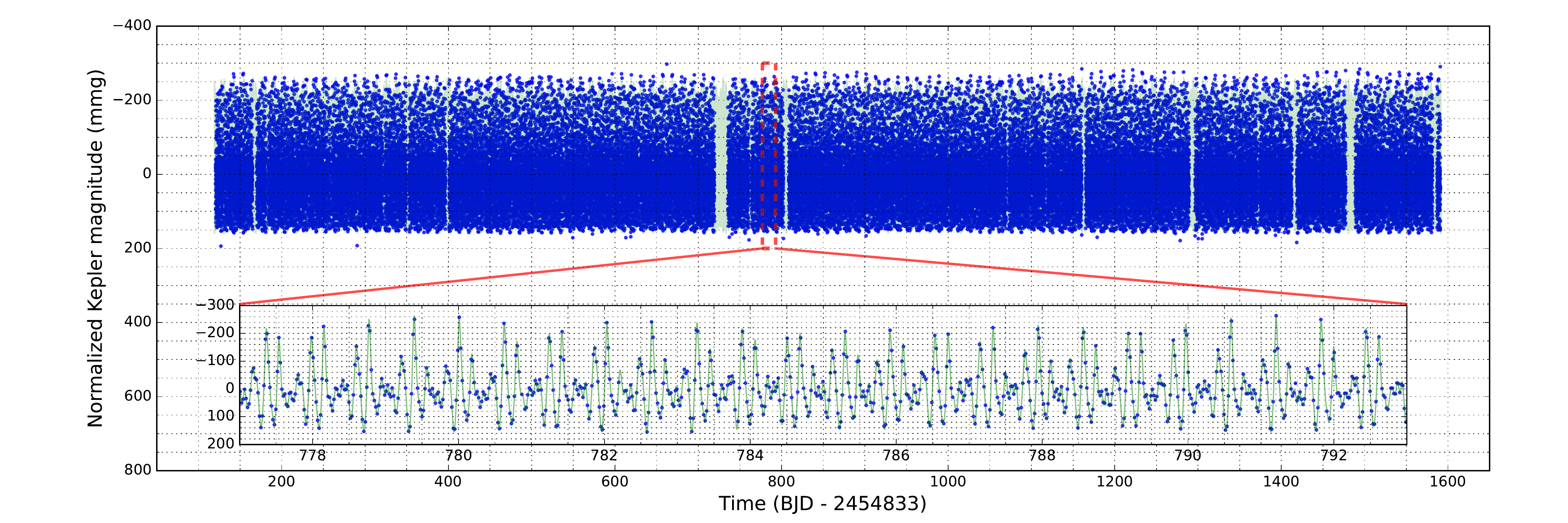}
  \includegraphics[width=0.9\textwidth]{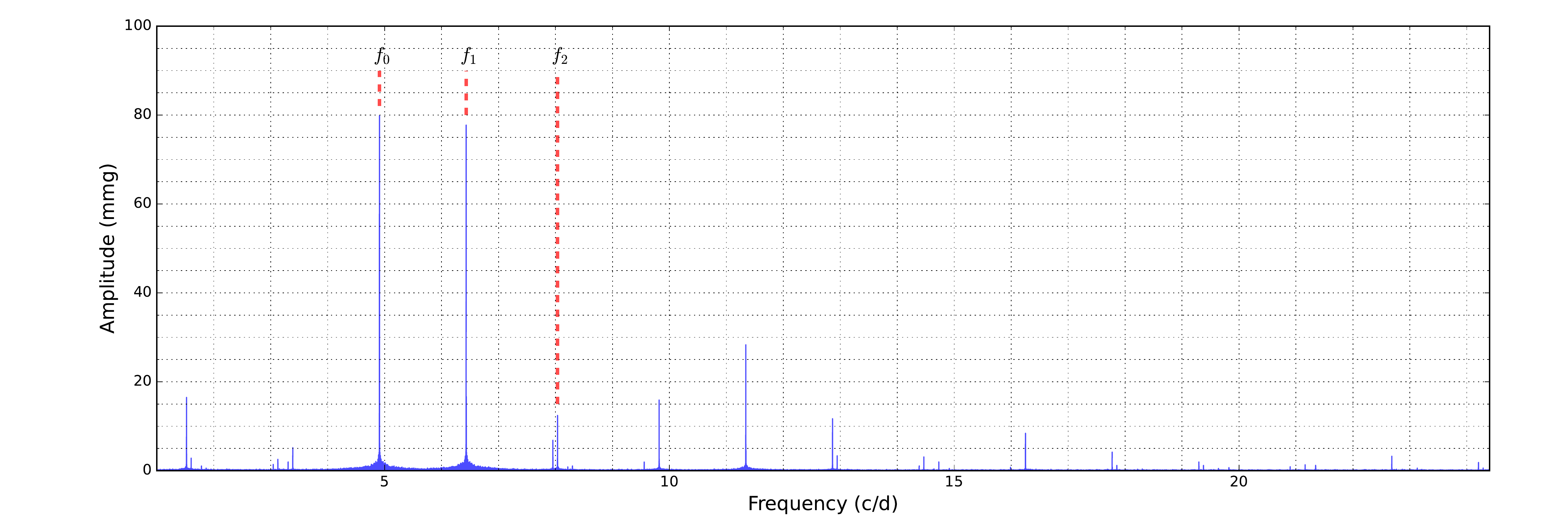}
  \caption{Overview of the normalized LC data in the time domain and frequency domain of KIC 6382916. The region surrounded by the black rectangular boxes is selected to zoom in the details of the observed light curves and the theoretical light curves, which are reproduced by the detected frequency solutions and represented by light green lines.}
  \label{fig:overview}
\end{figure*}

The long-cadence (LC) photometric data of KIC 6382916 from the {\it Kepler space telescope} were used in this work, which covers from BJD 2454953 to 2456424 (Quarter 0-17) (publicly available PDC data \citep{Kepler01,Kepler02}).
We downloaded the light curves (in the format of reduced BJD and magnitudes) of KIC 6382916 from the Mikulski Archive for Space Telescope (MAST) \footnote{\url{http://archive.stsci.edu/kepler}}, which were then normalized to be zero in the mean for each quarter. An overview of all the normalized LC data in the time domain and frequency domain are shown in Figure~\ref{fig:overview}.

All the above normalized LC data were prewhitened to extract the frequencies, amplitudes, and phases of the pulsation modes. This process was cutoff until an amplitude smaller than 1.3 mmg, and we obtained a total of 23 pulsation modes (see in Table \ref{tab:freq_solution}), which is much higher than the typical noise level in {\it Kepler} data of this star ($\sim$ 0.03 mmag). This choice ensured that the uncertainties of the amplitudes and phases were typically smaller than the variations in some pulsation modes of interest in 4 years, which was determined by the subsequent short-time Fourier transformation results. In each step in the prewhitening process, the Lomb-Scargle algorithm \citep{Lomb_Scargle} was used to help to find the initial value of frequency with largest amplitude, and then a non-linear least square fitting was performed to get the final values of the frequency, amplitude, and phase. The frequencies within $1.0 \le f \le 24.4\ \cd$ were considered in this work, which was determined by the LC Nyquist frequency \citep{Bowman2016}.

\begin{deluxetable}{c|ccccc|cc|rr}
  \label{tab:freq_solution}
\centering
\tablecaption{Frequency Solution of the First 23 Frequencies with the Largest Amplitudes. $\delta$ is defined as the difference between the observed frequency of F0-F22 mode and the marked one decomposed by $f_{0}$, $f_{1}$, and $f_{2}$. $\diff A/ \diff t$ denotes the linear variation rate of the amplitudes.}
\tabletypesize{\scriptsize}
\tablehead{
  \colhead{ID}\vline & \colhead{Frequency}& \colhead{$\sigma_{f}$} & \colhead{Amplitude} & \colhead{$\sigma_{A}$} & \colhead{S/N}\vline & \colhead{Mark} &  \colhead{$\delta$}\vline & \colhead{$(1/P)(\diff P/ \diff t)$} & \colhead{$(\diff A/ \diff t)$} \\
  \colhead{}\vline& \colhead{$(\cd)$} & \colhead{$(\cd)$} & \colhead{$(\mathrm{mmag})$} &  \colhead{$(\mathrm{mmag})$} &\colhead{}\vline  &\colhead{}  & \colhead{$(\cd)$}\vline & \colhead{$(\mathrm{yr}^{-1})$} & \colhead{$(\mathrm{mmag\ days}^{-1})$} 
}
\startdata
F0  & 4.909842  & 0.000001 & 79.9  & 0.2  & 334.9 & $f_{0}$              & $0$                  & $(3.0 \pm 1.2) \times 10^{-7}$ & $(12 \pm 12) \times 10^{-5}$\\
F1  & 6.431890  & 0.000001 & 77.8  & 0.2  & 377.2 & $f_{1}$              & $0$                  & $(3.2 \pm 1.0) \times 10^{-7}$ & $(-45 \pm 12)  \times 10^{-5}$\\
F2  & 11.341730 & 0.000001 & 28.38 & 0.09 & 320.3 & $f_{0}+f_{1}$        & $-1.9\times10^{-6}$ & $(2.4 \pm 1.6) \times 10^{-7}$ & $(-7 \pm 12) \times 10^{-5}$\\
F3  & 1.522039  & 0.000002 & 16.53 & 0.07 & 221.1 & $-f_{0}+f_{1}$       & $-8.9\times10^{-6}$ & $(4.6 \pm 1.8) \times 10^{-6}$ & $(-1 \pm 12) \times 10^{-5}$\\
F4  & 9.819693  & 0.000001 & 16.11 & 0.06 & 272.0 & $2f_{0}$             & $8.6\times10^{-6}$  & $(4.0 \pm 3.2) \times 10^{-7}$ & $(-3 \pm 12) \times 10^{-5}$\\
F5  & 8.035410  & 0.000002 & 12.54 & 0.06 & 226.5 & $f_{2}$              & $0$                  & $(2.4 \pm 0.6) \times 10^{-6}$ & $(206 \pm 12) \times 10^{-5}$\\
F6  & 12.863778 & 0.000002 & 11.71 & 0.05 & 246.2 & $2f_{1}$             & $-1.9\times10^{-6}$ & $(3.8 \pm 3.0) \times 10^{-7}$ & $(-16 \pm 12) \times 10^{-5}$\\
F7  & 16.251580 & 0.000002 & 8.51  & 0.04 & 200.9 & $2f_{0}+f_{1}$       & $5.7\times10^{-6}$  & $(3.4 \pm 3.2) \times 10^{-7}$ & $(-11 \pm 12) \times 10^{-5}$\\
F8  & 7.953927  & 0.000002 & 6.94  & 0.04 & 162.2 & $-f_{0}+2f_{1}$      & $-1.1\times10^{-5}$ & $(6.8 \pm 8.4) \times 10^{-7}$ & $(10 \pm 12) \times 10^{-5}$\\
F9  & 3.387805  & 0.000004 & 5.26  & 0.06 & 84.5  & $2f_{0}-f_{1}$       & $1.1\times10^{-5}$  & $-(0.6 \pm 3.0) \times 10^{-6}$& $(11 \pm 12) \times 10^{-5}$\\
F10 & 17.773617 & 0.000003 & 4.23  & 0.04 & 114.8 & $f_{0}+2f_{1}$       & $-5.7\times10^{-6}$ & $(8.0 \pm 6.2) \times 10^{-7}$ & $(-20 \pm 12) \times 10^{-5}$\\
F11 & 12.945261 & 0.000004 & 3.37  & 0.03 & 103.0 & $f_{0}+f_{2}$        & $9.5\times10^{-6}$  & $(7.8 \pm 10.6) \times 10^{-7}$ & $(58 \pm 15) \times 10^{-5}$\\
F12 & 22.683468 & 0.000004 & 3.24  & 0.03 & 98.3  & $2f_{0}+2f_{1}$      & $3.8\times10^{-6}$  & $(4.6 \pm 6.2) \times 10^{-7}$ & $(-2 \pm 12) \times 10^{-5}$\\
F13 & 14.467309 & 0.000004 & 3.19  & 0.03 & 92.7  & $f_{1}+f_{2}$        & $8.6\times10^{-6}$  & $(16.0 \pm 10.0) \times 10^{-7}$ & $(53 \pm 12) \times 10^{-5}$\\
F14 & 1.603523  & 0.000007 & 2.98  & 0.05 & 55.2  & $-f_{1}+f_{2}$       & $3.1\times10^{-6}$  & $(3.8 \pm 11.4) \times 10^{-6}$ & $(35 \pm 12) \times 10^{-5}$\\
F15 & 3.125582  & 0.000008 & 2.58  & 0.06 & 45.4  & $-f_{0}+f_{2}$       & $1.4\times10^{-5}$  & $(7.6 \pm 5.4) \times 10^{-6}$ & $(54 \pm 12) \times 10^{-5}$\\
F16 & 3.30630   & 0.00001  & 2.18  & 0.06 & 39.2  & $f_{0}+f_{1}-f_{2}$  & $-2.2\times10^{-5}$ & $-(12.0 \pm 6.0) \times 10^{-6}$& $(55 \pm 15) \times 10^{-5}$\\
F17 & 14.729543 & 0.000006 & 2.10  & 0.03 & 65.4  & $3f_{0}$             & $1.6\times10^{-5}$  & $-(5.0 \pm 14.8) \times 10^{-7}$& $(8 \pm 12) \times 10^{-5}$\\
F18 & 9.557459  & 0.000006 & 1.99  & 0.03 & 60.7  & $-f_{0}+f_{1}+f_{2}$ & $9.5\times10^{-7}$  & $(7.0 \pm 23.0) \times 10^{-7}$ & $(17 \pm 12) \times 10^{-5}$\\
F19 & 19.295654 & 0.000007 & 1.93  & 0.03 & 57.9  & $3f_{1}$             & $-1.5\times10^{-5}$ & $(3.6 \pm 12.0) \times 10^{-7}$ & $(2 \pm 12) \times 10^{-5}$\\
F20 & 24.205505 & 0.000006 & 1.87  & 0.03 & 59.5  & $f_{0}+3f_{1}$       & $-5.7\times10^{-6}$ & $(4.4 \pm 9.8) \times 10^{-7}$ & $(-4 \pm 12) \times 10^{-5}$\\
F21 & 21.161431 & 0.000009 & 1.37  & 0.03 & 43.7  & $3f_{0}+f_{1}$       & $1.3\times10^{-5}$  & $(0.4 \pm 15.4) \times 10^{-7}$ & $(2 \pm 12) \times 10^{-5}$\\
F22 & 3.04408   & 0.00001  & 1.38  & 0.05 & 25.8  & $-2f_{0}+2f_{1}$     & $-1.6\times10^{-5}$ & $-(2.4 \pm 1.2) \times 10^{-5}$& $(12 \pm 12) \times 10^{-5}$\\
\enddata
\end{deluxetable}

\begin{figure*}[htp]
  \centering
  \includegraphics[width=0.495\textwidth]{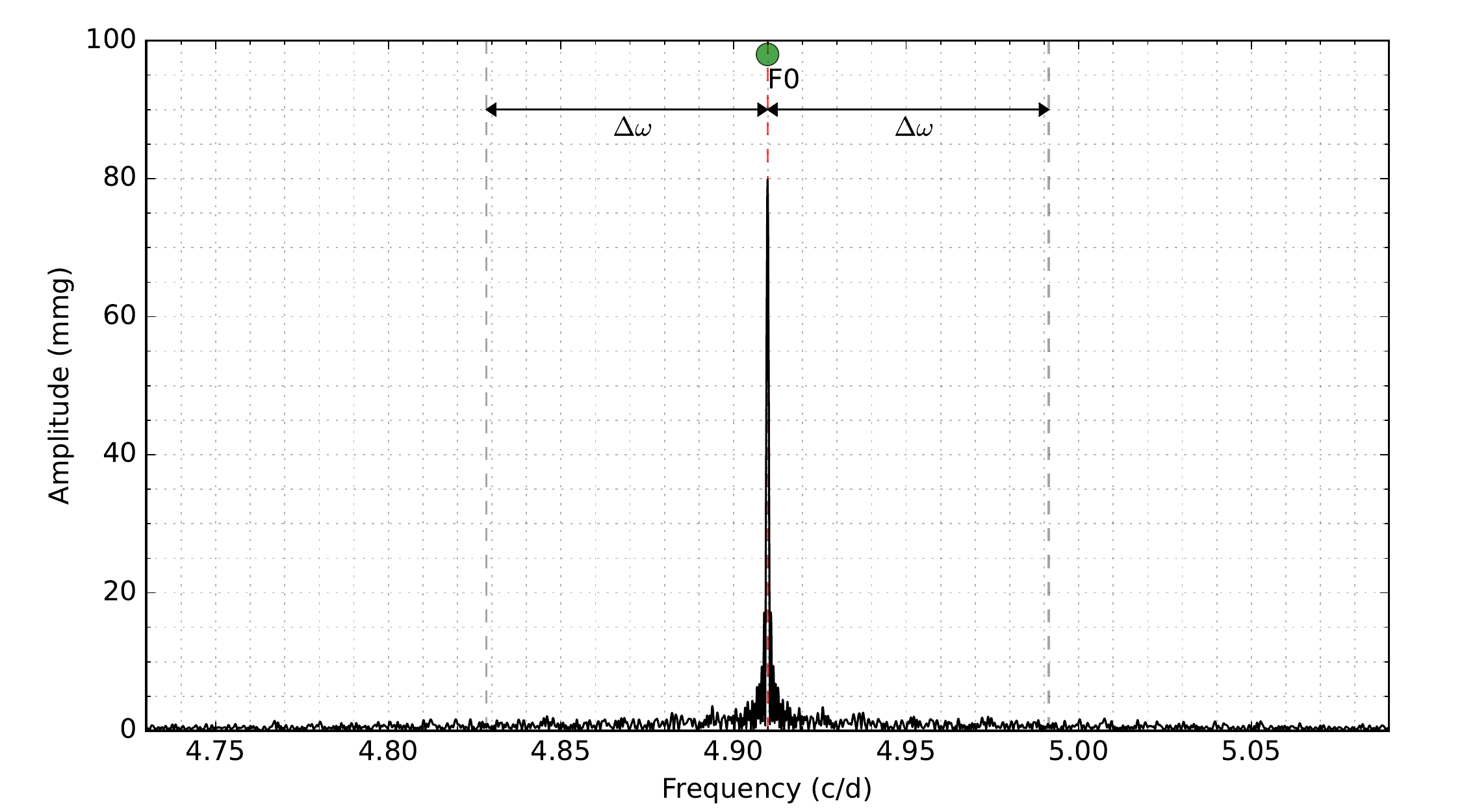}
  \includegraphics[width=0.495\textwidth]{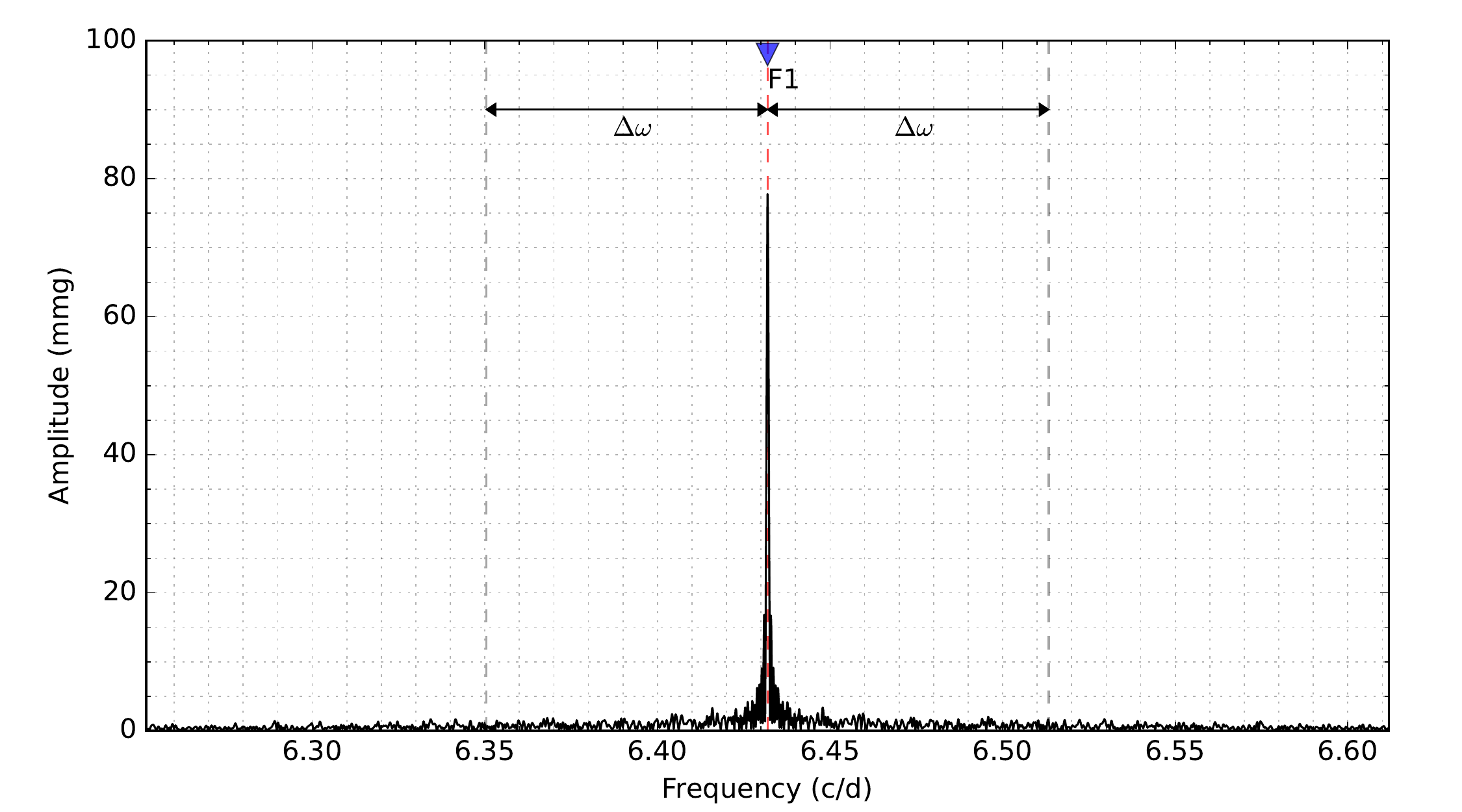}
  \includegraphics[width=0.495\textwidth]{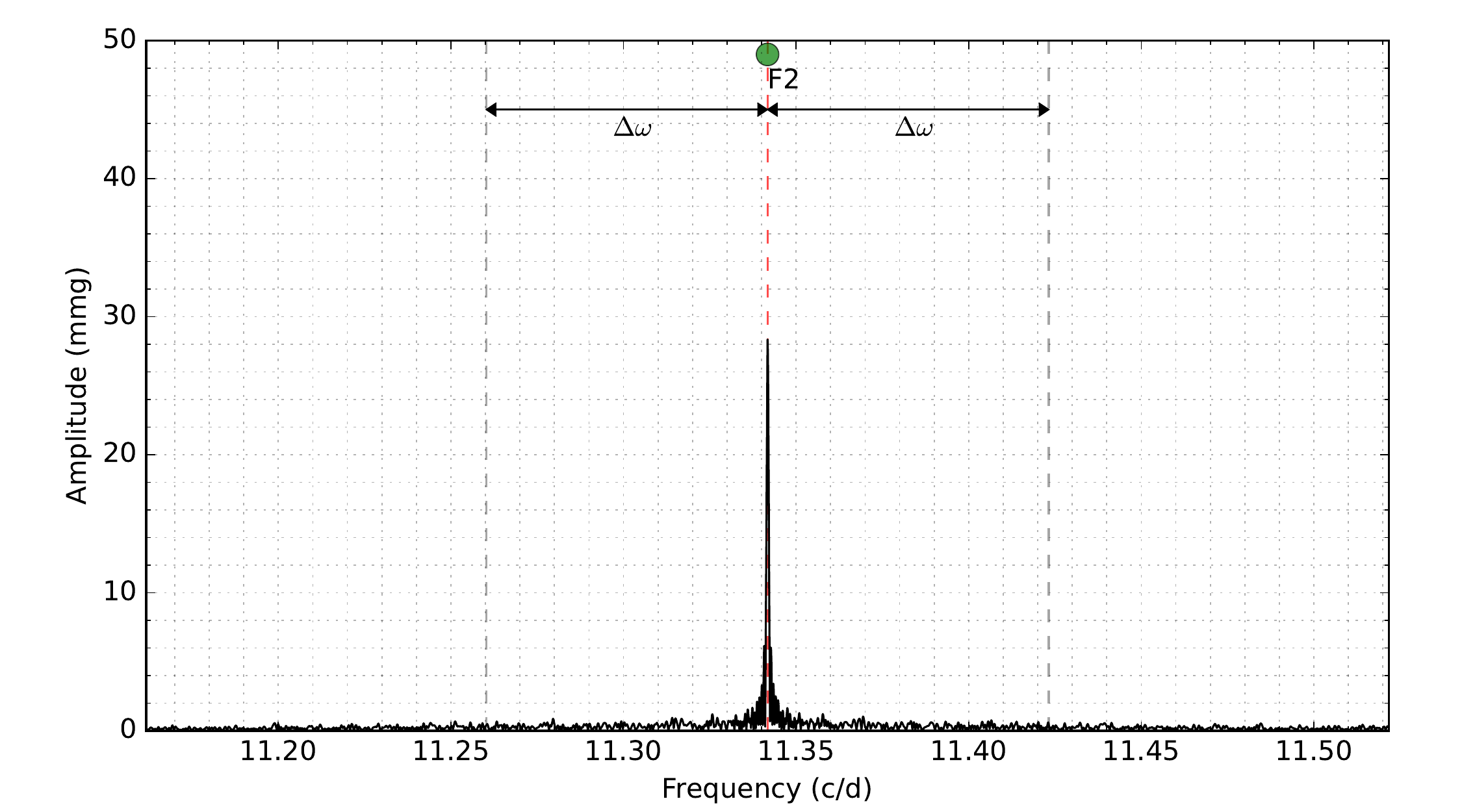}
  \includegraphics[width=0.495\textwidth]{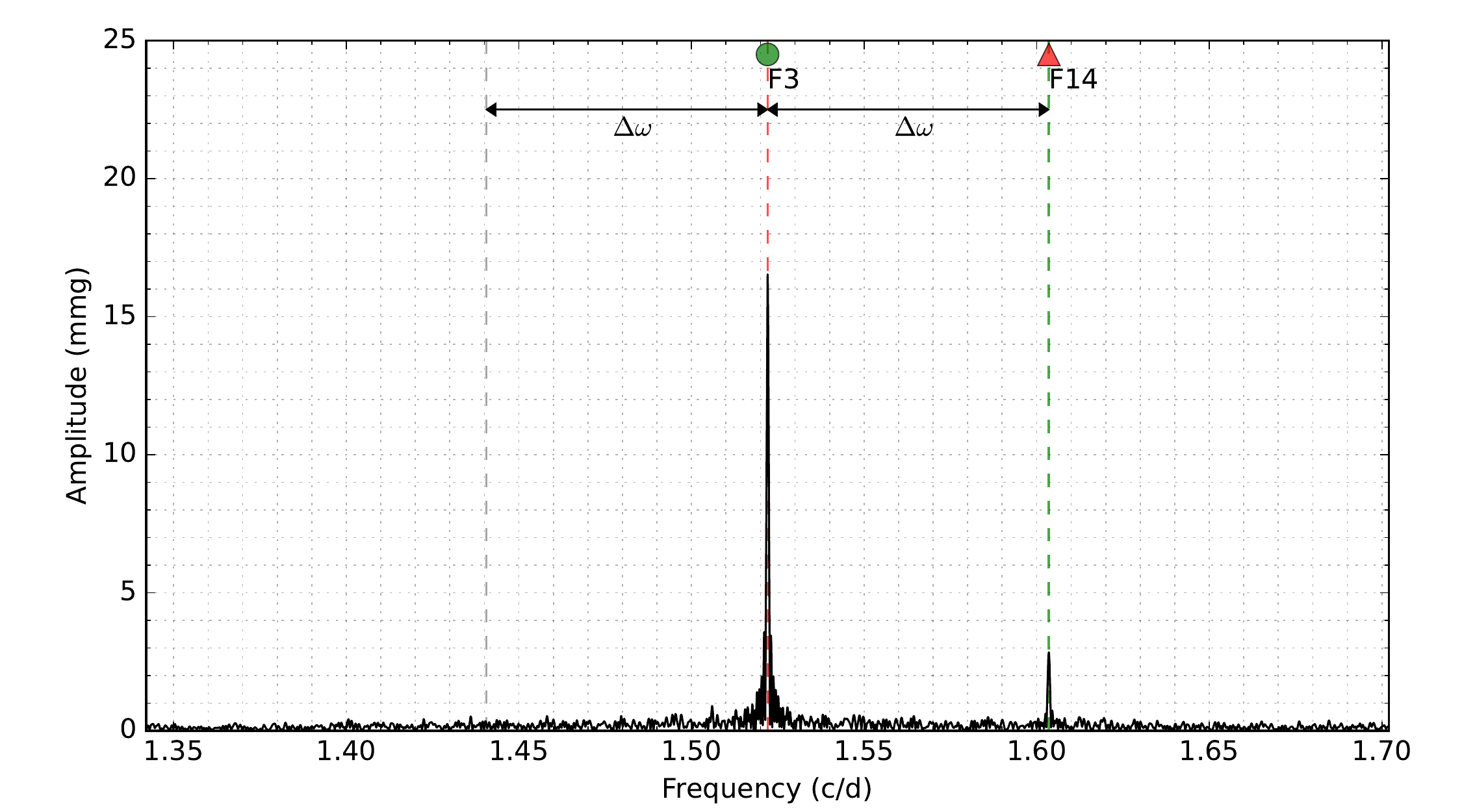}
  \includegraphics[width=0.495\textwidth]{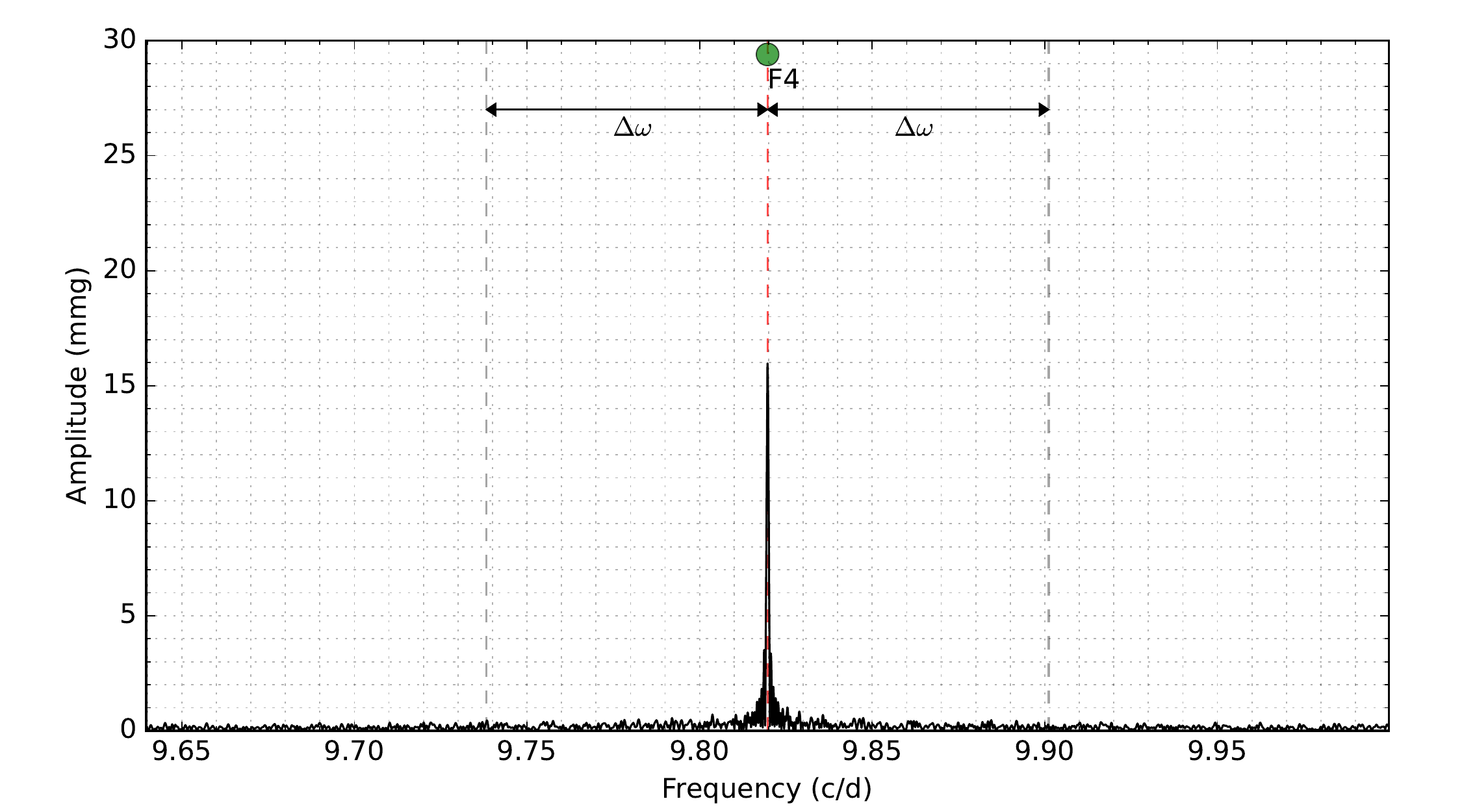}
  \includegraphics[width=0.495\textwidth]{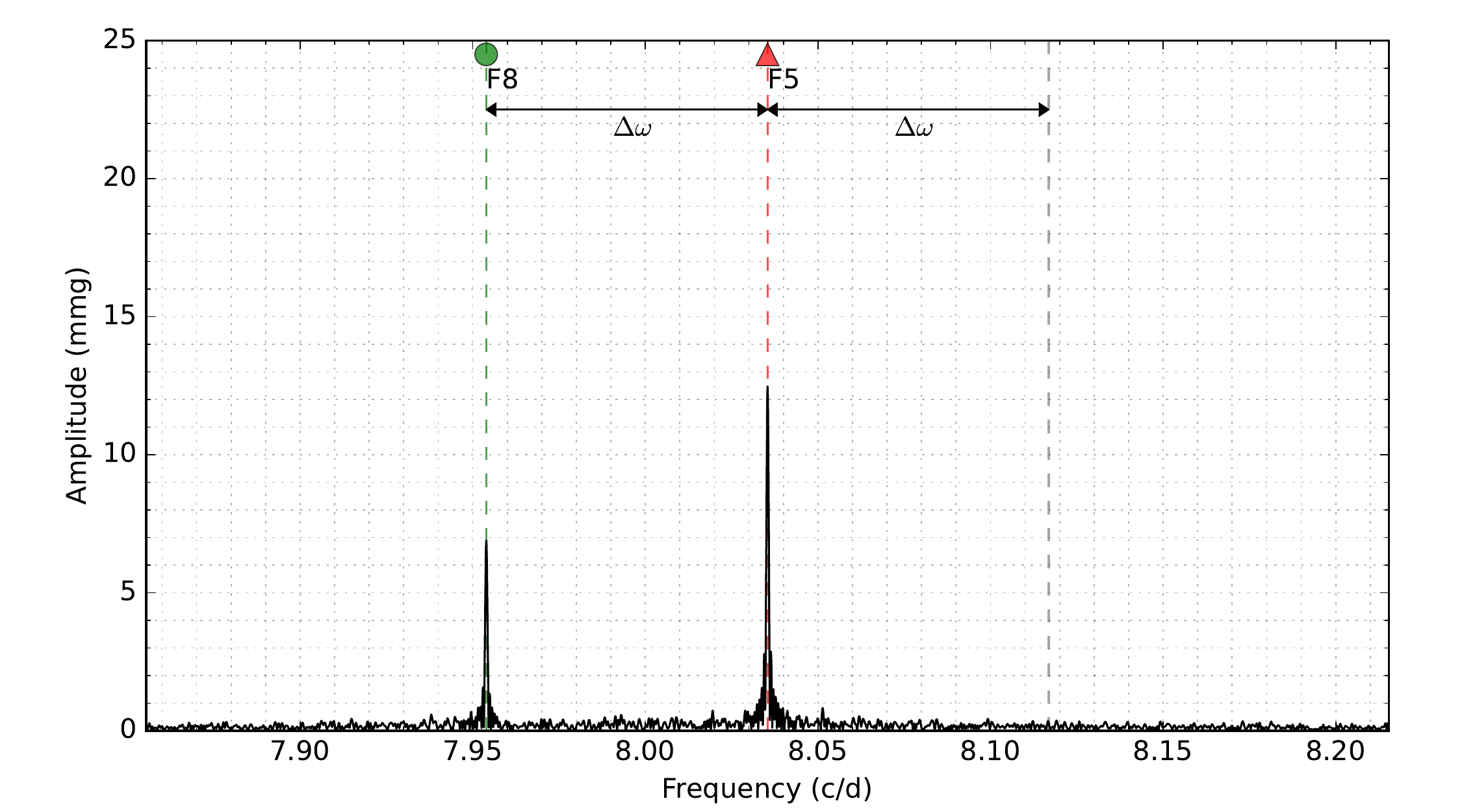}
  \caption{Zoom-in of the 23 pulsation modes in the frequency domain, Part I. The vertical dash lines mark the position of the 23 frequencies and that are $\Delta \omega = 0.0815$ away from them on both sides. Gray indicates there is no obvious peaks in that position; red indicates the pulsation mode in that position shows a positive value of $(1/P)(\diff P/ \diff t)$; blue indicates the pulsation mode in that position shows a negative value of $(1/P)(\diff P/ \diff t)$; green indicates the pulsation mode in that position shows an uncertain sign of $(1/P)(\diff P/ \diff t)$. The blue circle on the top indicates the amplitude of the pulsation mode is uncertain; the upward-pointing red triangle indicates the amplitude of the pulsation mode is increasing; the downward-pointing blue triangle indicates the amplitude of the pulsation mode is decreasing.}
  \label{fig:spec01}
\end{figure*}

\begin{figure*}[htp]
  \centering
  \includegraphics[width=0.495\textwidth]{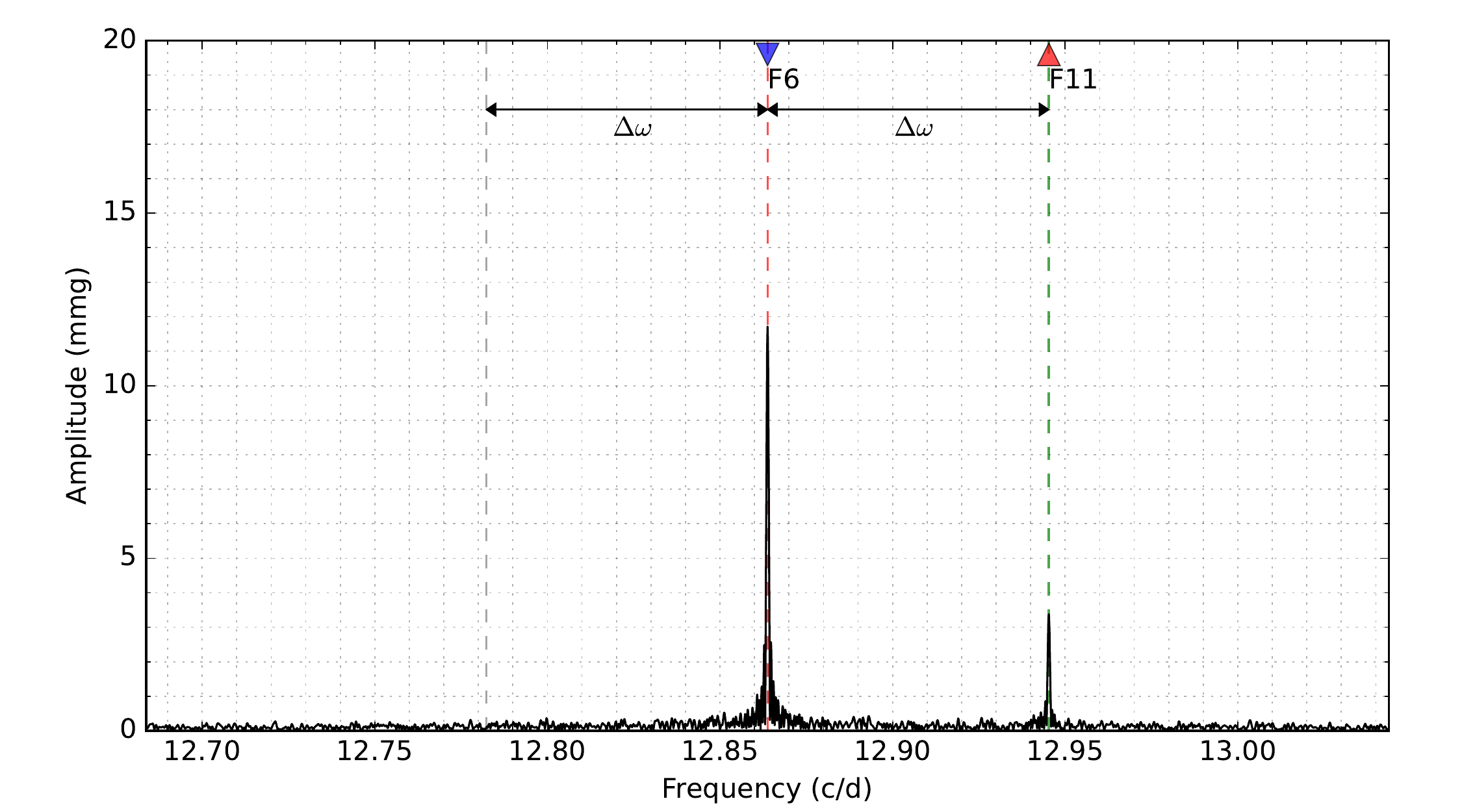}
  \includegraphics[width=0.495\textwidth]{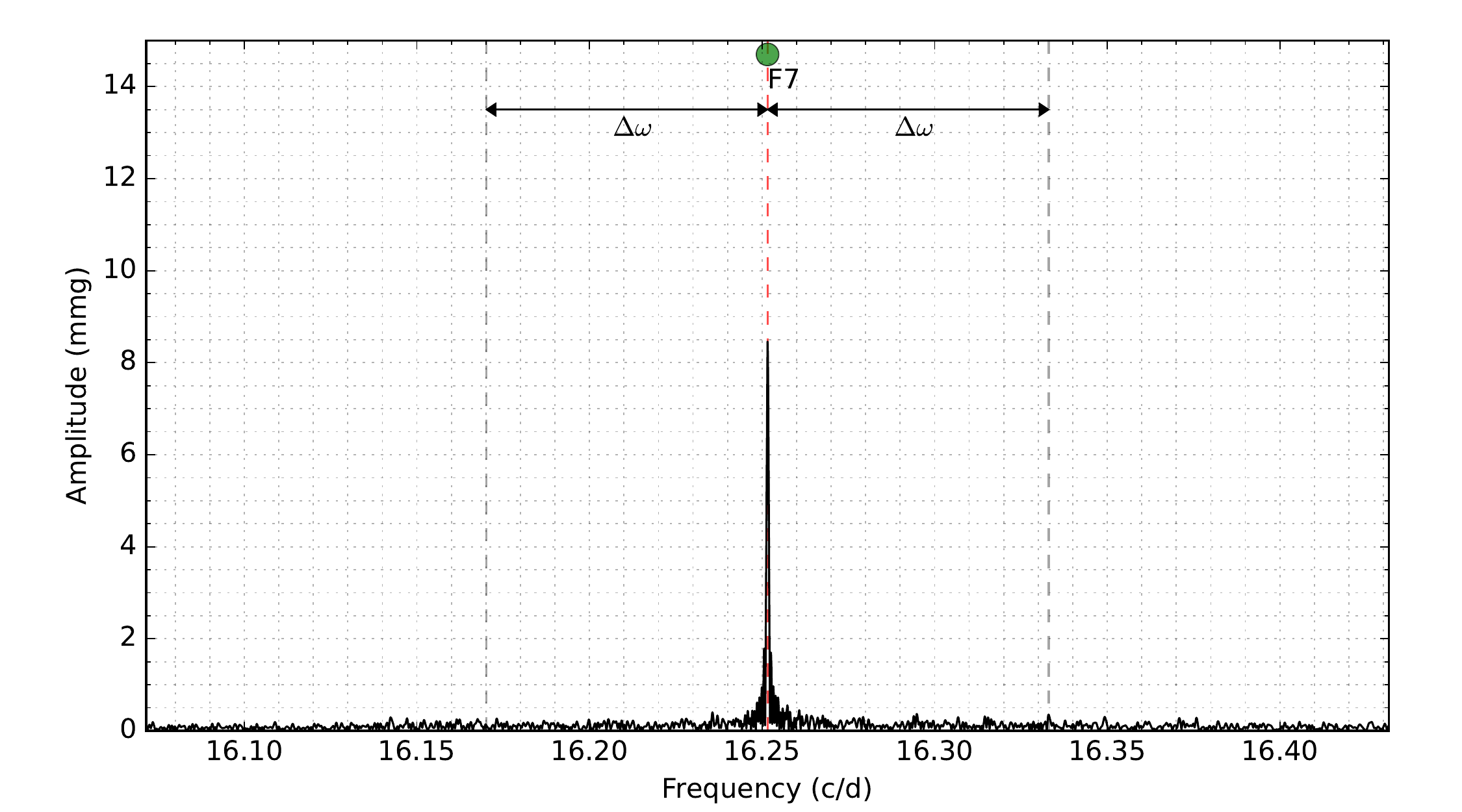}
  \includegraphics[width=0.495\textwidth]{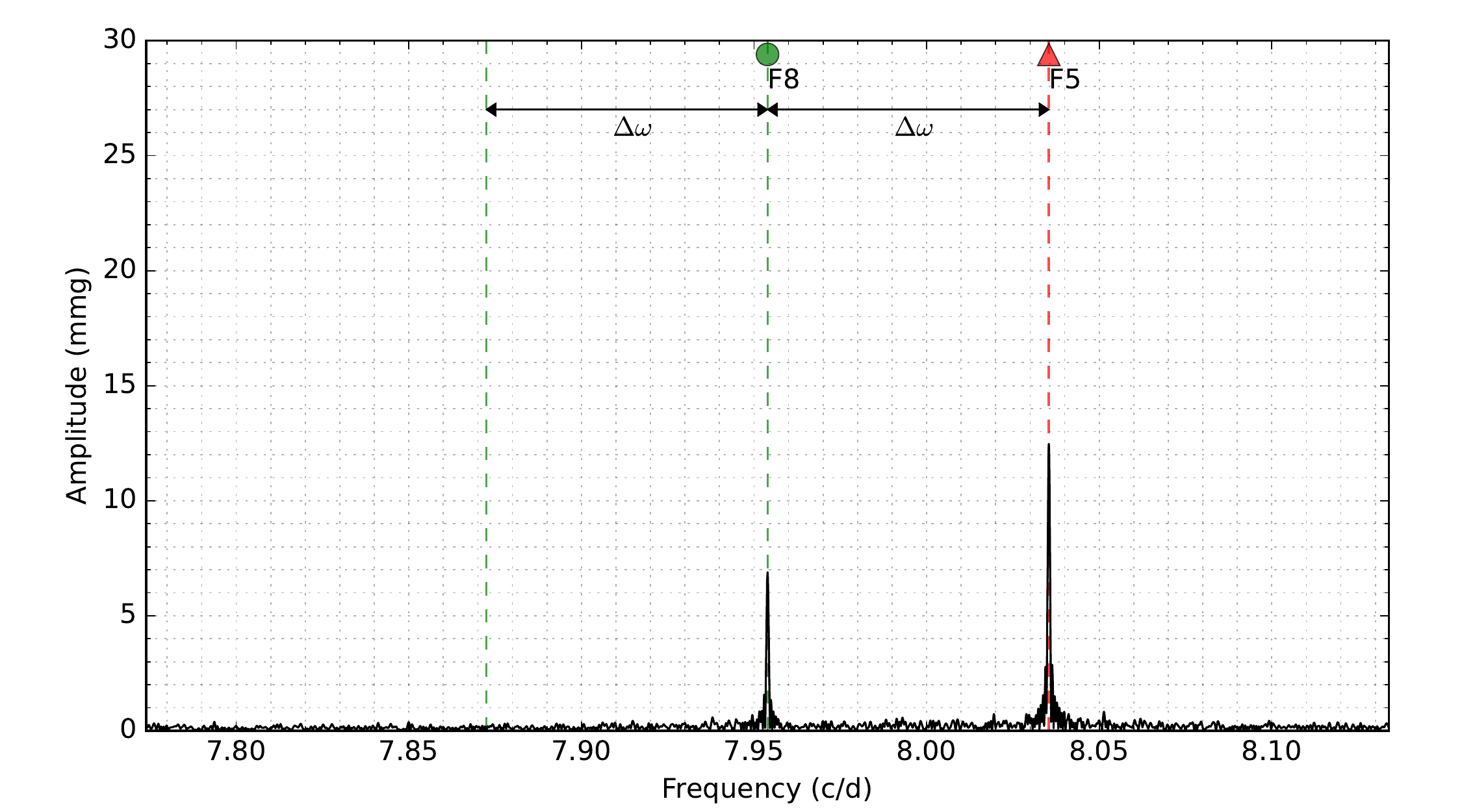}
  \includegraphics[width=0.495\textwidth]{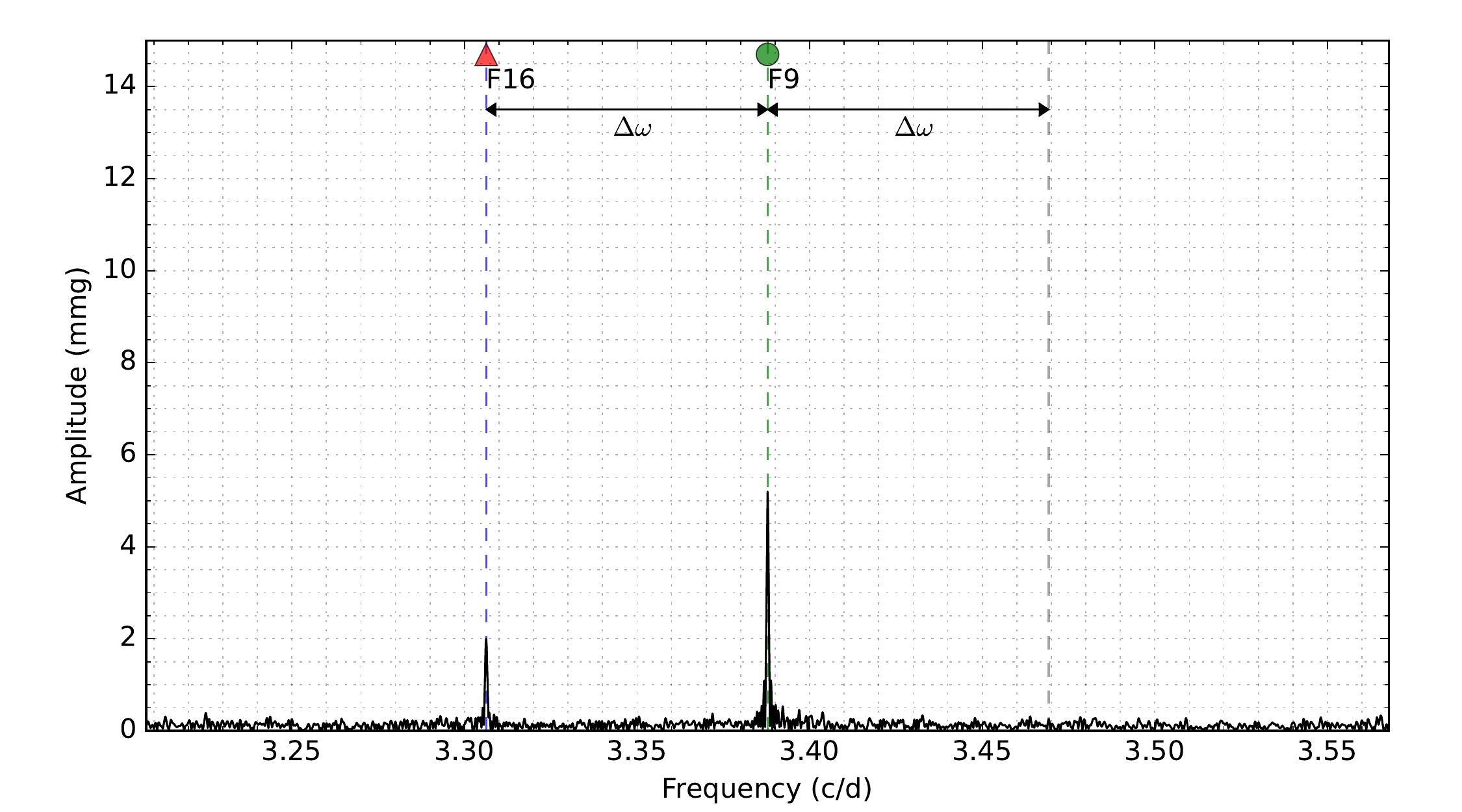}
  \includegraphics[width=0.495\textwidth]{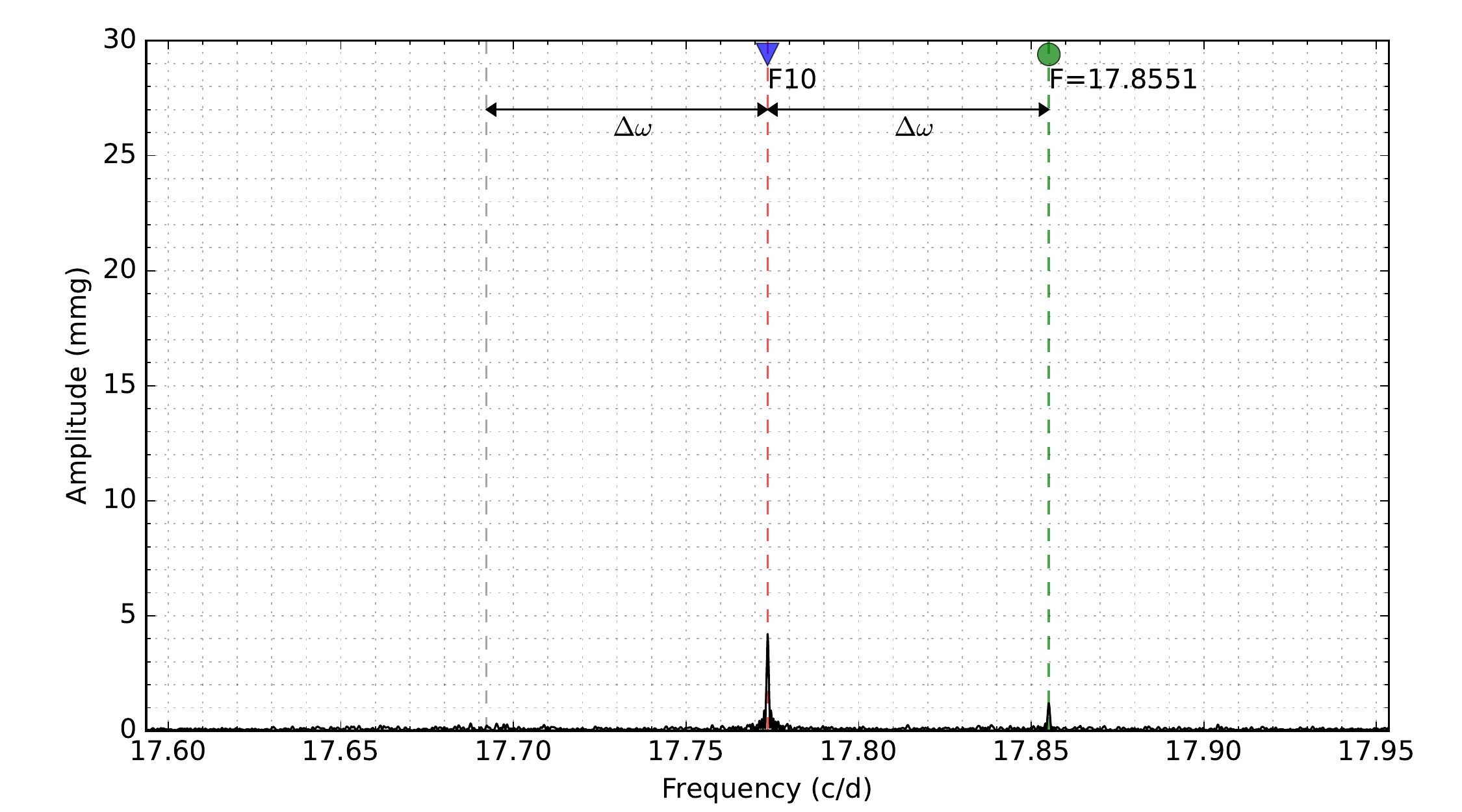}
  \includegraphics[width=0.495\textwidth]{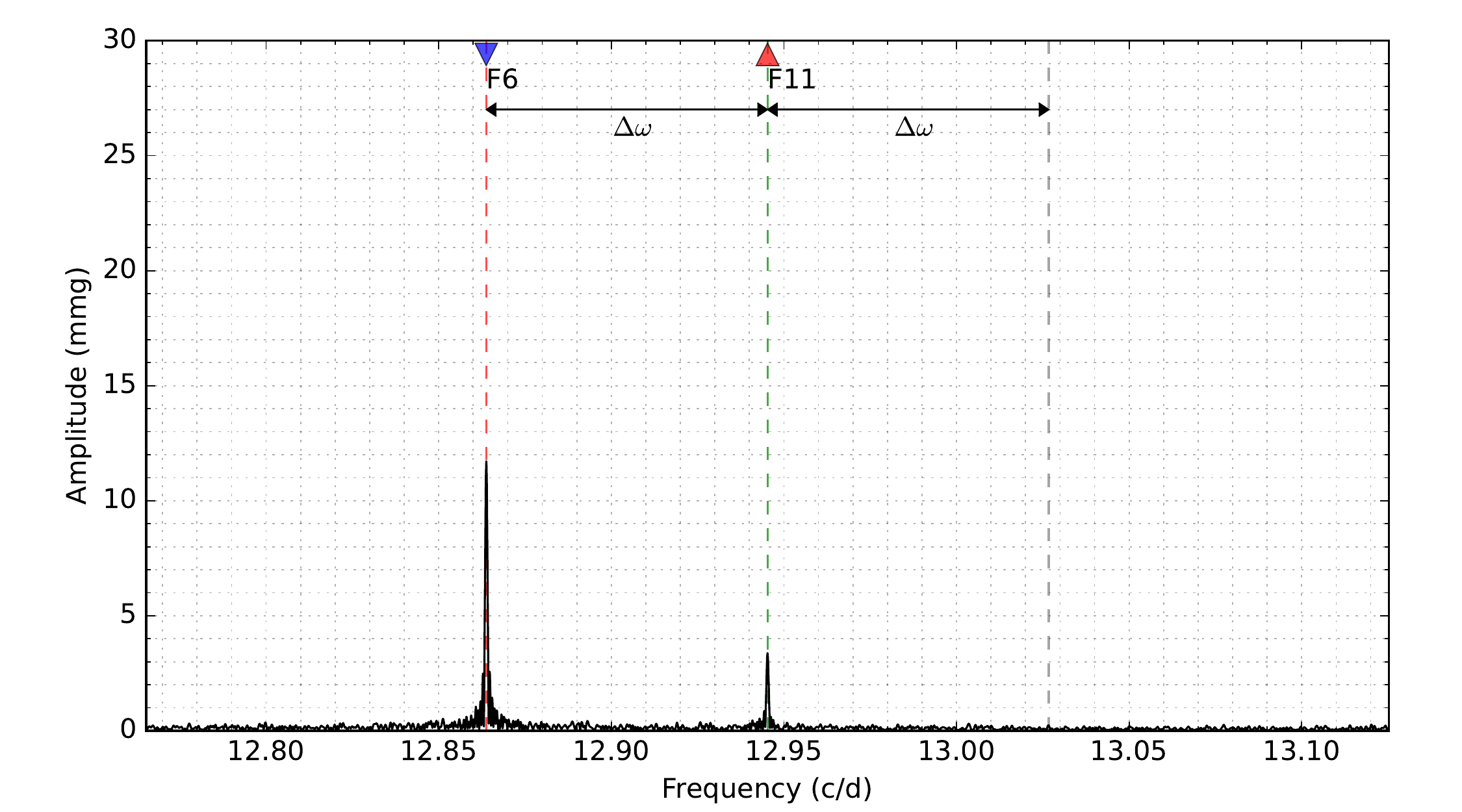}
  \includegraphics[width=0.495\textwidth]{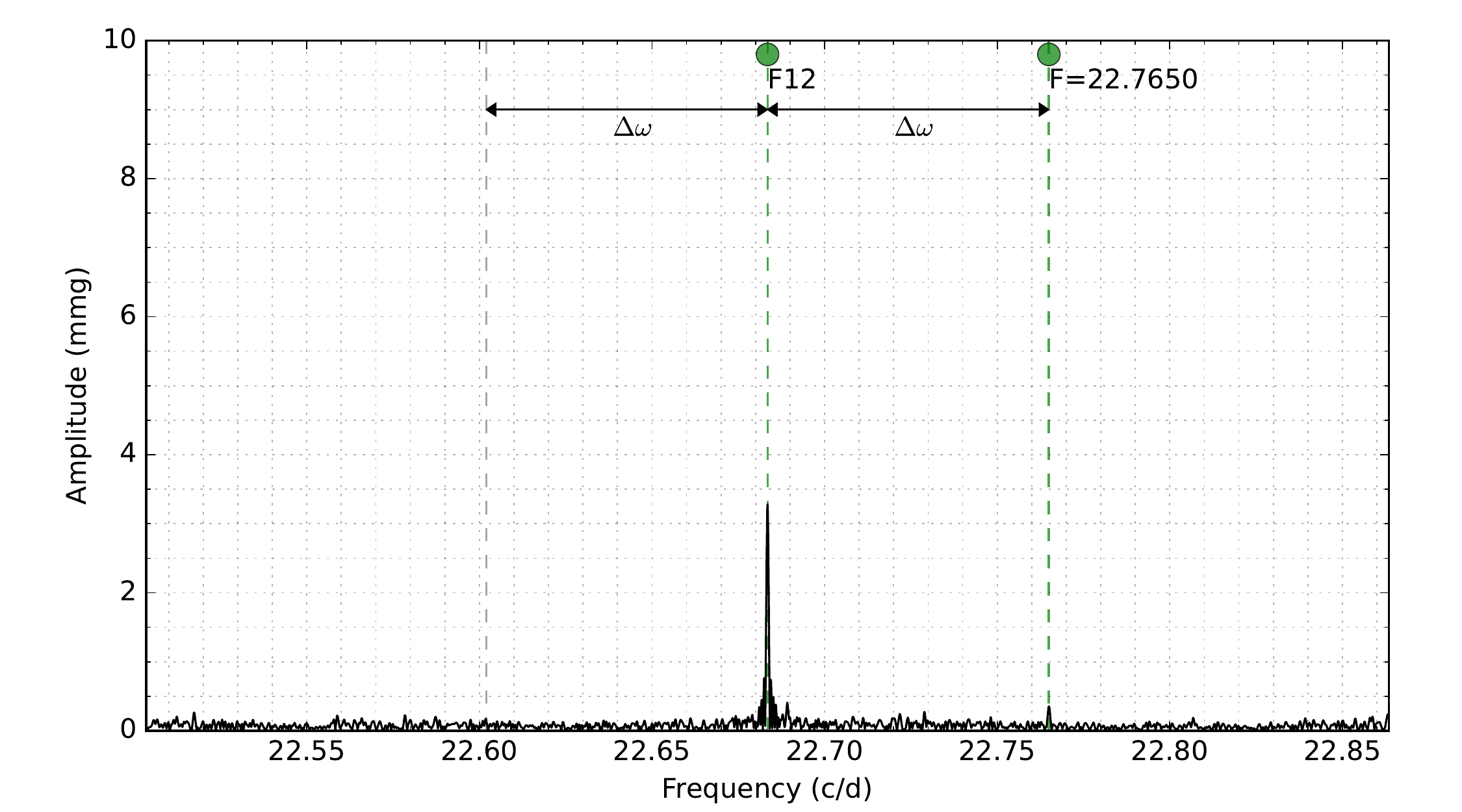}
  \includegraphics[width=0.495\textwidth]{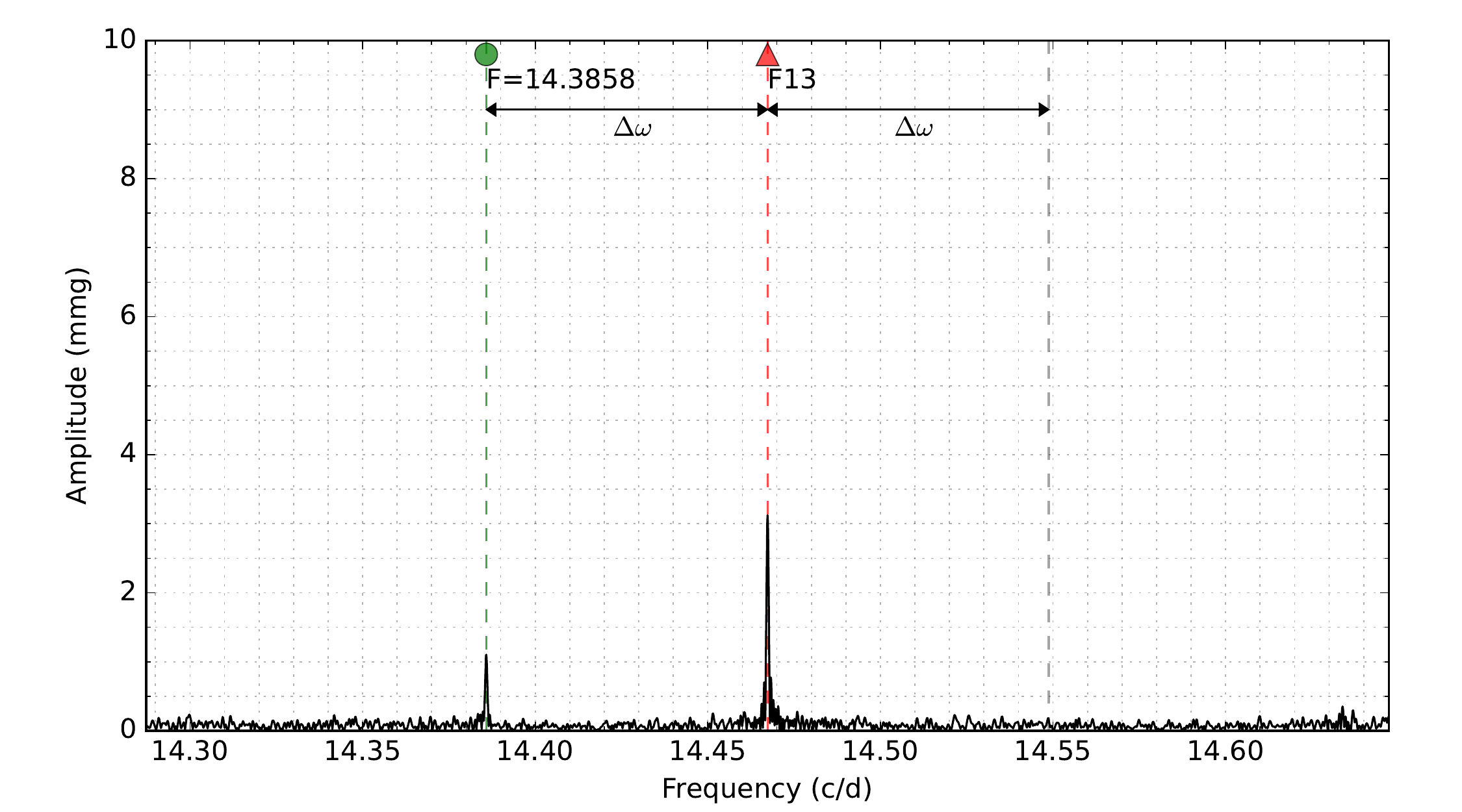}
  \caption{Zoom-in of the 23 pulsation modes in the frequency domain, Part II.}
  \label{fig:spec02}
\end{figure*}

\begin{figure*}[htp]
  \centering
  \includegraphics[width=0.495\textwidth]{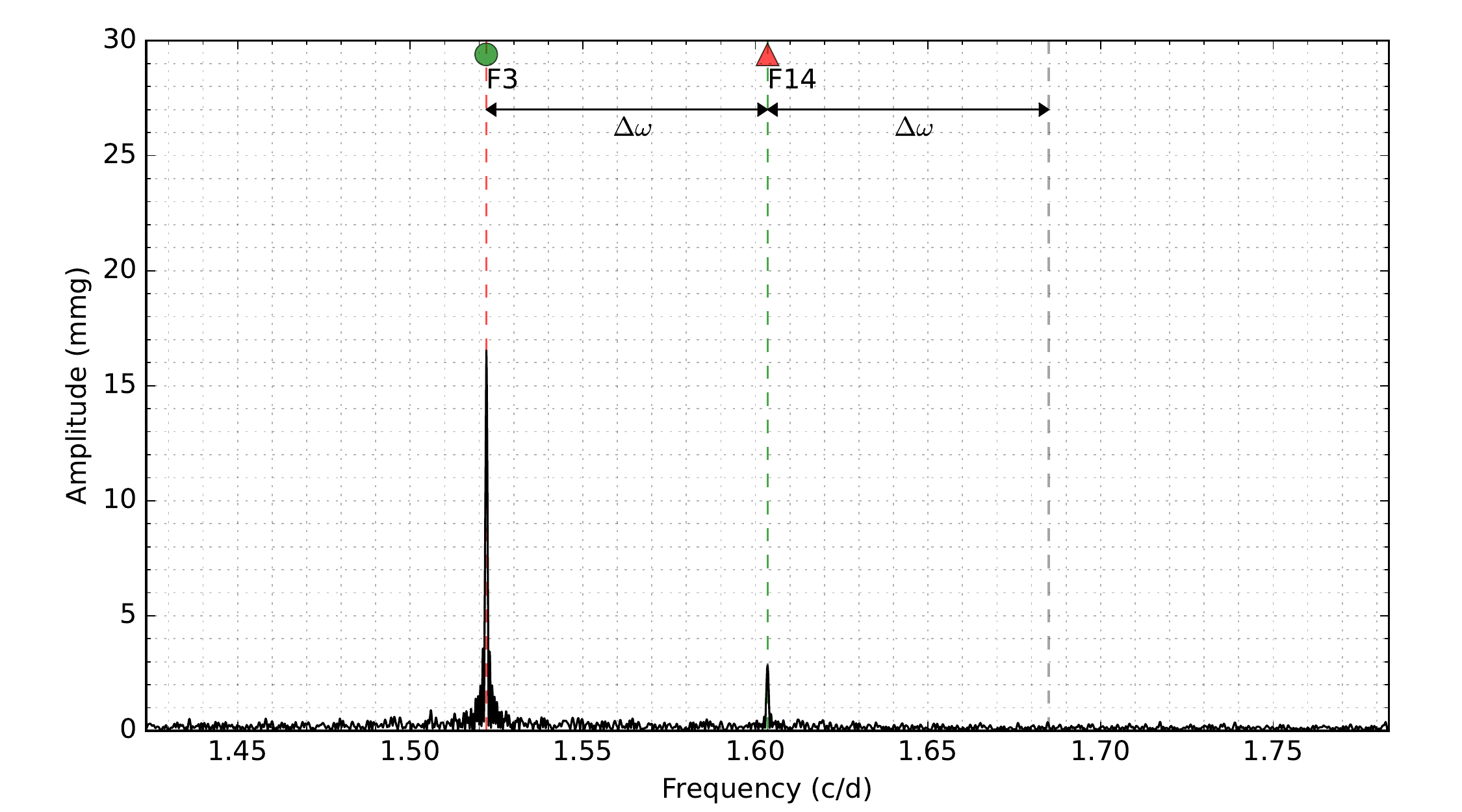}
  \includegraphics[width=0.495\textwidth]{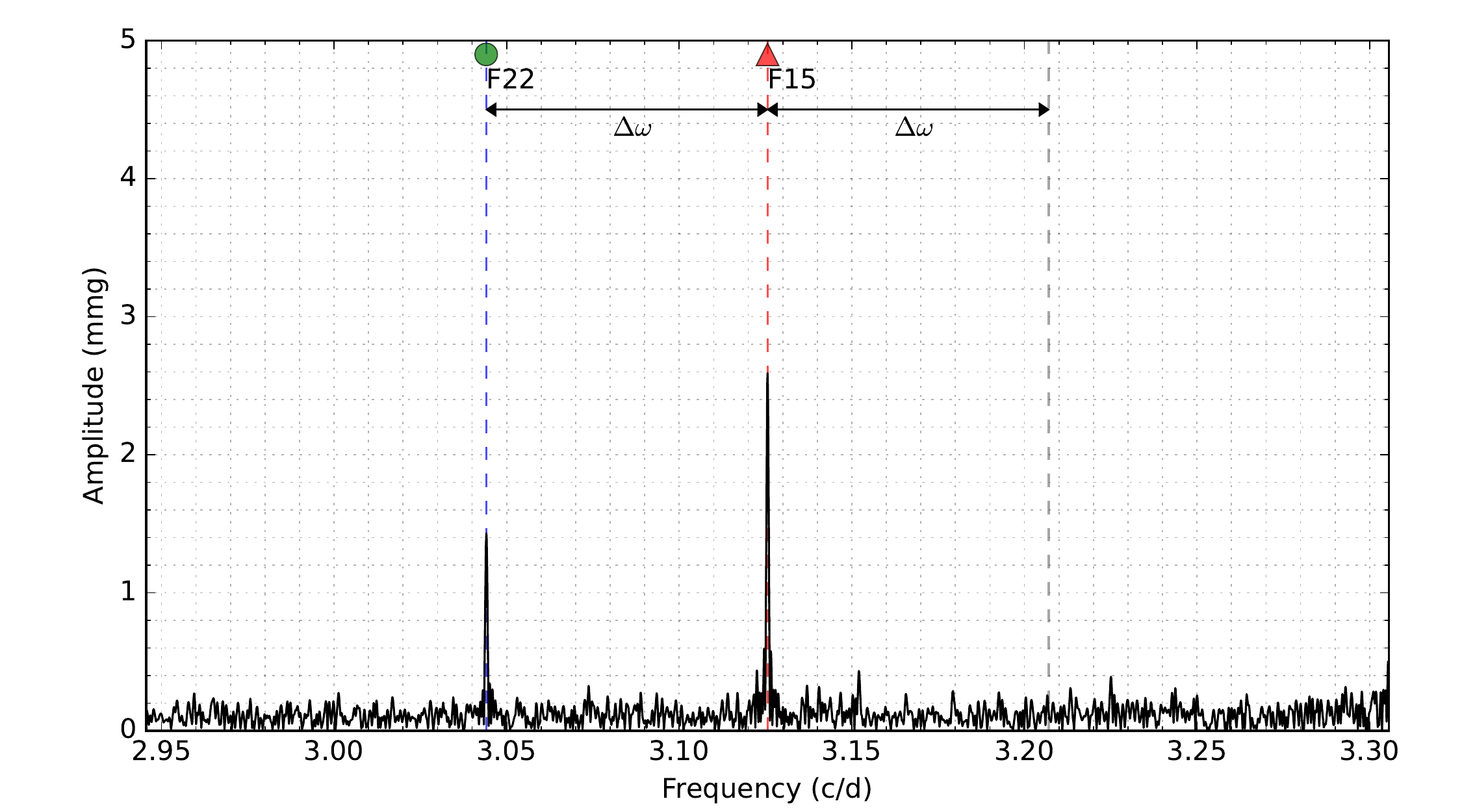}
  \includegraphics[width=0.495\textwidth]{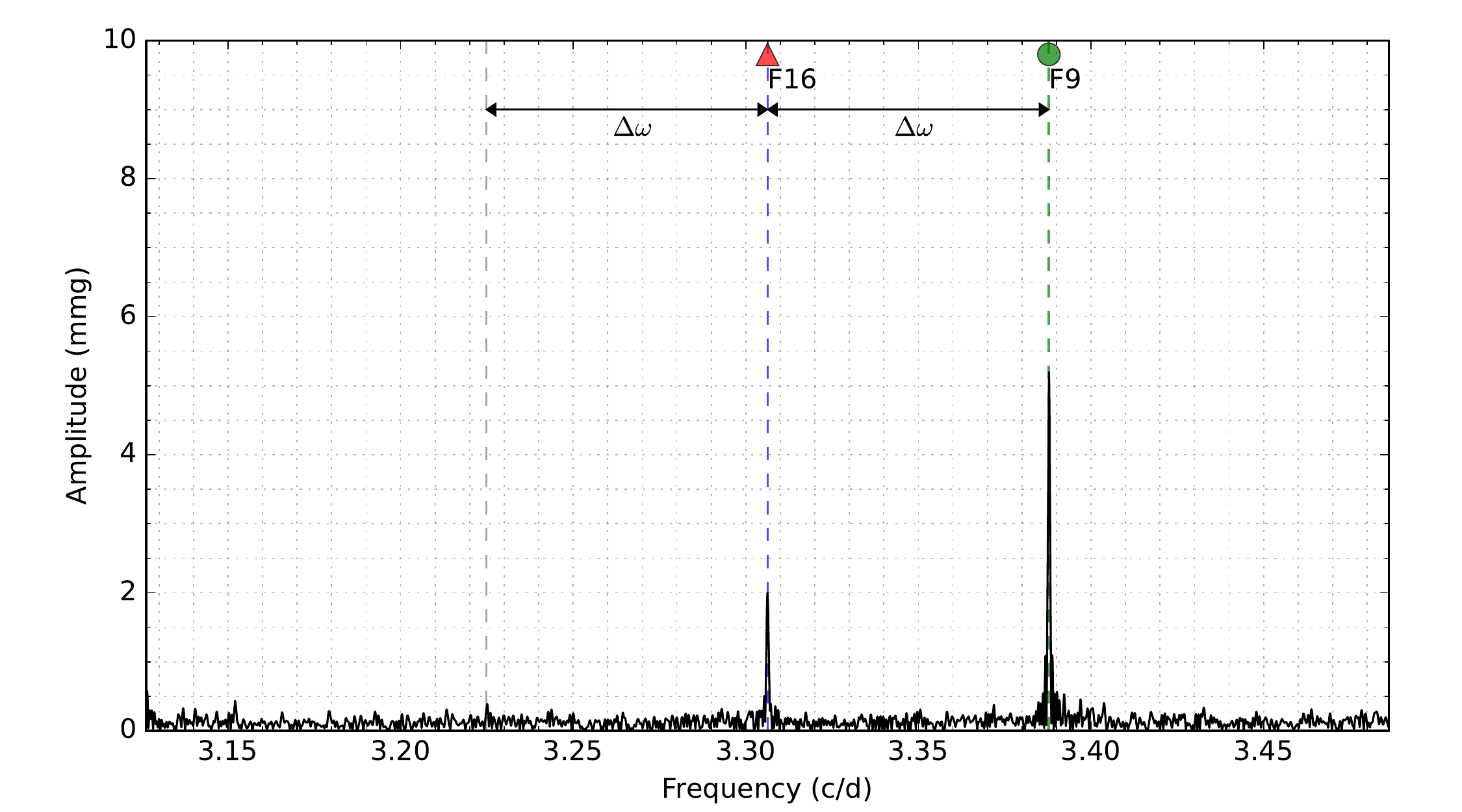}
  \includegraphics[width=0.495\textwidth]{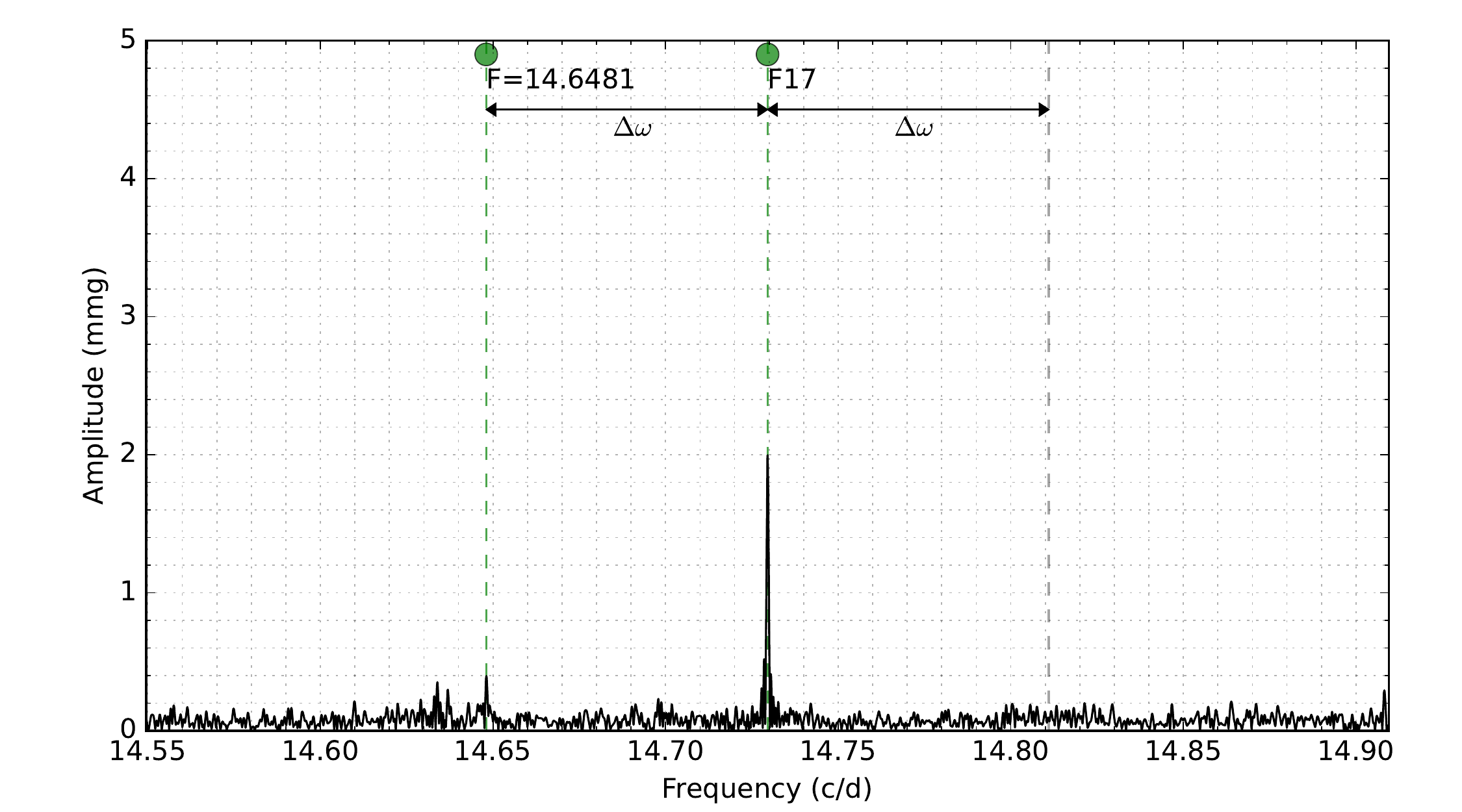}
  \includegraphics[width=0.495\textwidth]{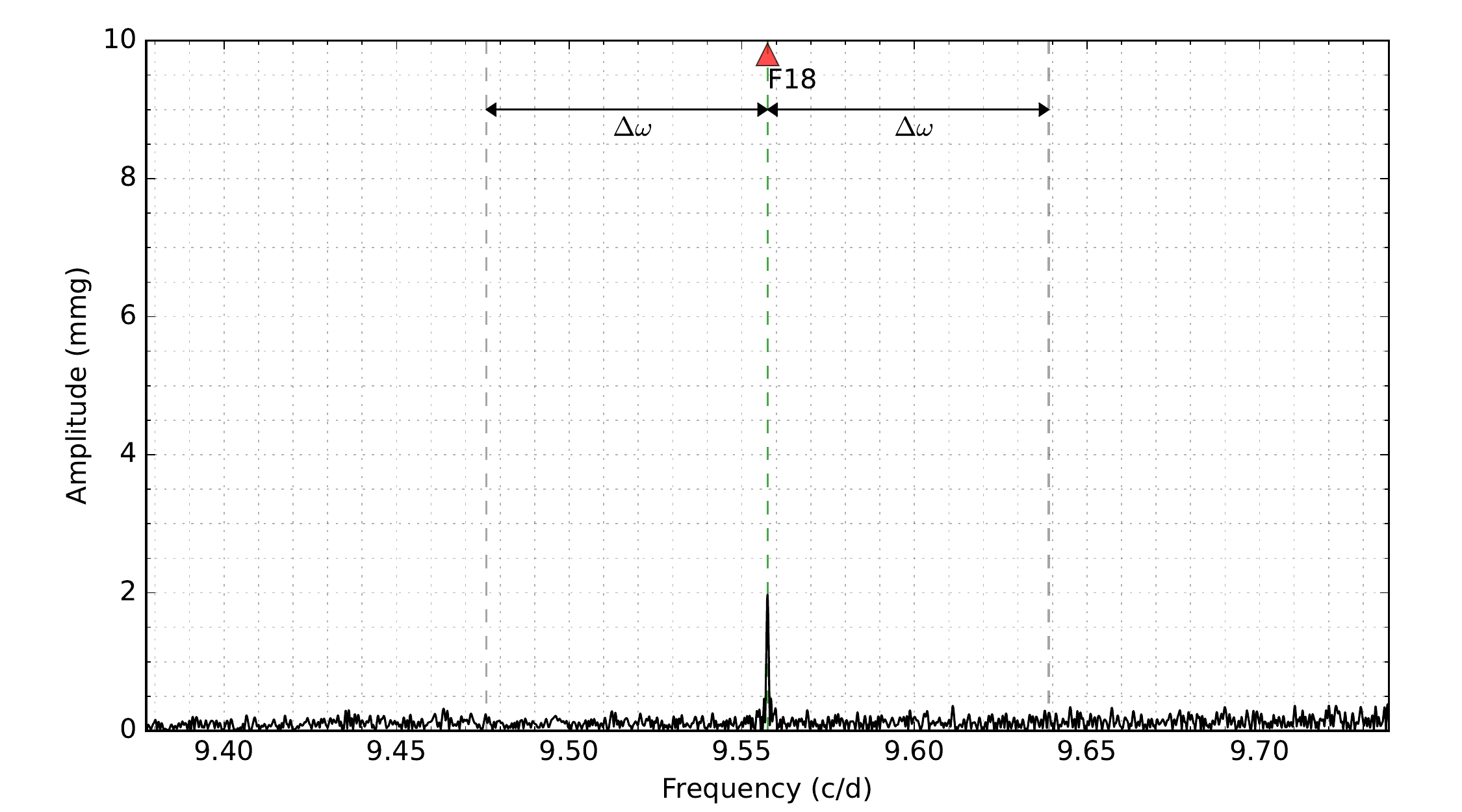}
  \includegraphics[width=0.495\textwidth]{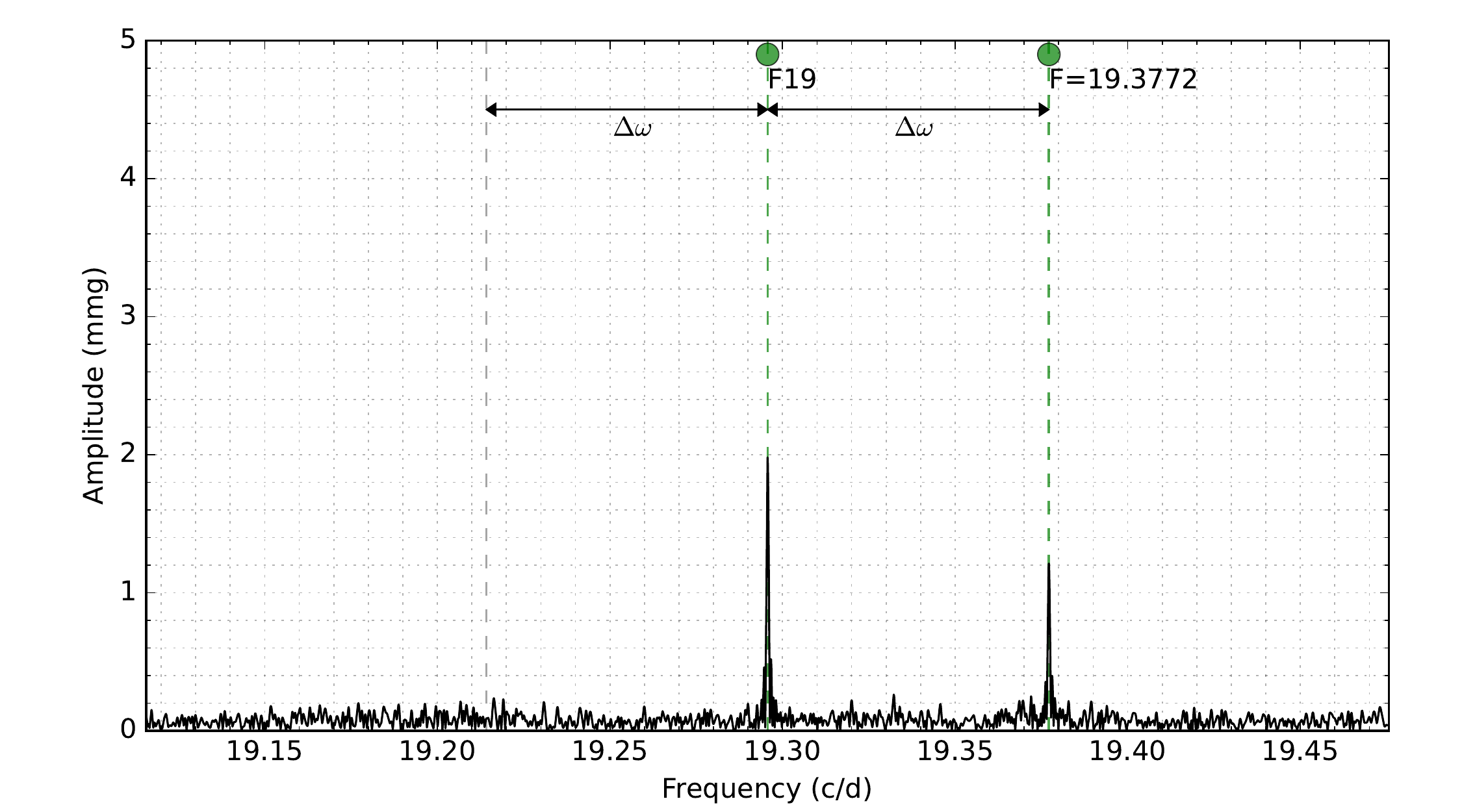}
  \includegraphics[width=0.495\textwidth]{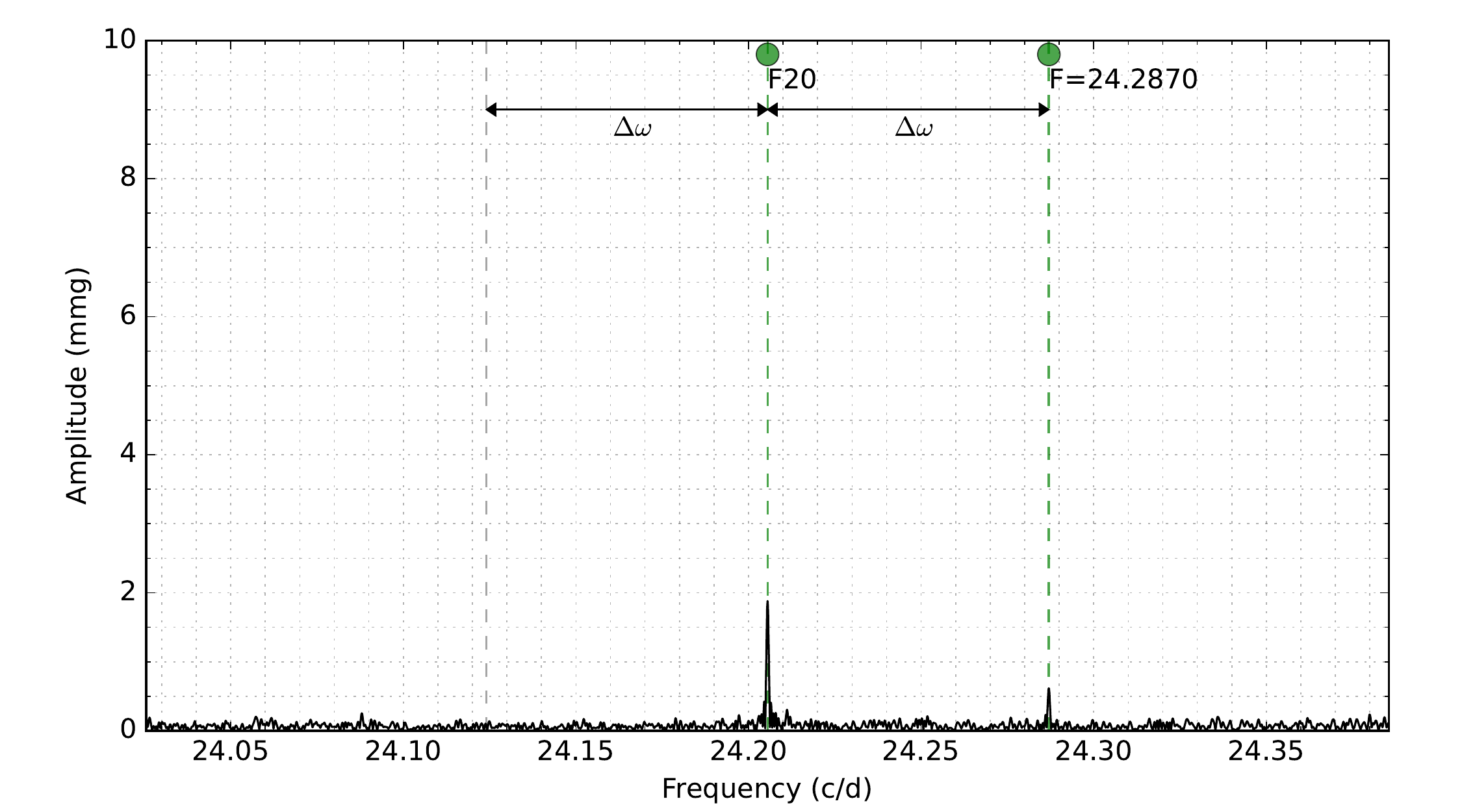}
  \includegraphics[width=0.495\textwidth]{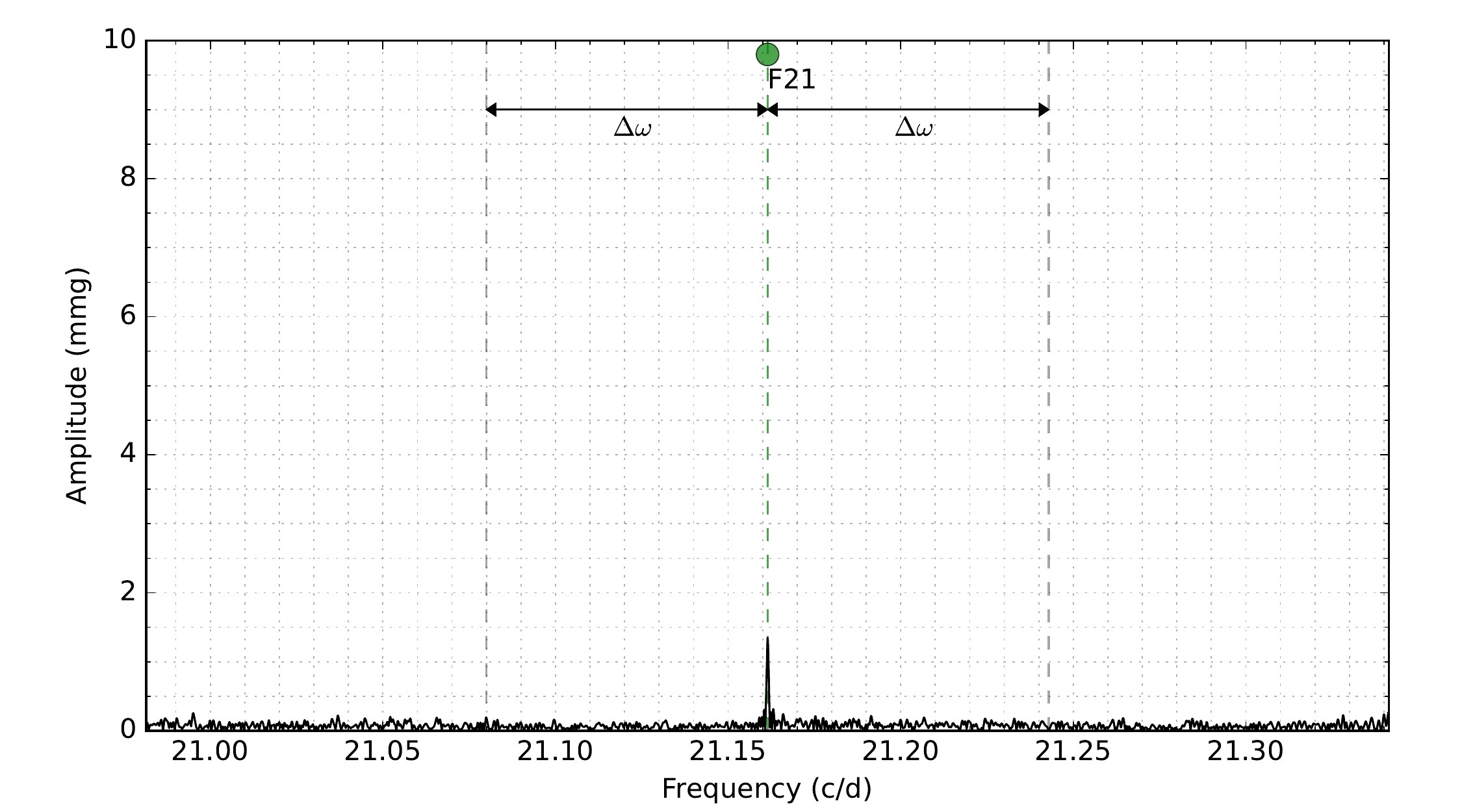}
  \caption{Zoom-in of the 23 pulsation modes in the frequency domain, Part III.}
  \label{fig:spec03}
\end{figure*}

\begin{figure*}[htp]
  \centering
  \includegraphics[width=0.495\textwidth]{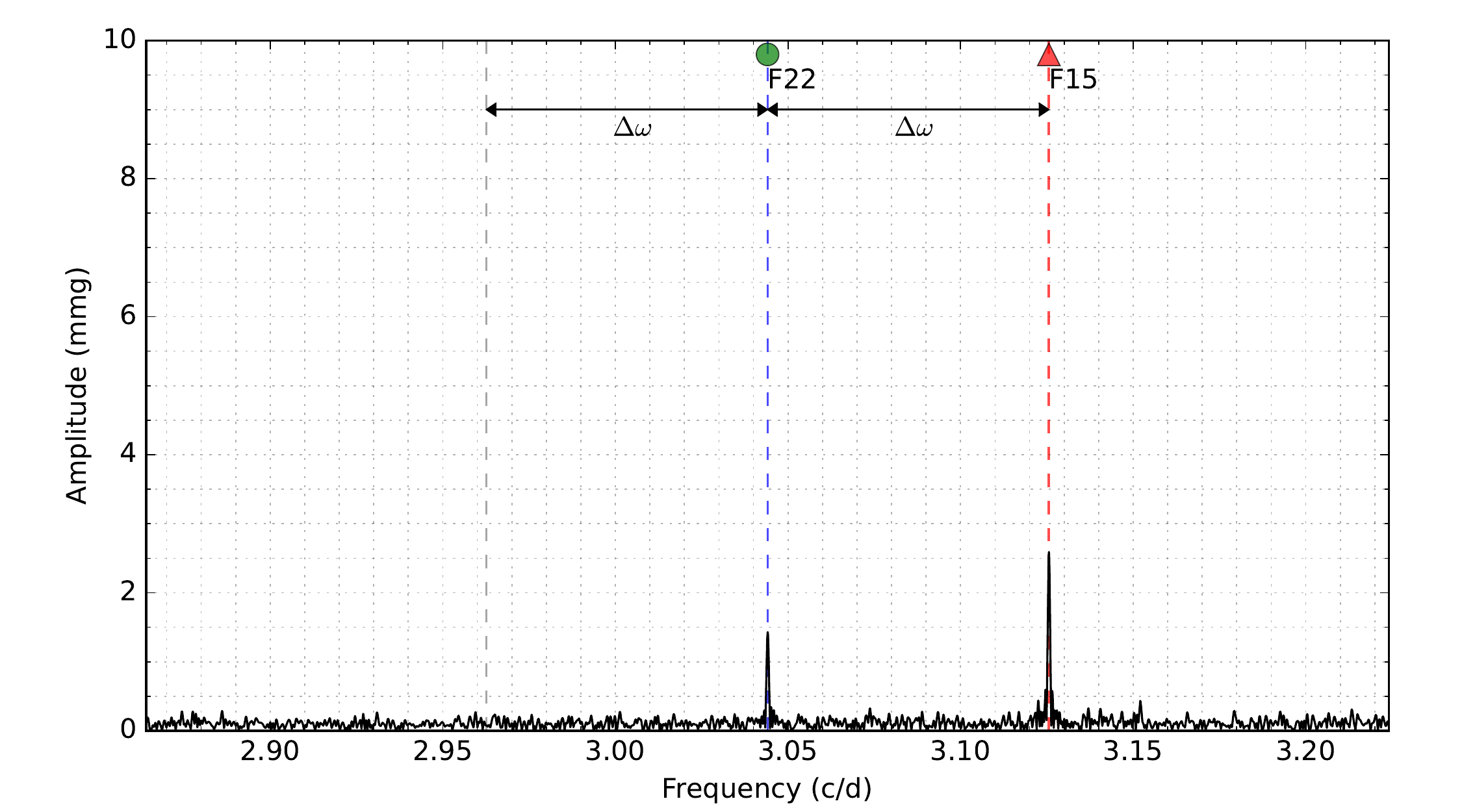}
  \caption{Zoom-in of the 23 pulsation modes in the frequency domain, Part IV.}
  \label{fig:spec04}
\end{figure*}

In order to show the detailed structures in the 23 pulsation modes in frequency domain, a zoom-in of them is shown in Figure \ref{fig:spec01}, \ref{fig:spec02}, \ref{fig:spec03}, and \ref{fig:spec04}.

The short-time Fourier transformation \citep{Bowman2016,Zong2016,Zong2018} was then performed to the normalized LC data to get the variation in the amplitudes and phases. In this process, a time window of 120 days was moving from the start to the end time of the LC data, with a step of 20 days. In each step, the prewhitening process was performed to extract the amplitudes and phases of the specific 23 pulsation modes, while the frequencies were fixed as the values obtained in the complete LC data in Table \ref{tab:freq_solution}.

\begin{figure*}[htp]
  \centering
  \includegraphics[width=0.495\textwidth]{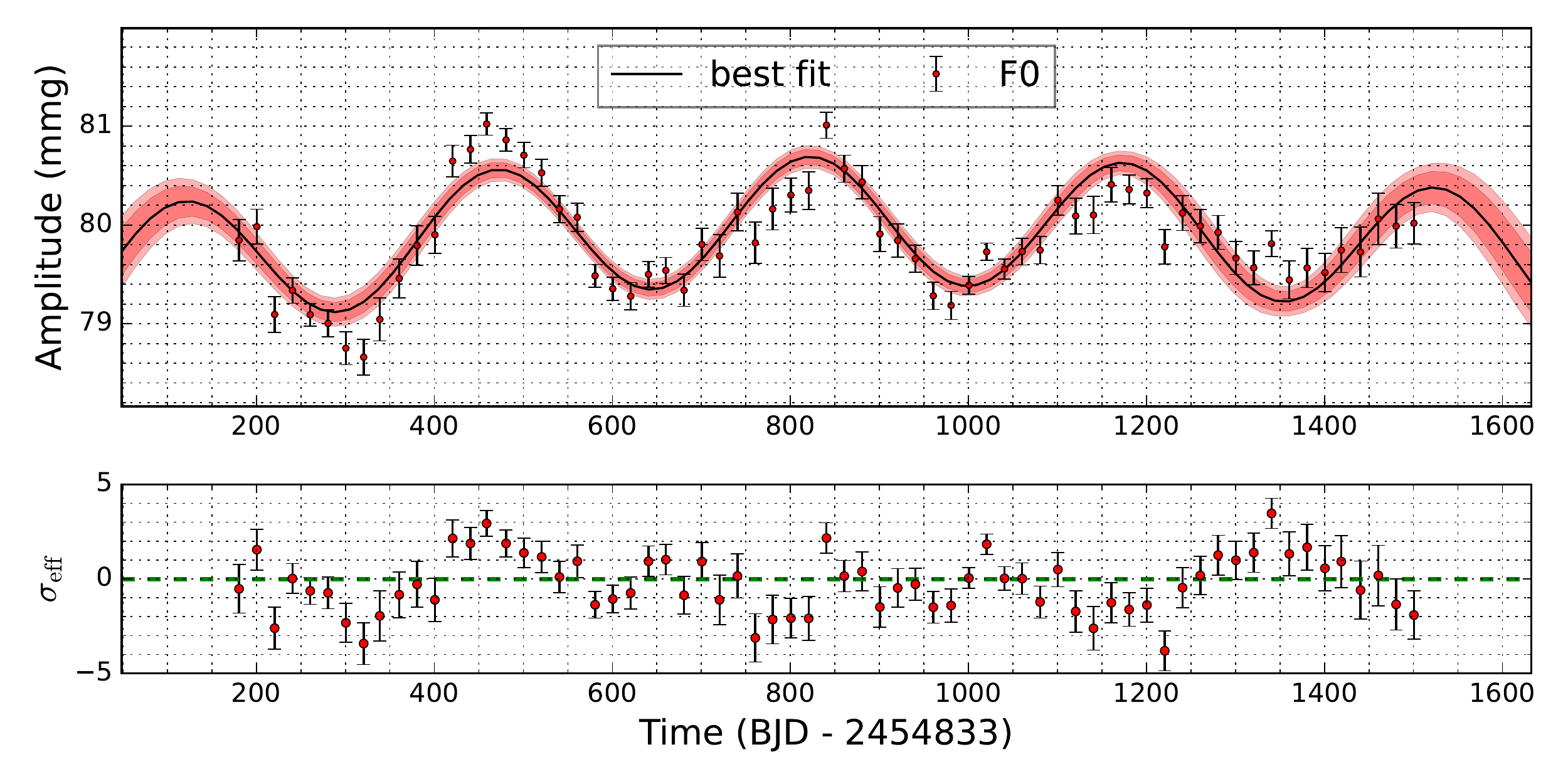}
  \includegraphics[width=0.495\textwidth]{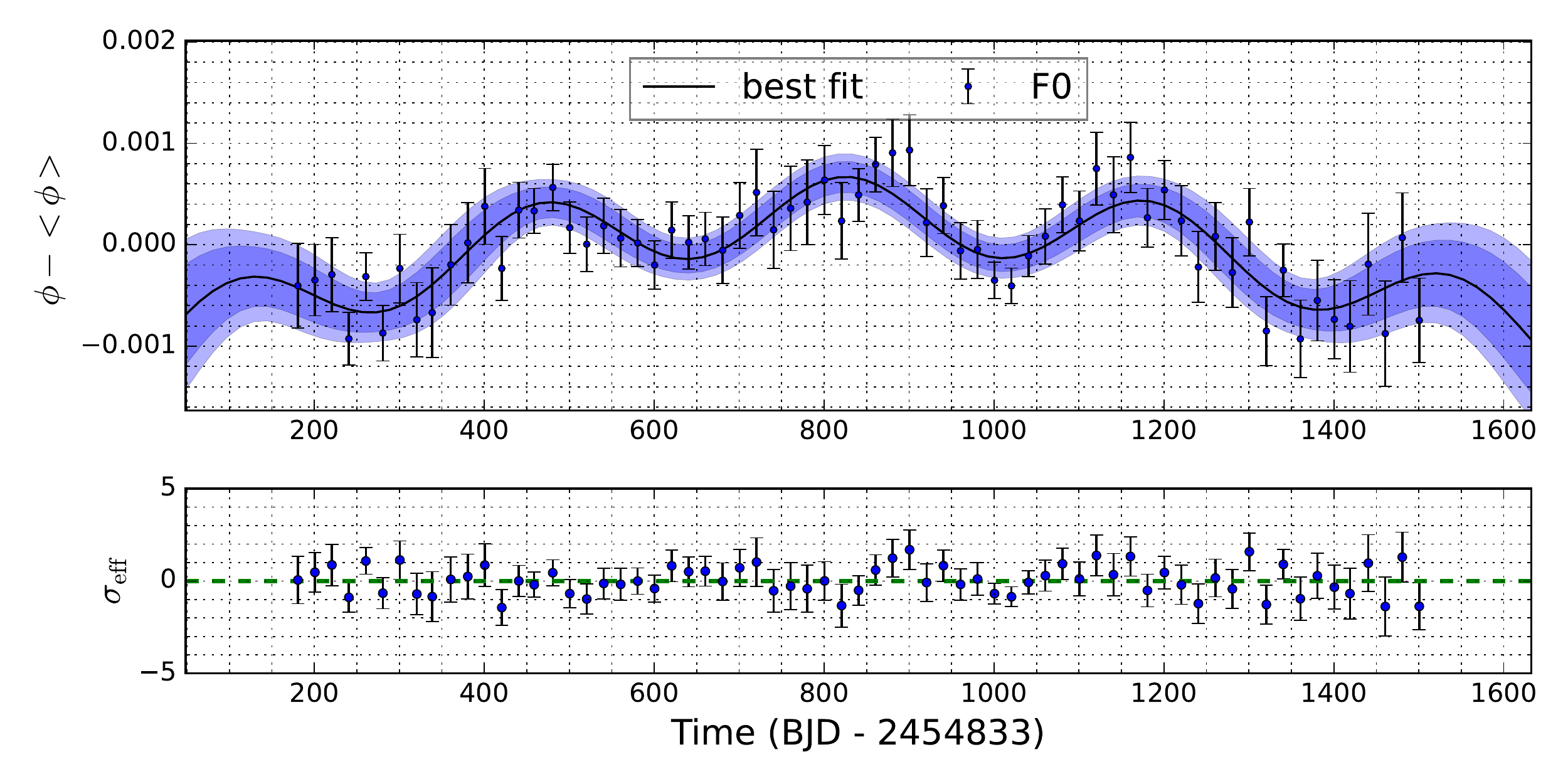}
  \includegraphics[width=0.495\textwidth]{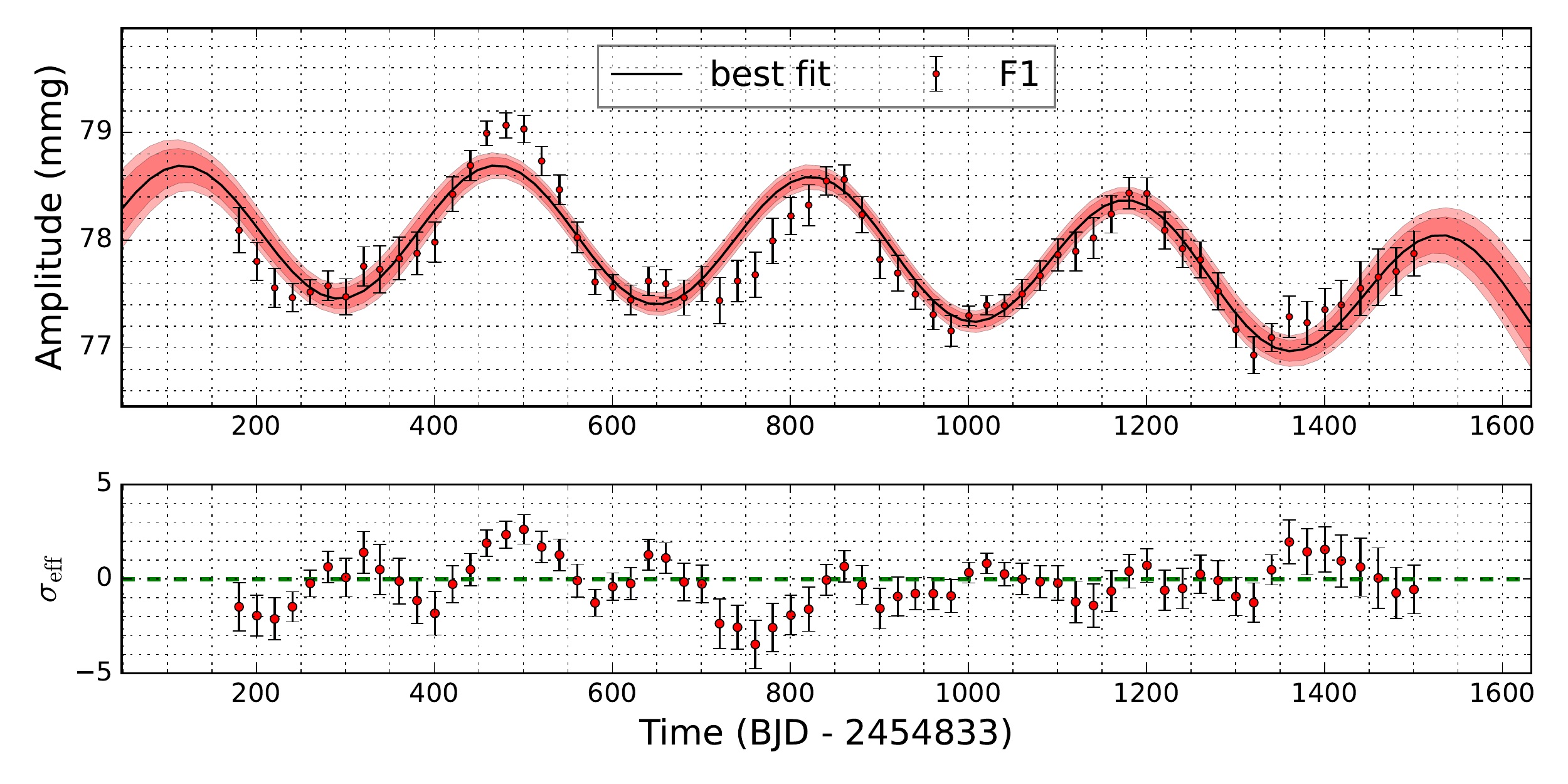}
  \includegraphics[width=0.495\textwidth]{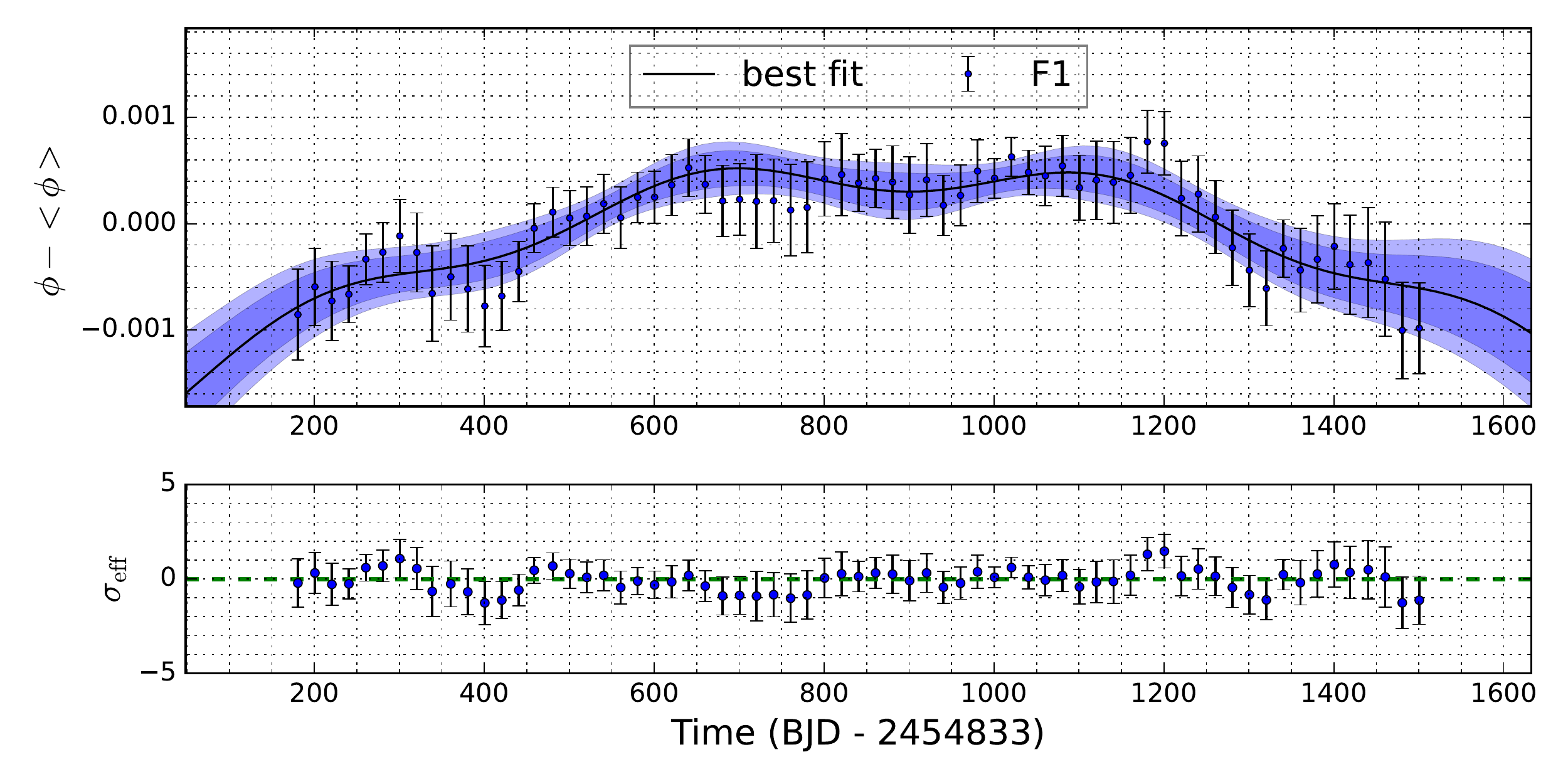}
  \includegraphics[width=0.495\textwidth]{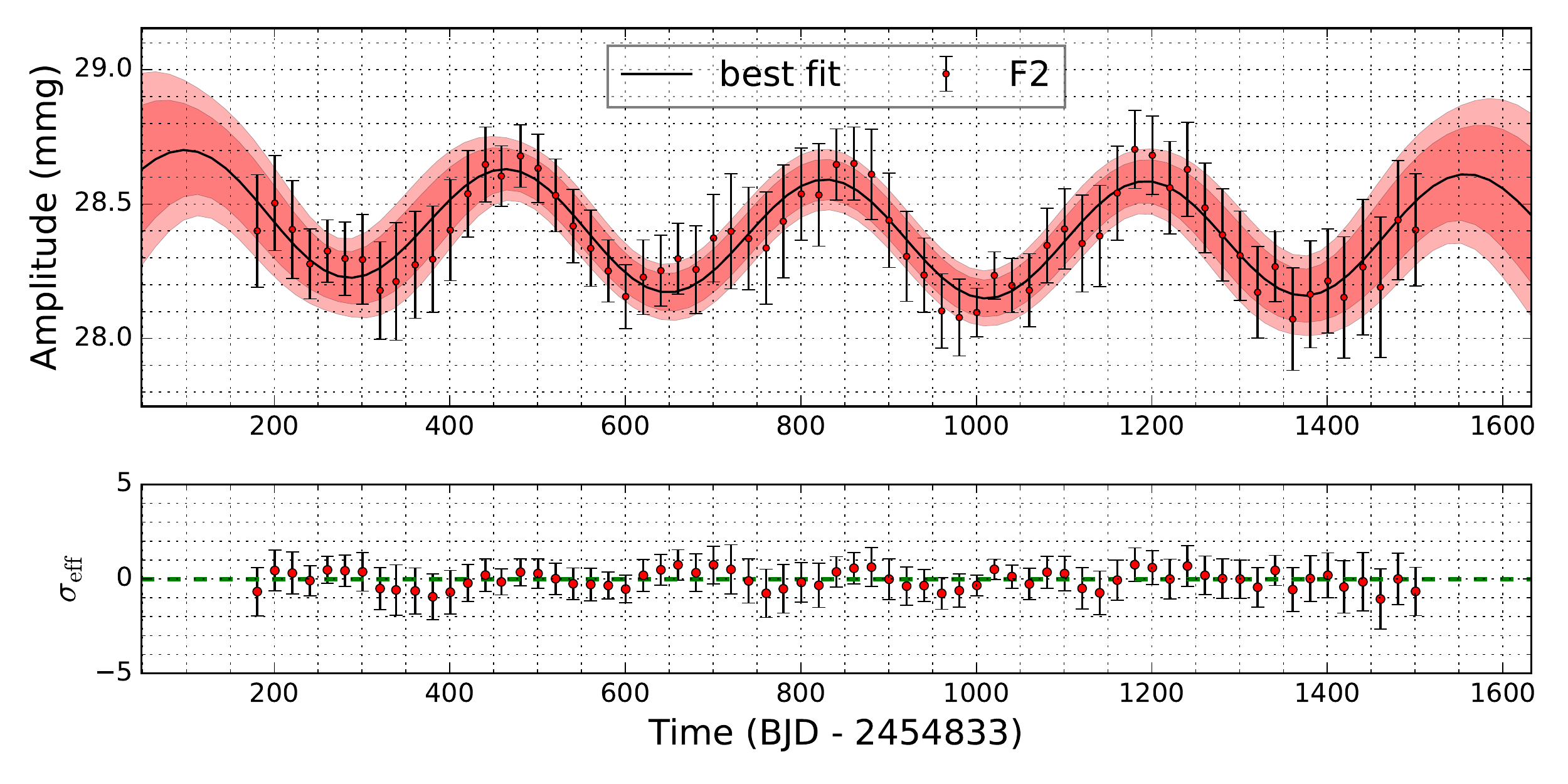}
  \includegraphics[width=0.495\textwidth]{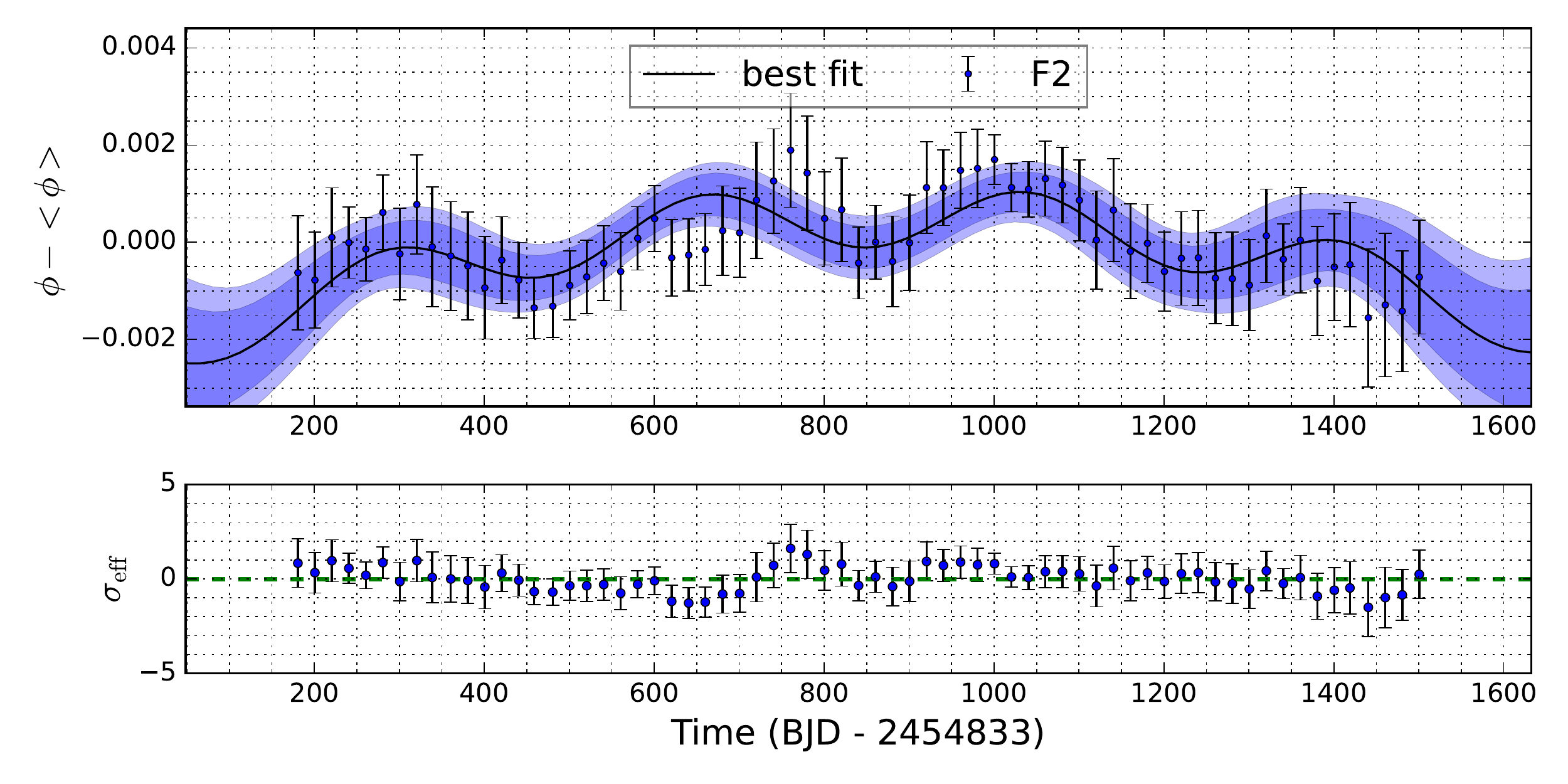}
  \includegraphics[width=0.495\textwidth]{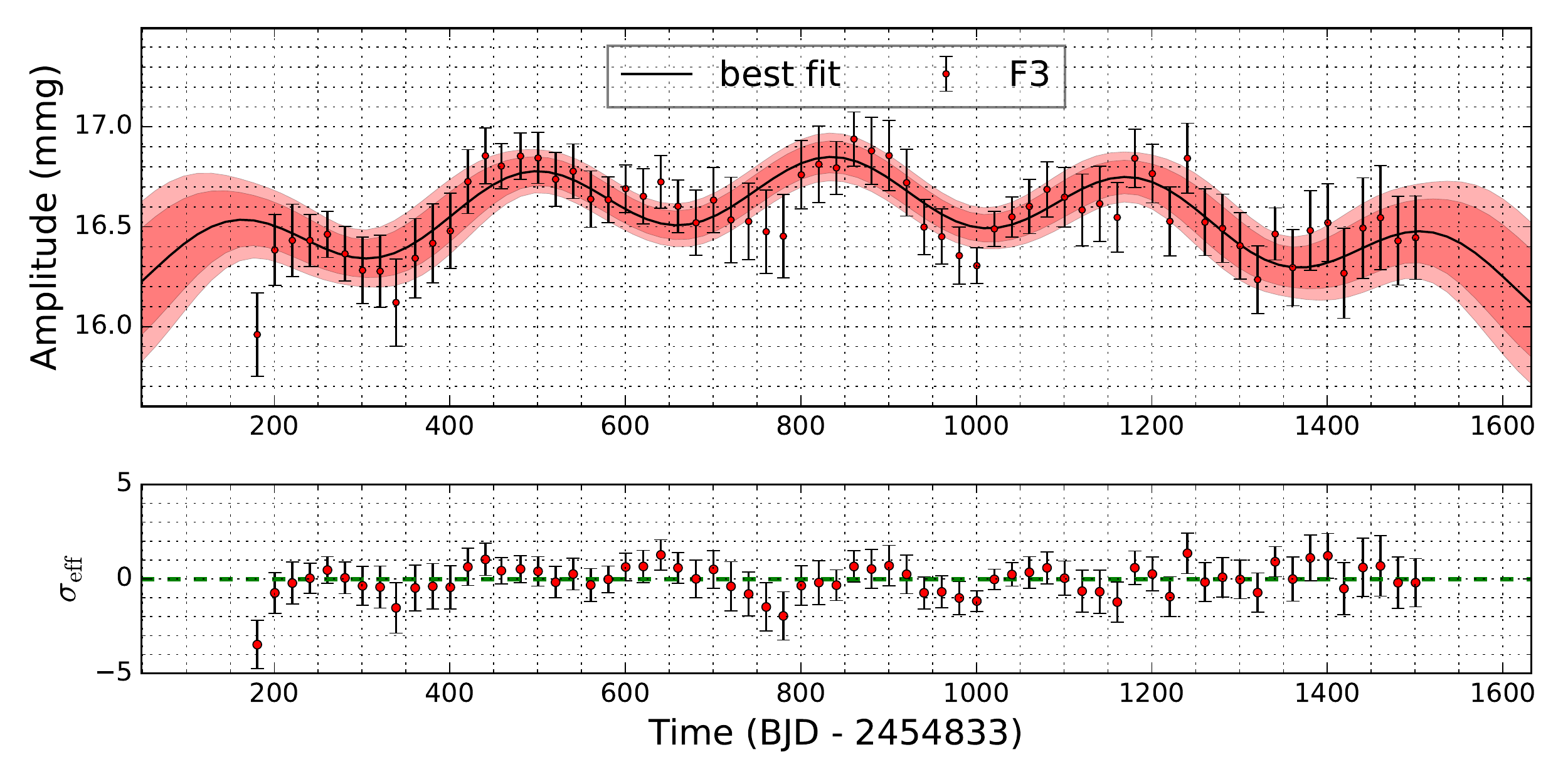}
  \includegraphics[width=0.495\textwidth]{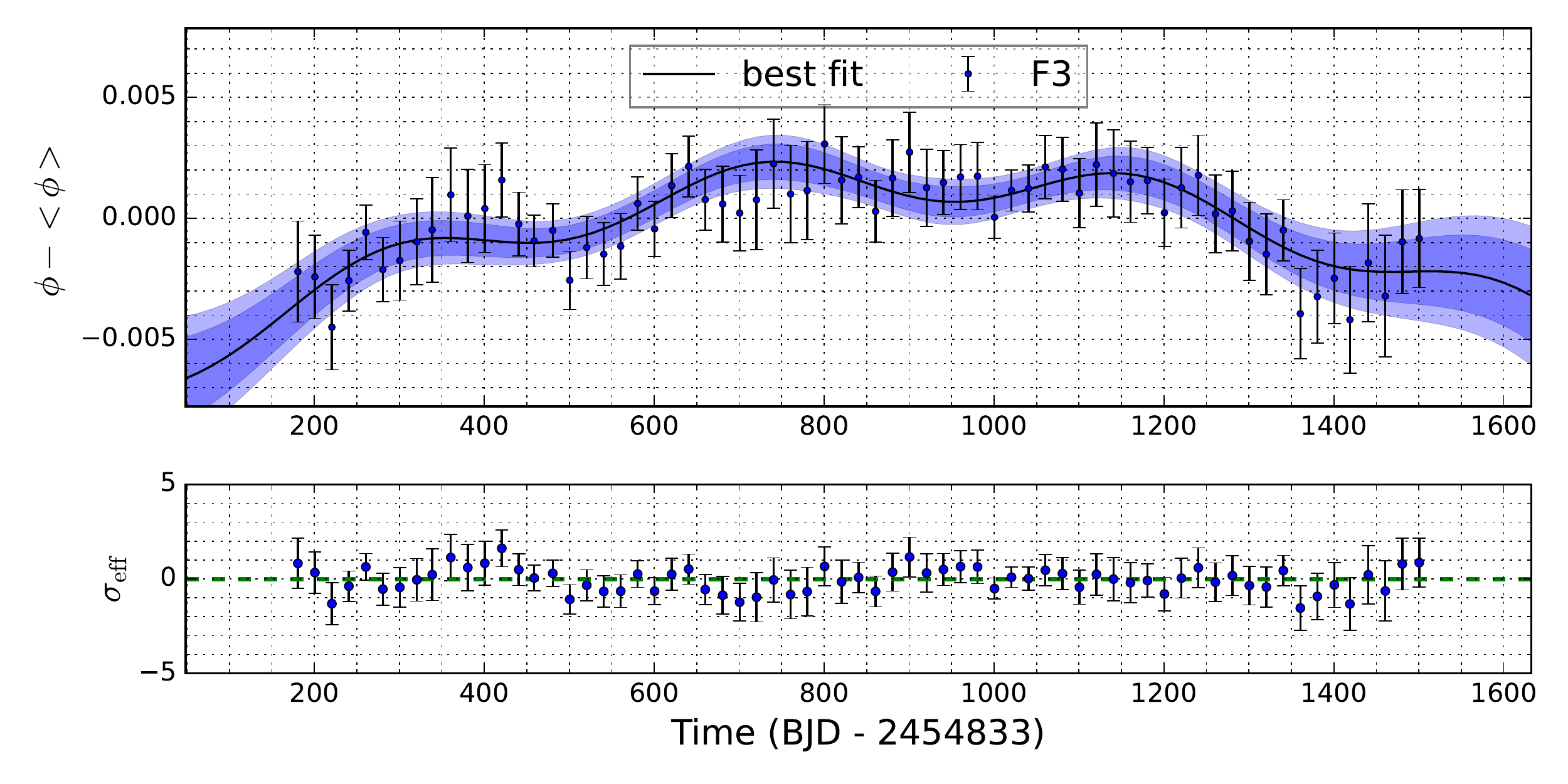}
  \caption{Variation in the amplitudes and phases of the 23 pulsation modes, Part I. The best-fit results to Equations \ref{eq:fit_a} and \ref{eq:fit_p} are presented by the black solid lines. The $2\sigma$ (deep red and blue) and $3\sigma$ (light red and blue) bounds are also shown in the panels. In the lower panel of each subfigure, the $\sigma_{\mathrm{eff}}$ is defined as $(Q_{\mathrm{obs}} - Q_{\mathrm{cal}})/{\sigma}$, where $Q_\mathrm{obs}$ and $Q_\mathrm{cal}$ are the values that come from the observation and model calculation, respectively; $\sigma$ is the uncertainty of the observed points.}
  \label{fig:var_amp_phase01}
\end{figure*}

\begin{figure*}[htp]
  \centering
  \includegraphics[width=0.495\textwidth]{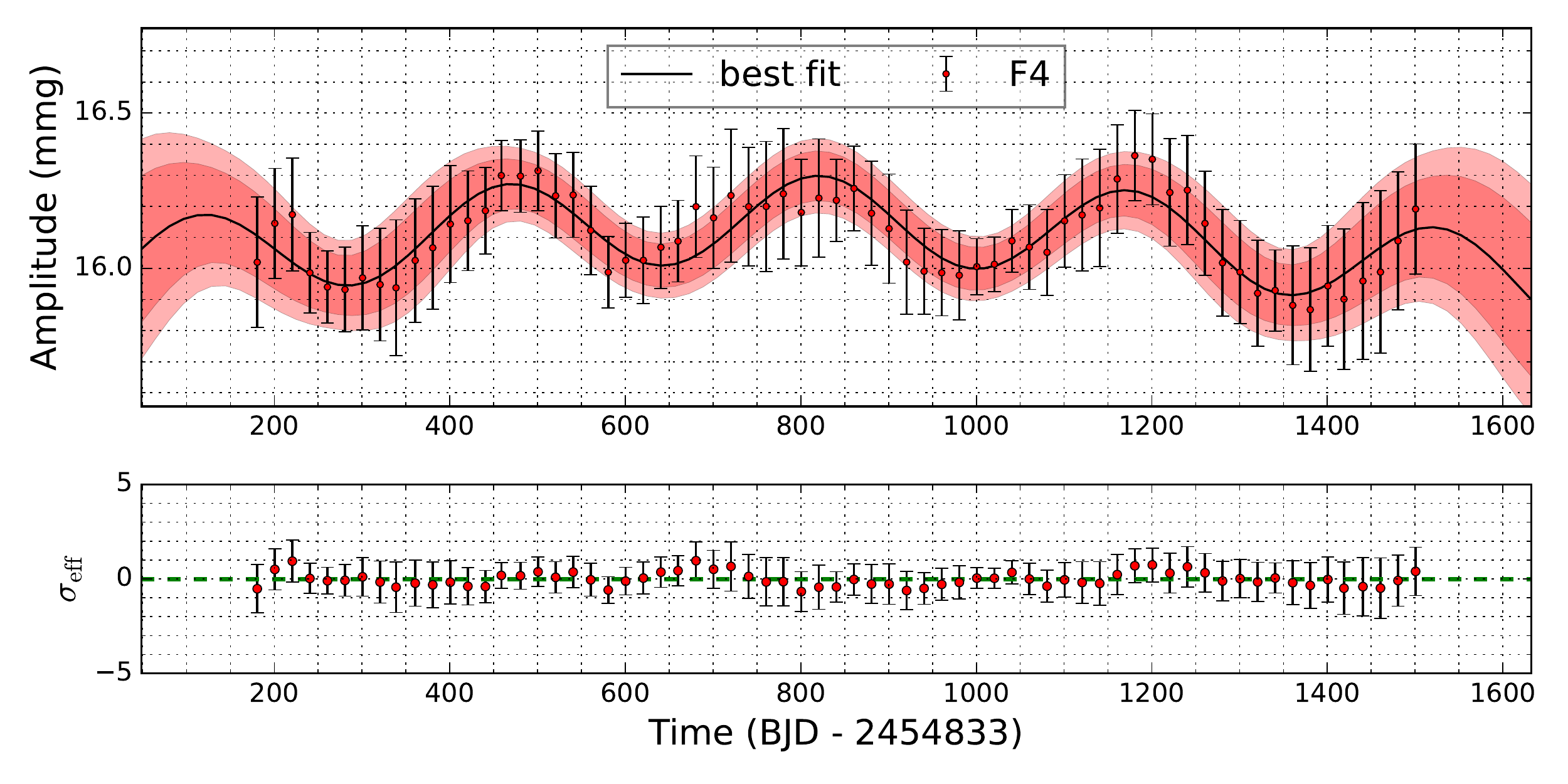}
  \includegraphics[width=0.495\textwidth]{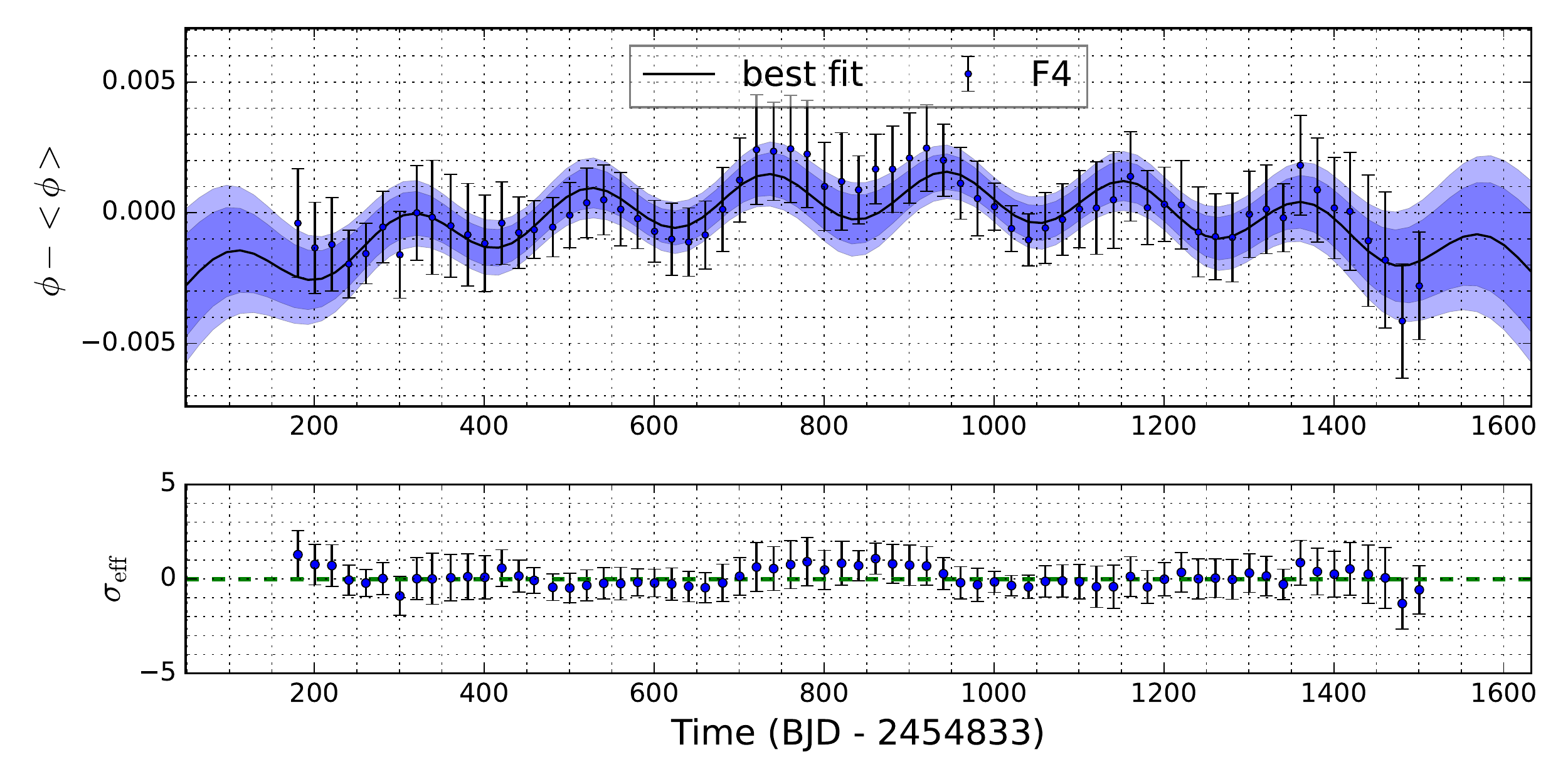}
  \includegraphics[width=0.495\textwidth]{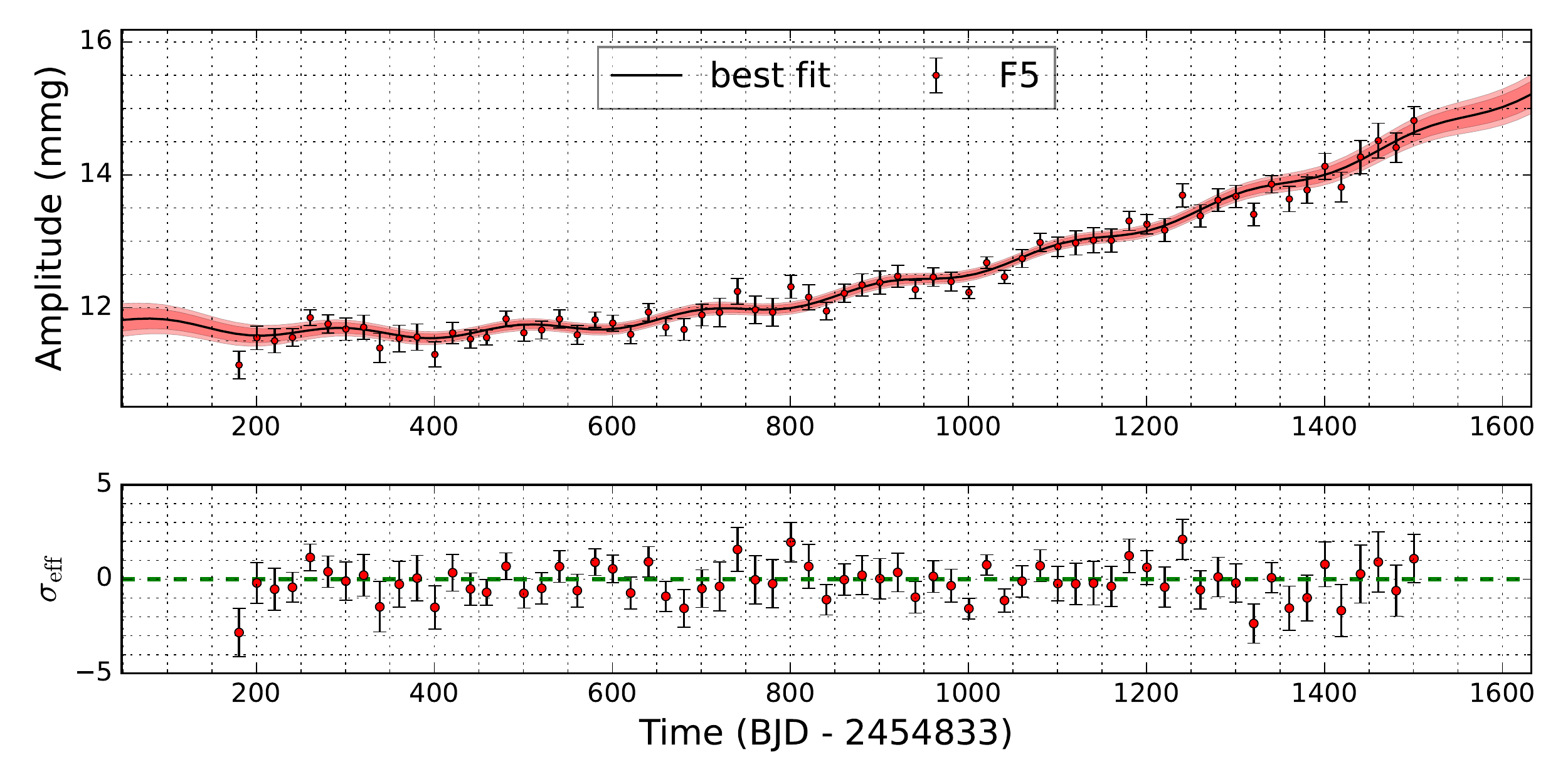}
  \includegraphics[width=0.495\textwidth]{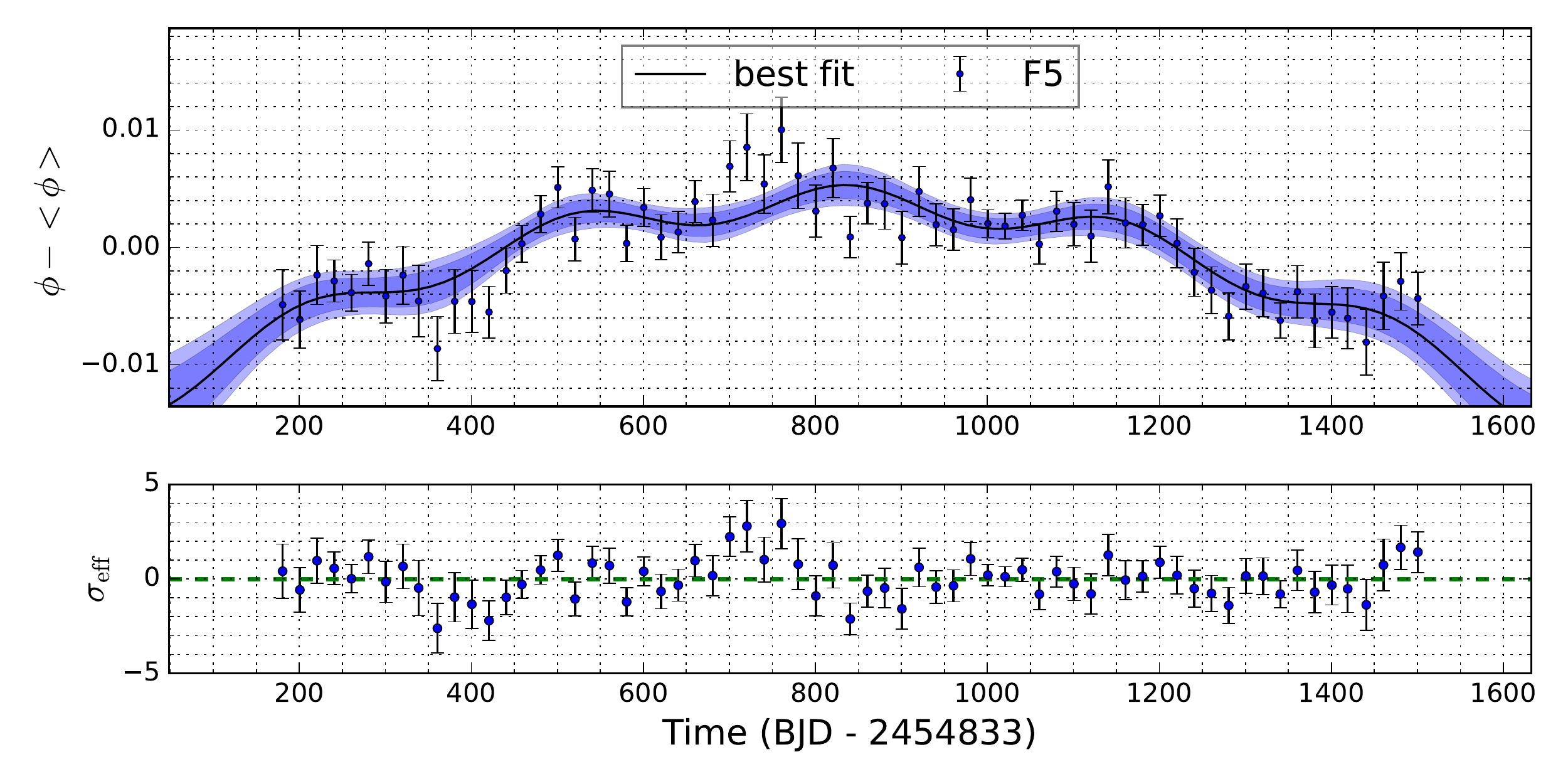}
  \includegraphics[width=0.495\textwidth]{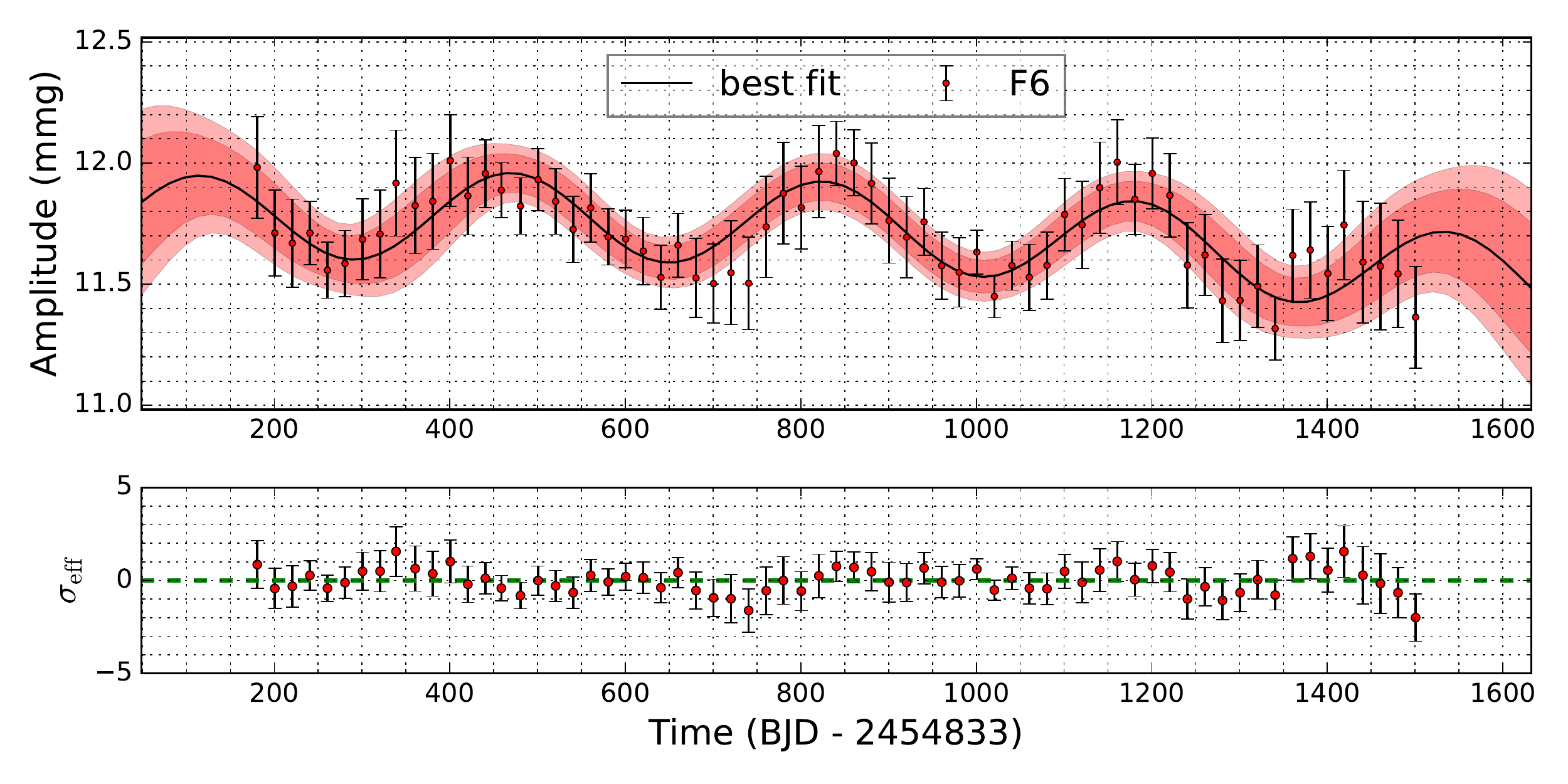}
  \includegraphics[width=0.495\textwidth]{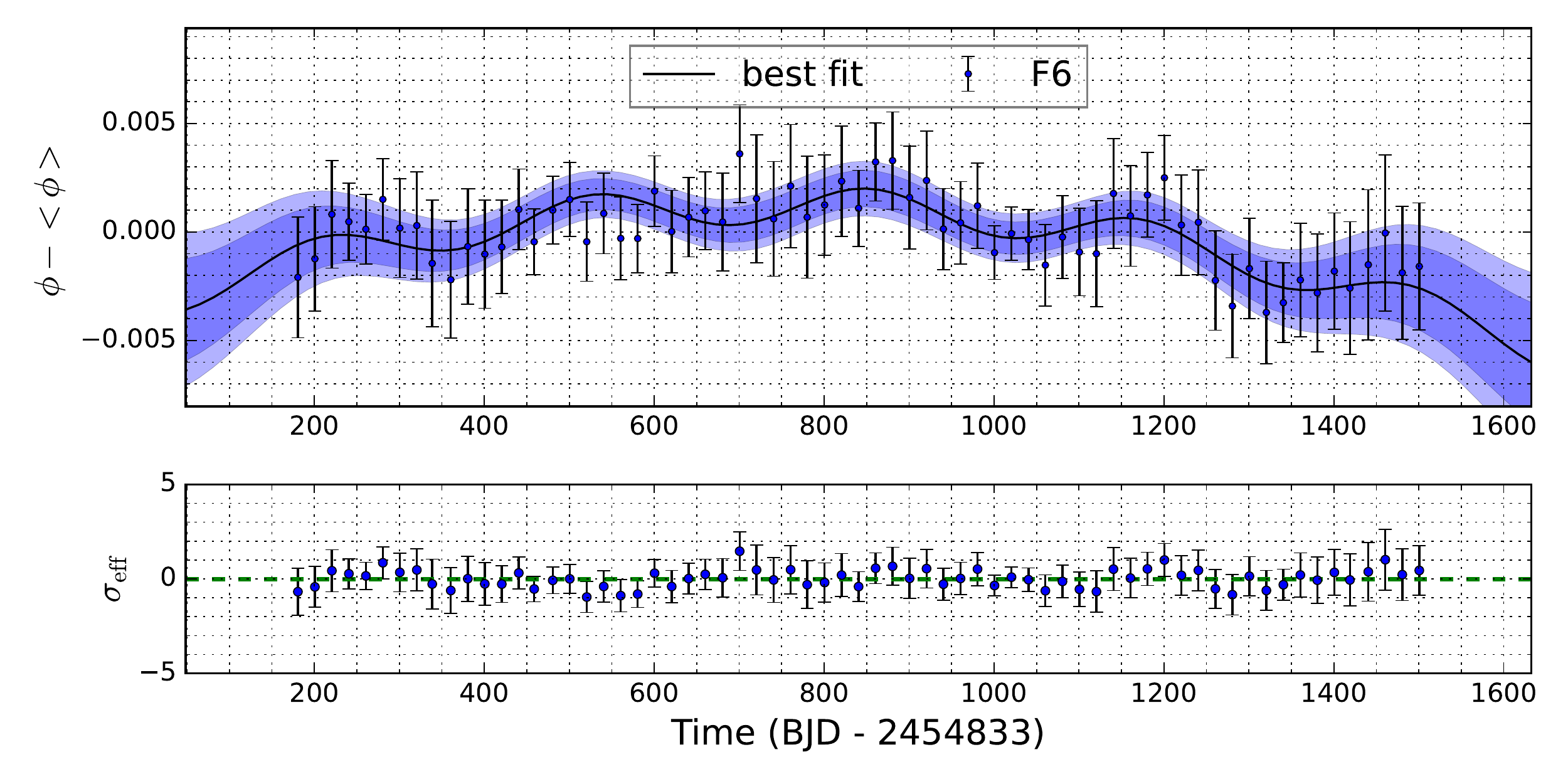}
  \includegraphics[width=0.495\textwidth]{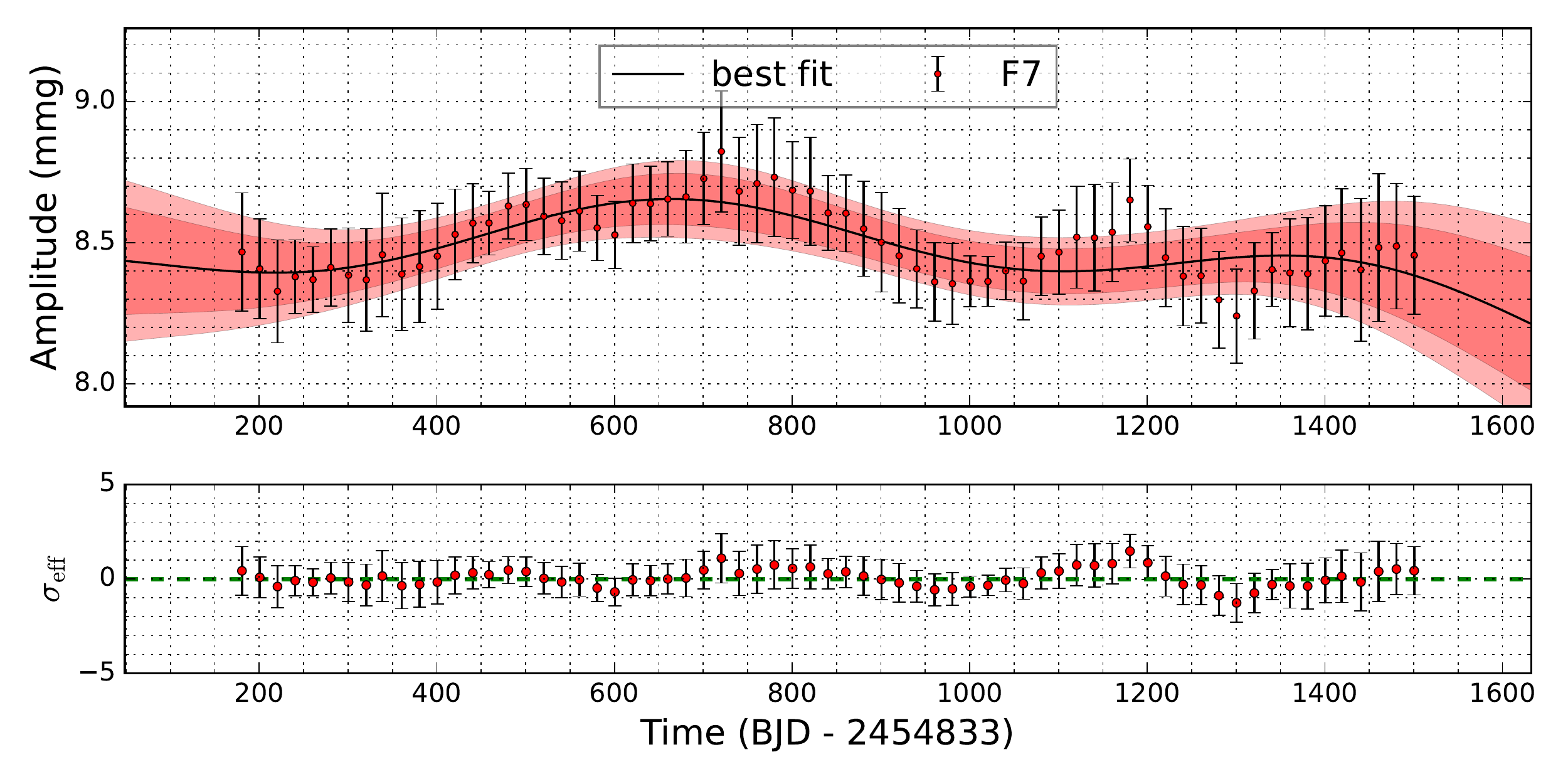}
  \includegraphics[width=0.495\textwidth]{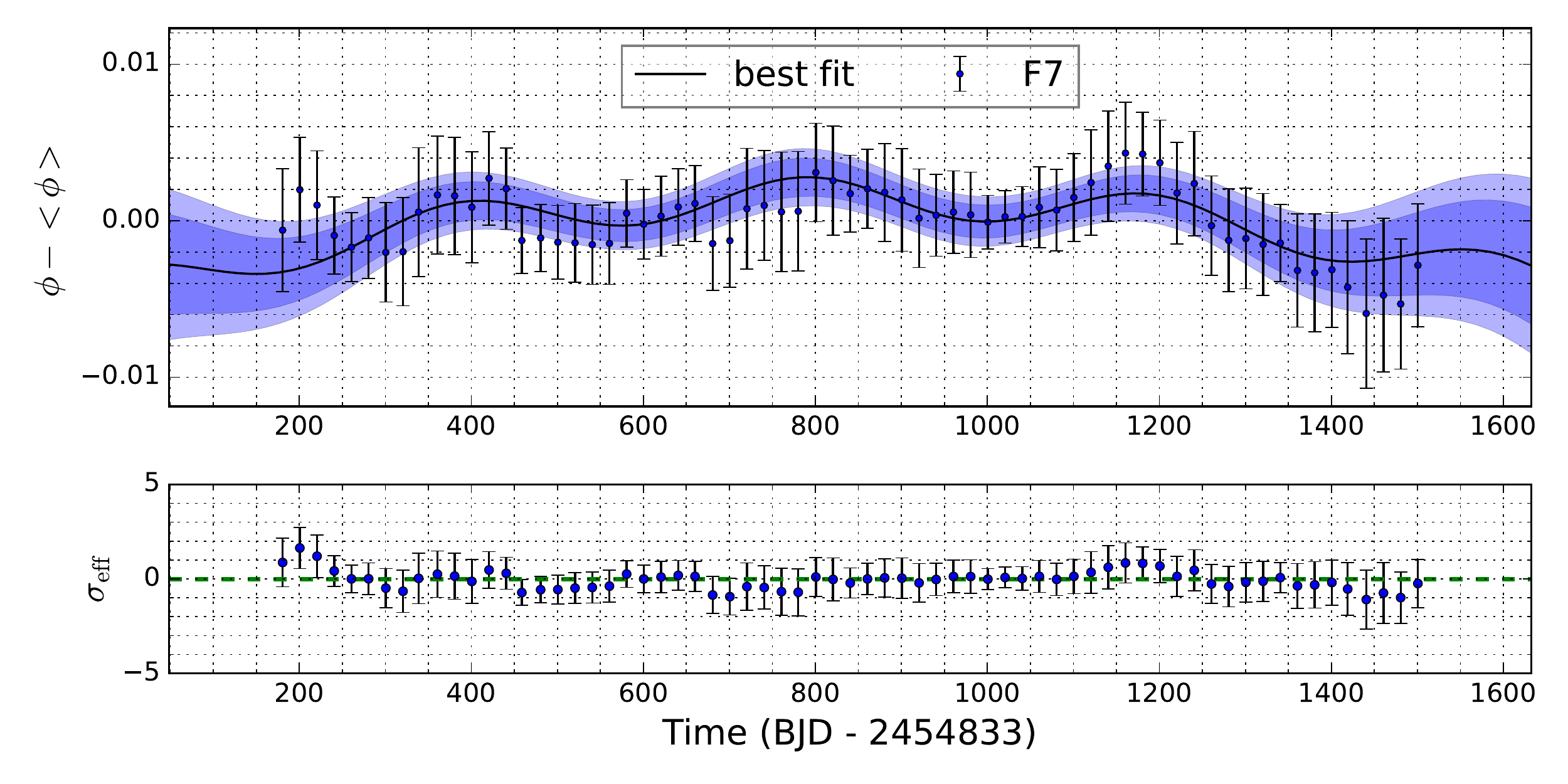}
  \includegraphics[width=0.495\textwidth]{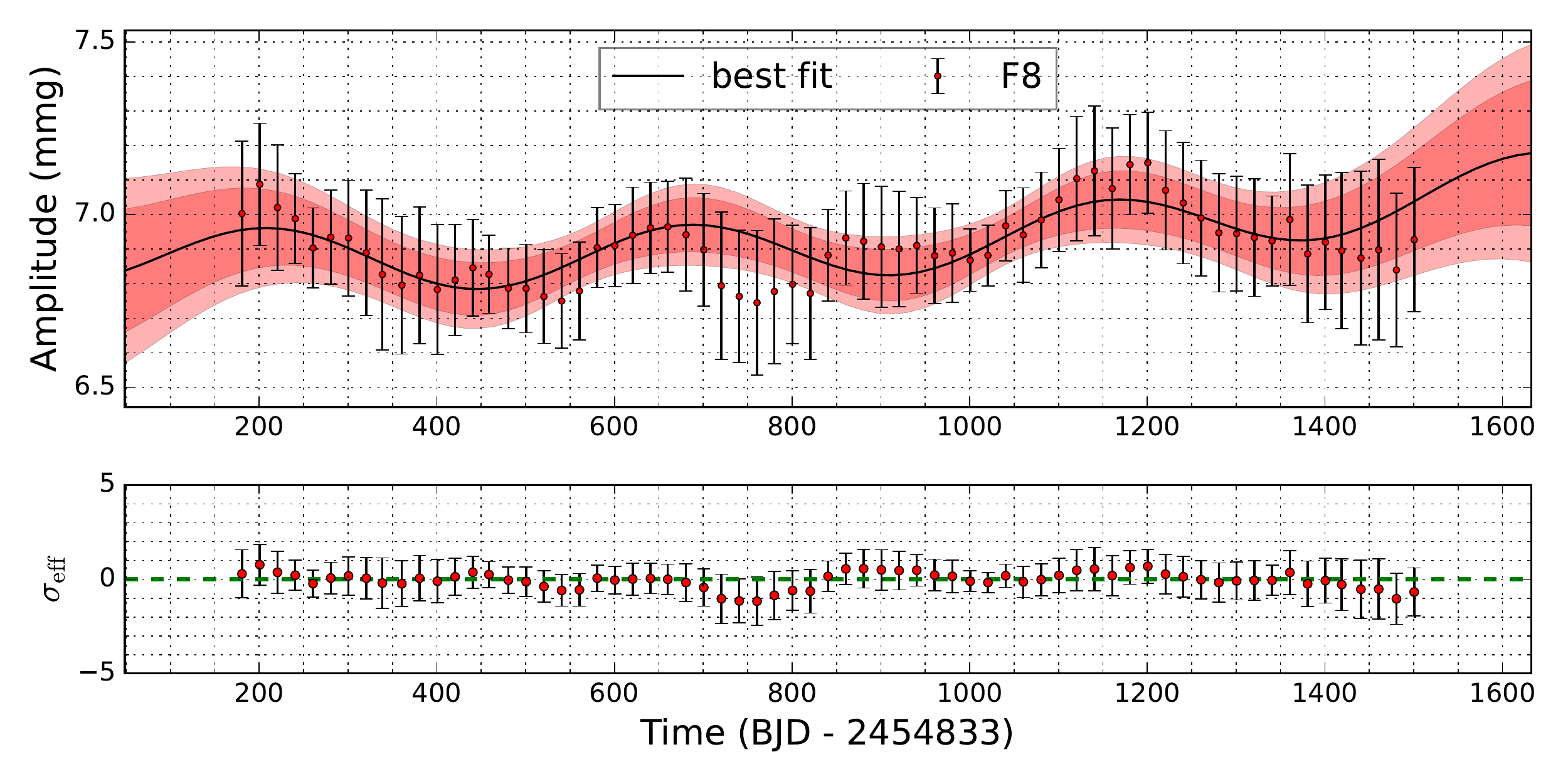}
  \includegraphics[width=0.495\textwidth]{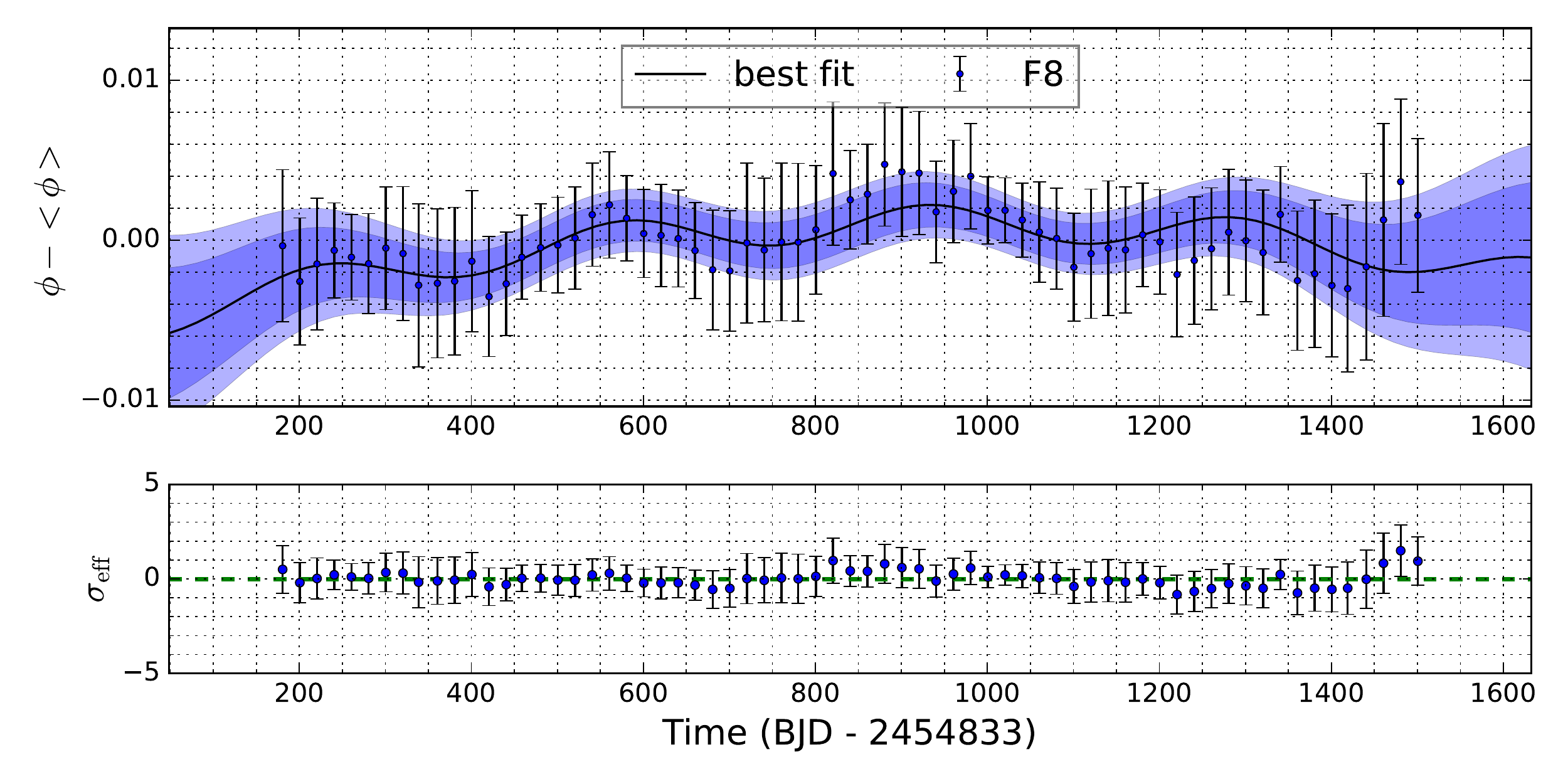}
  \caption{Variation in the amplitudes and phases of the 23 pulsation modes, Part II.}
  \label{fig:var_amp_phase02}
\end{figure*}

\begin{figure*}[htp]
  \centering
  \includegraphics[width=0.495\textwidth]{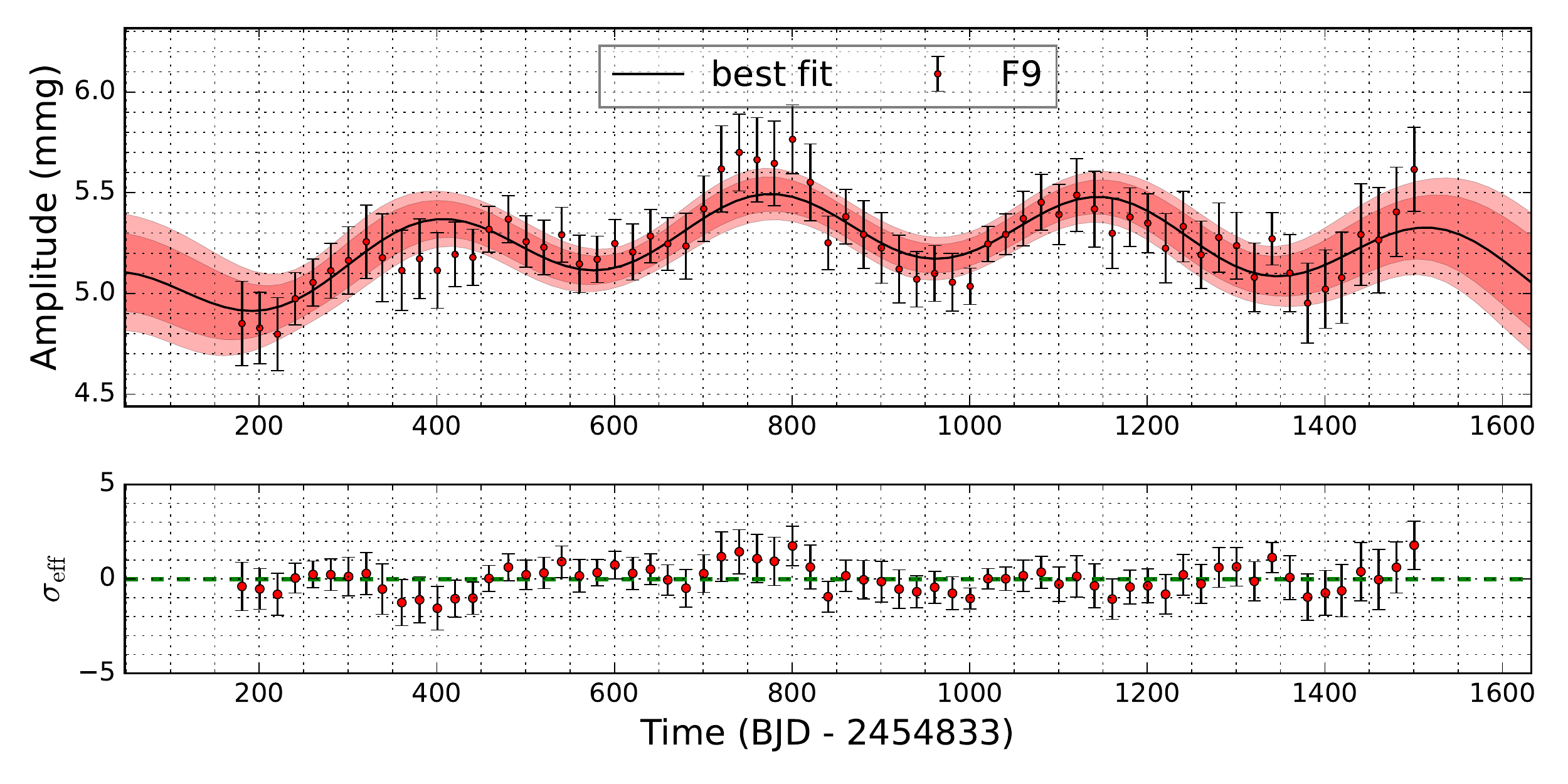}
  \includegraphics[width=0.495\textwidth]{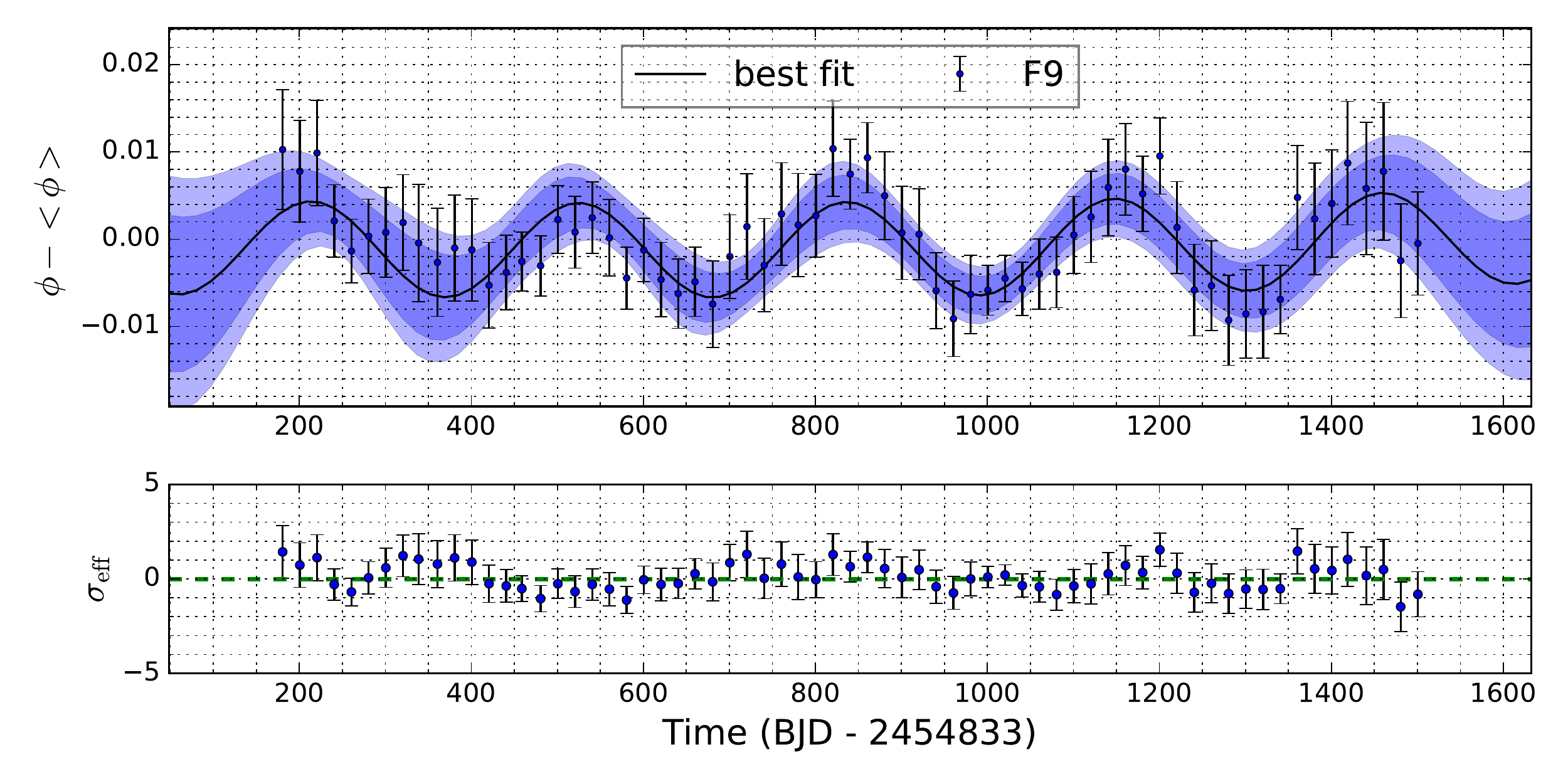}
  \includegraphics[width=0.495\textwidth]{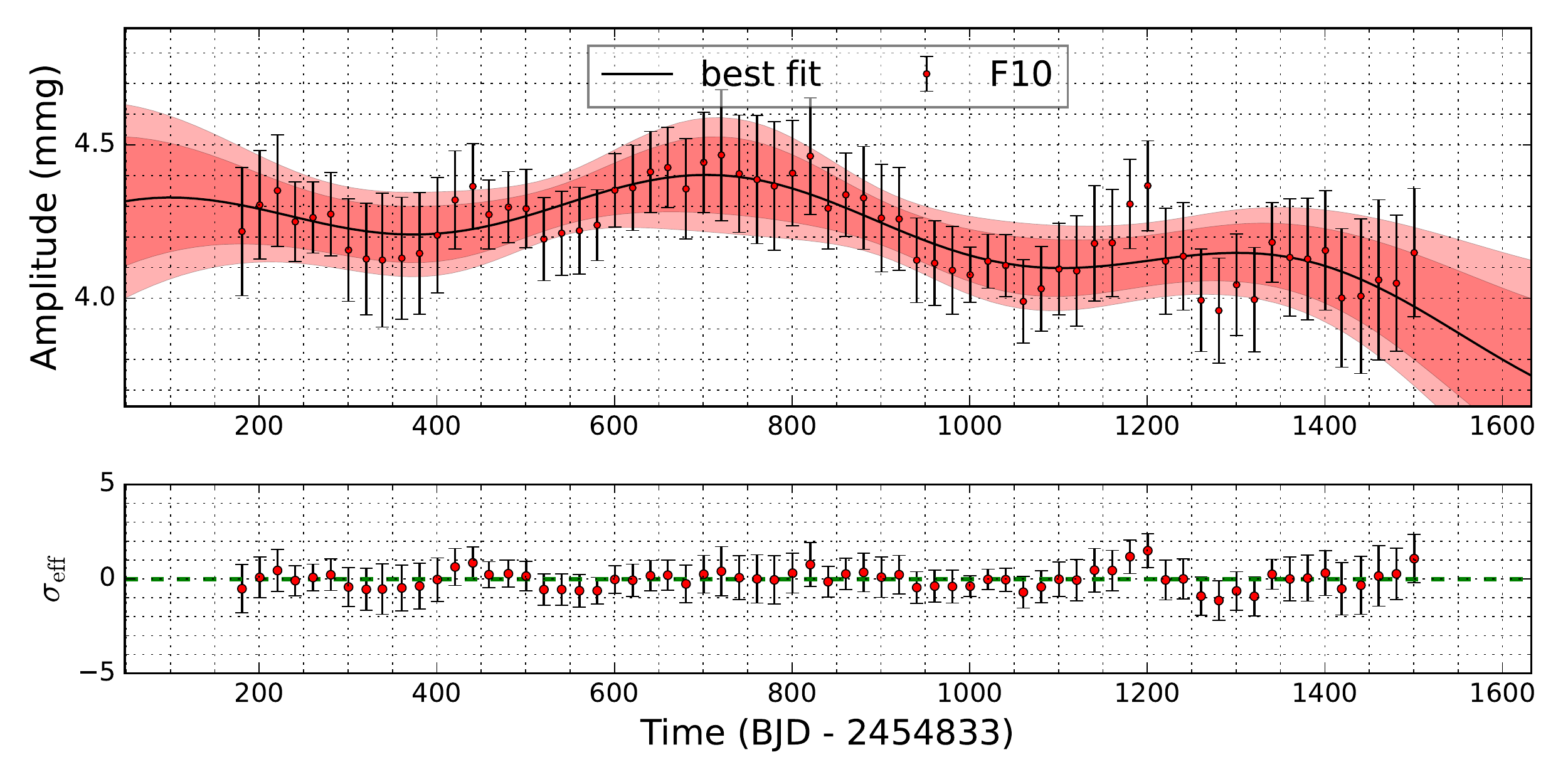}
  \includegraphics[width=0.495\textwidth]{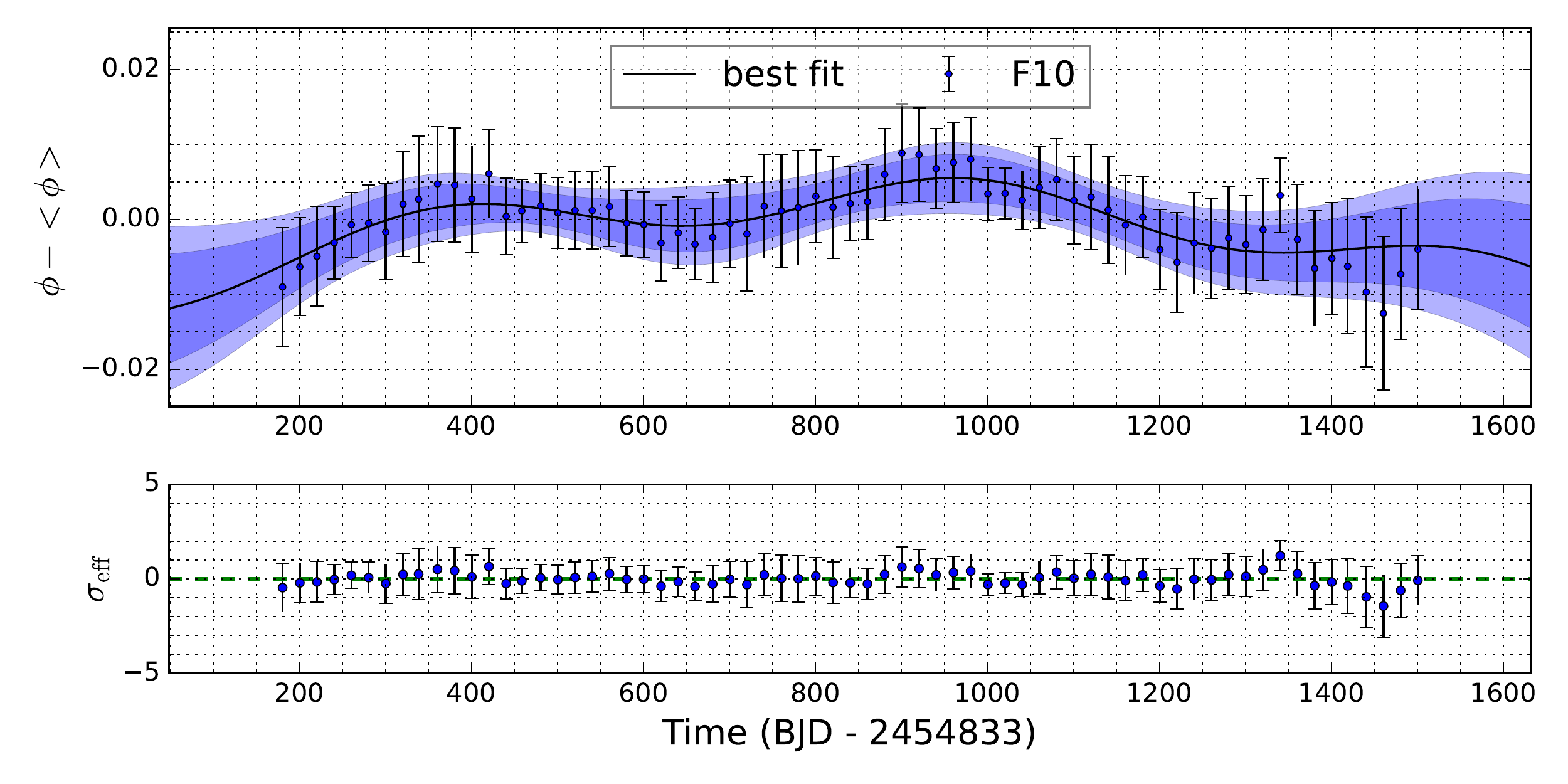}
  \includegraphics[width=0.495\textwidth]{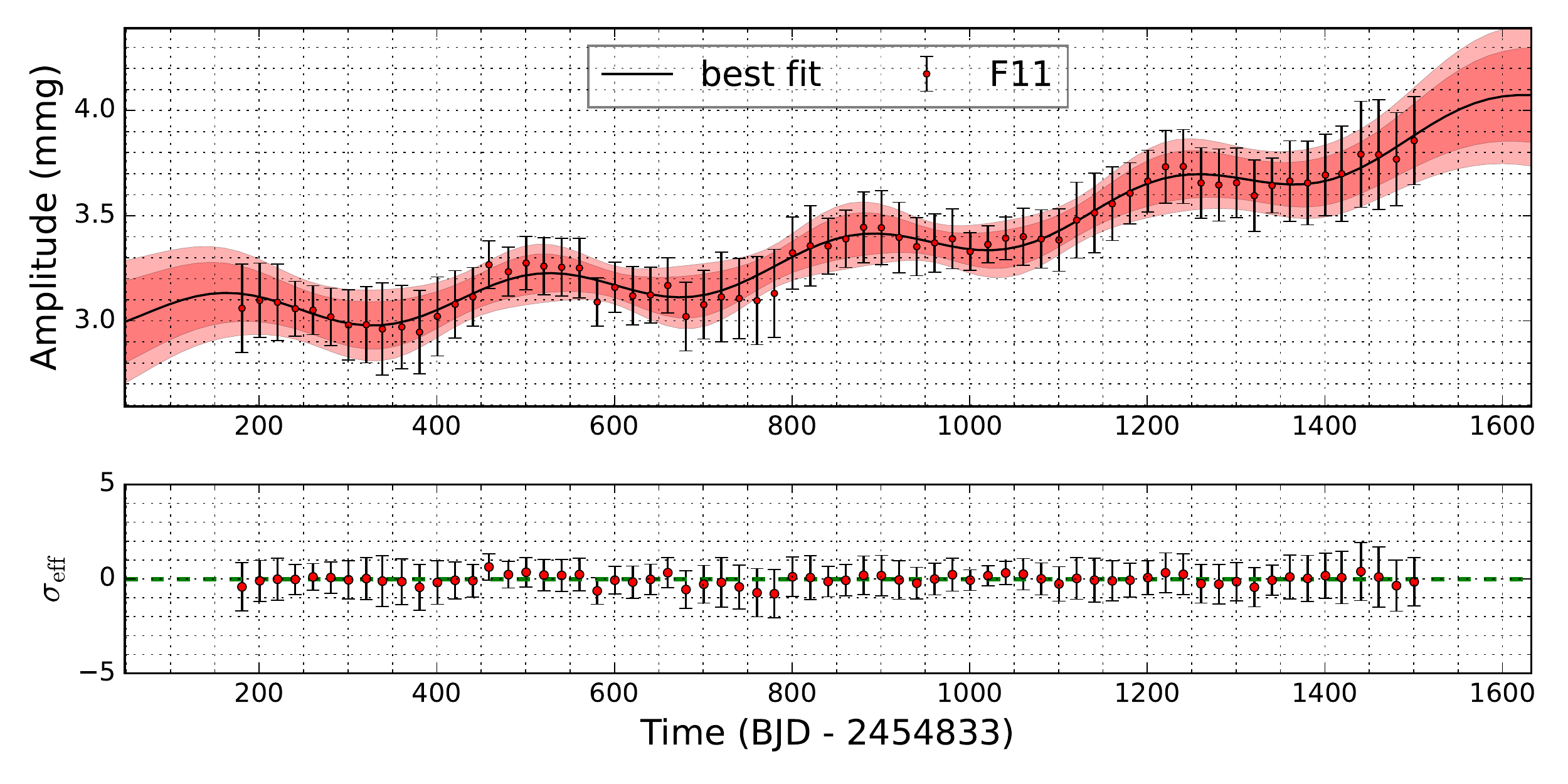}
  \includegraphics[width=0.495\textwidth]{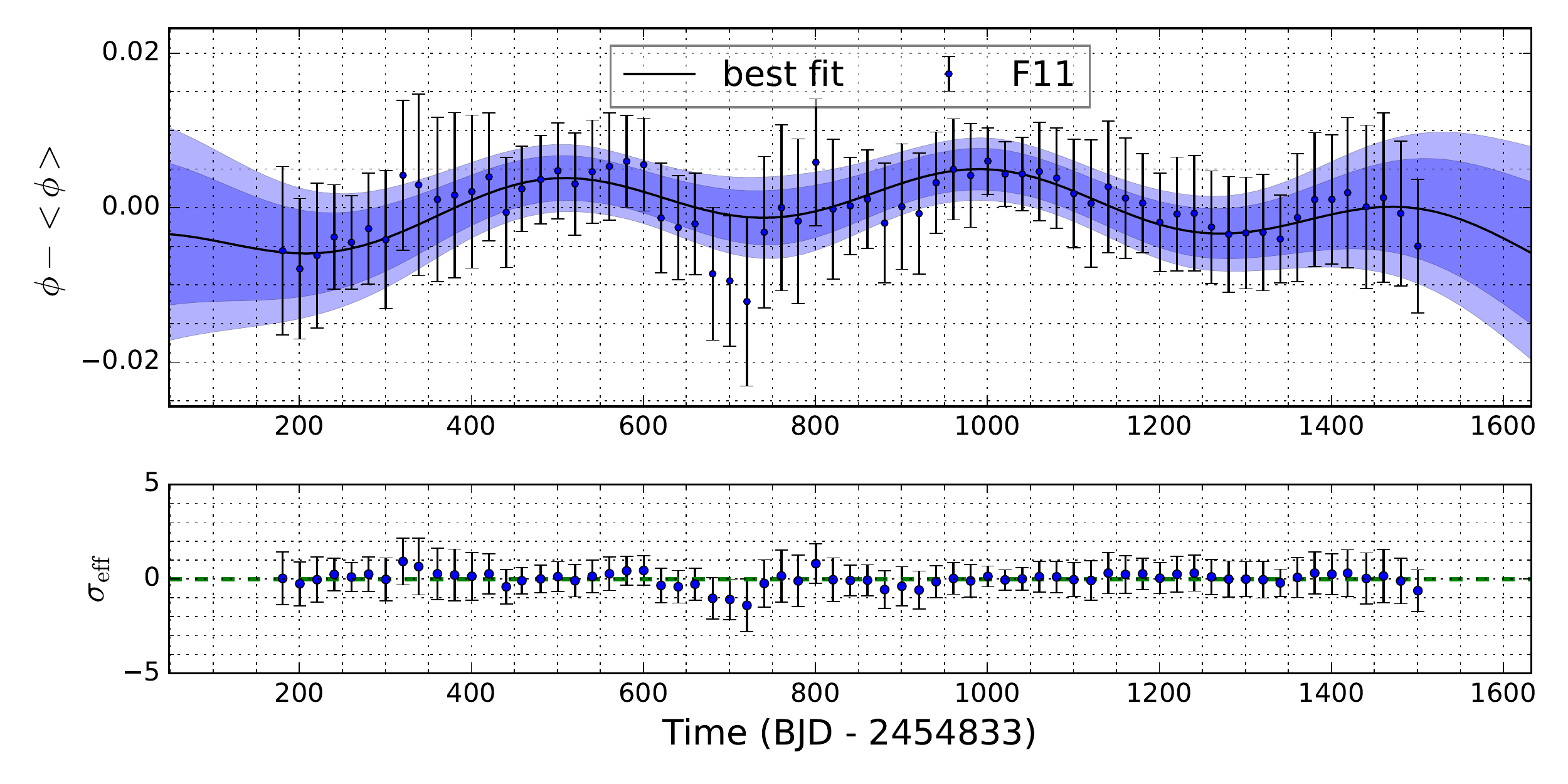}
  \includegraphics[width=0.495\textwidth]{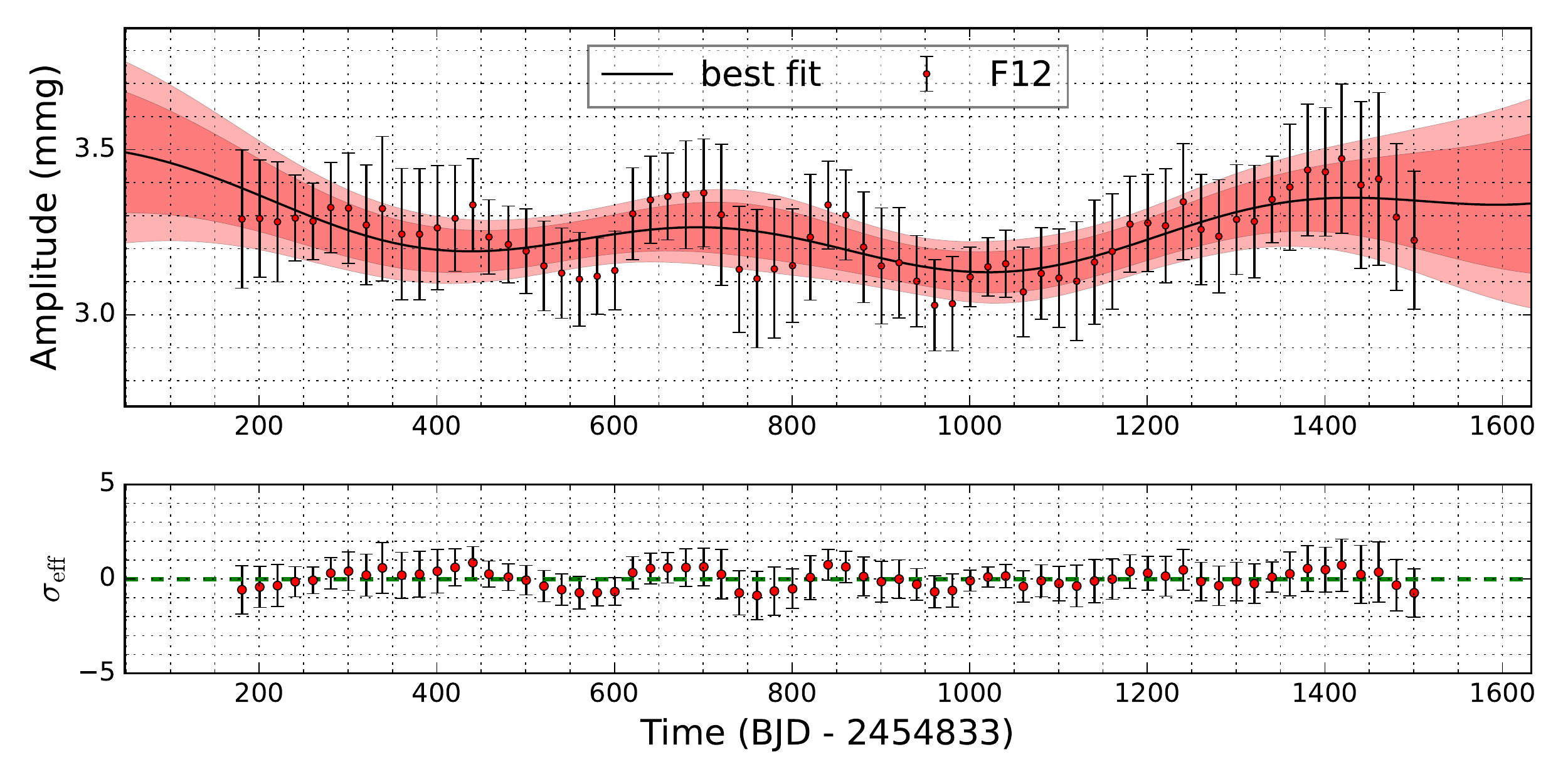}
  \includegraphics[width=0.495\textwidth]{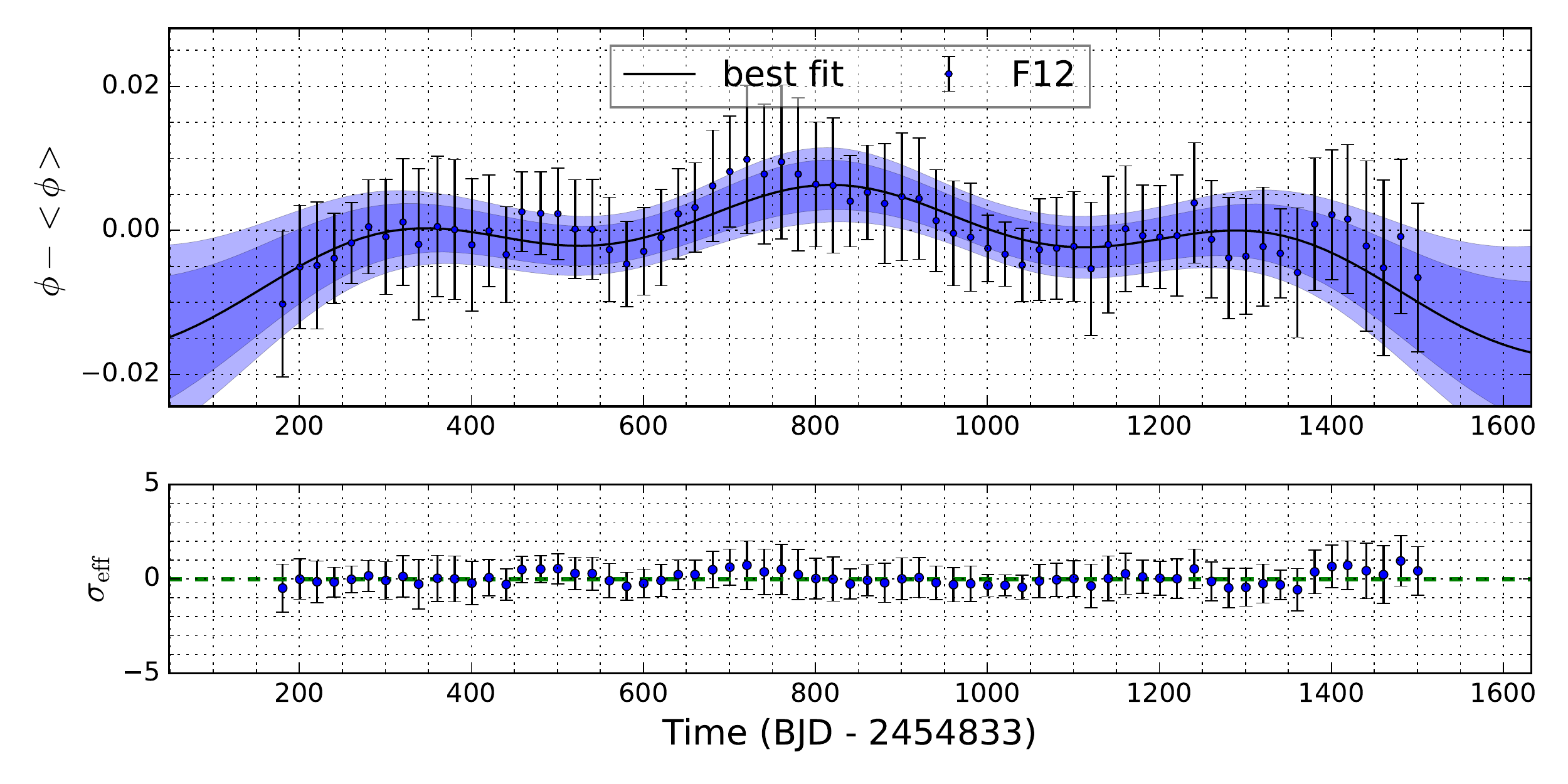}
  \includegraphics[width=0.495\textwidth]{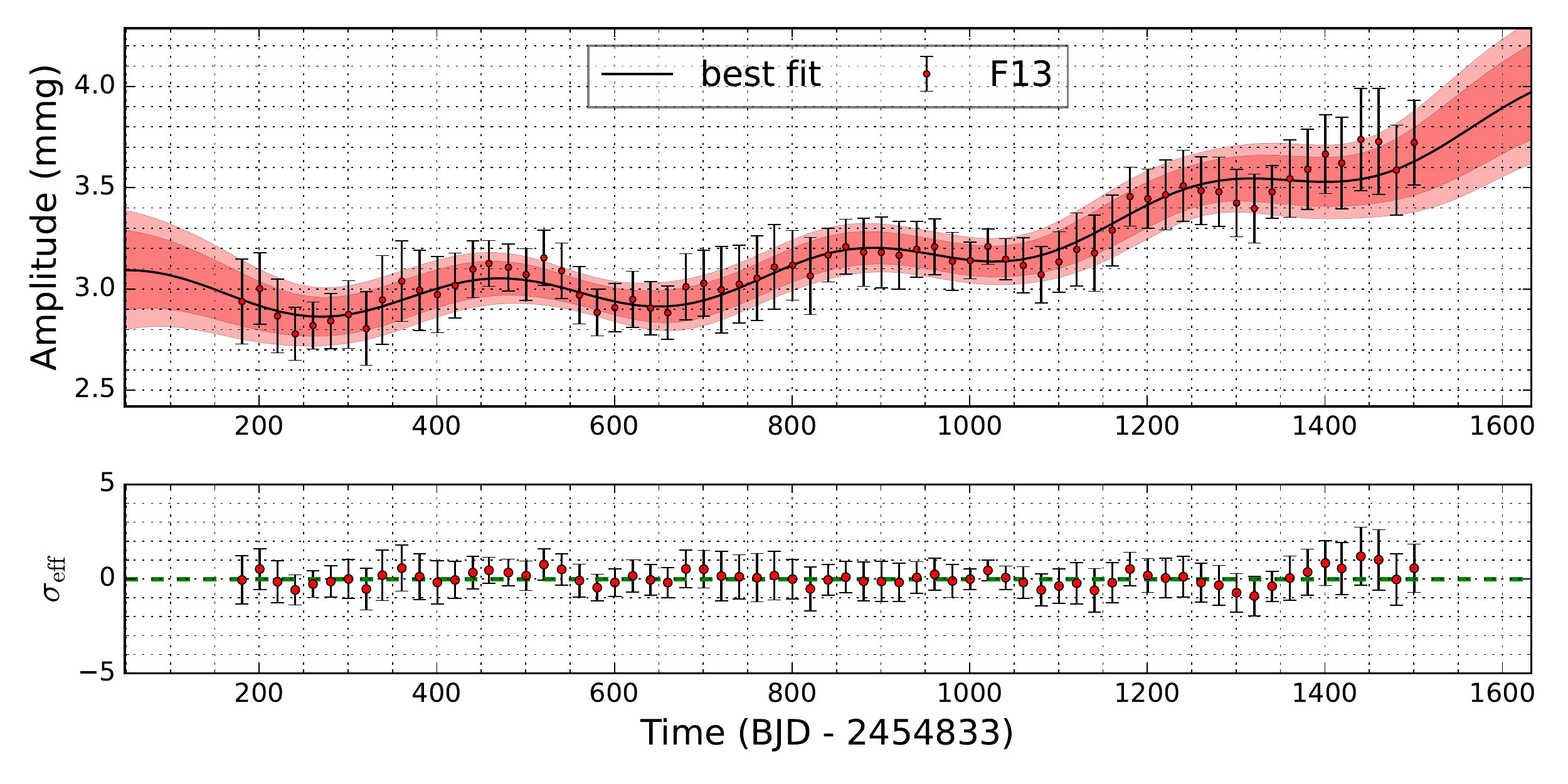}
  \includegraphics[width=0.495\textwidth]{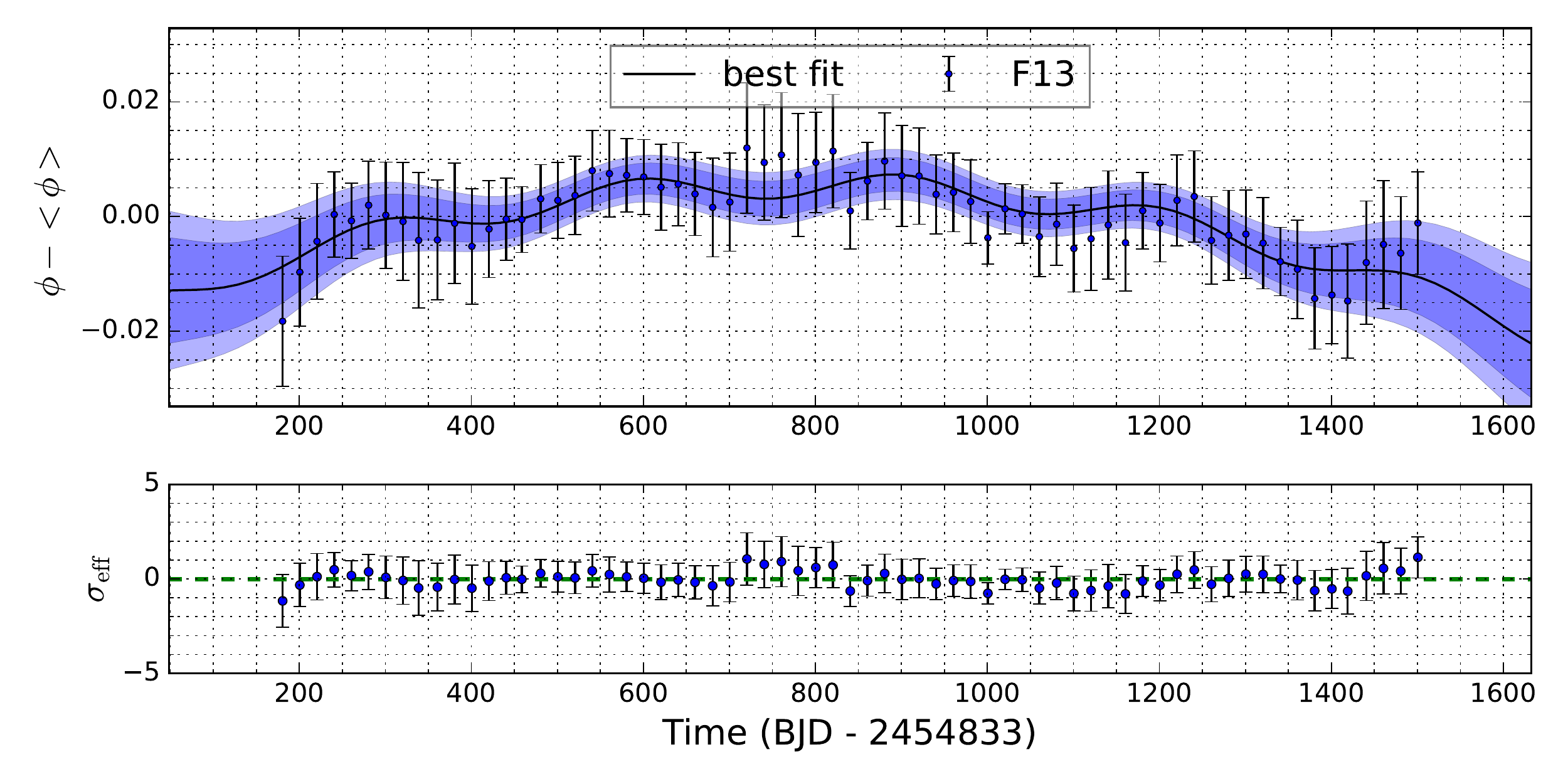}
  \caption{Variation in the amplitudes and phases of the 23 pulsation modes, Part III.}
  \label{fig:var_amp_phase03}
\end{figure*}

\begin{figure*}[htp]
  \centering
  \includegraphics[width=0.495\textwidth]{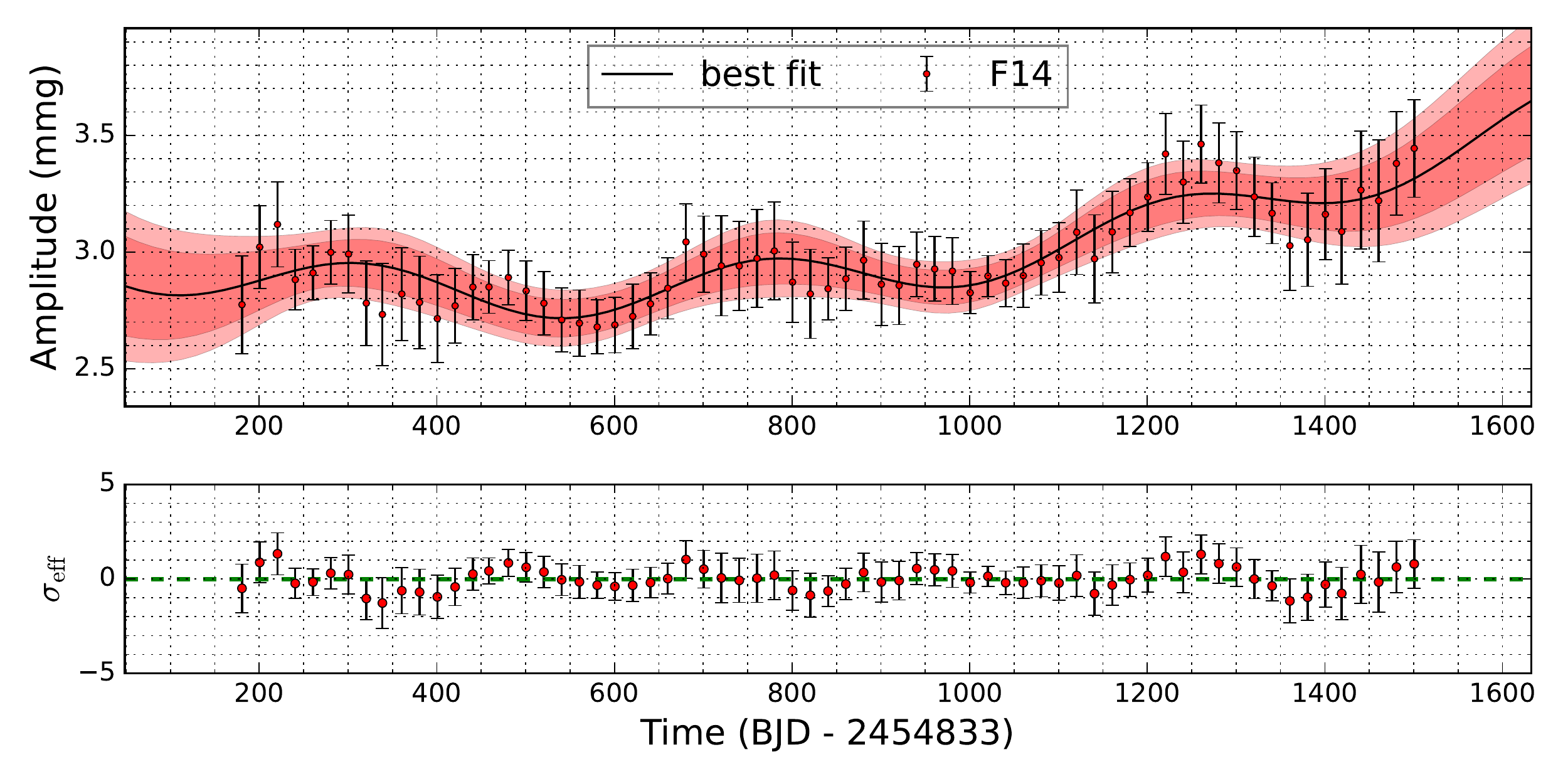}
  \includegraphics[width=0.495\textwidth]{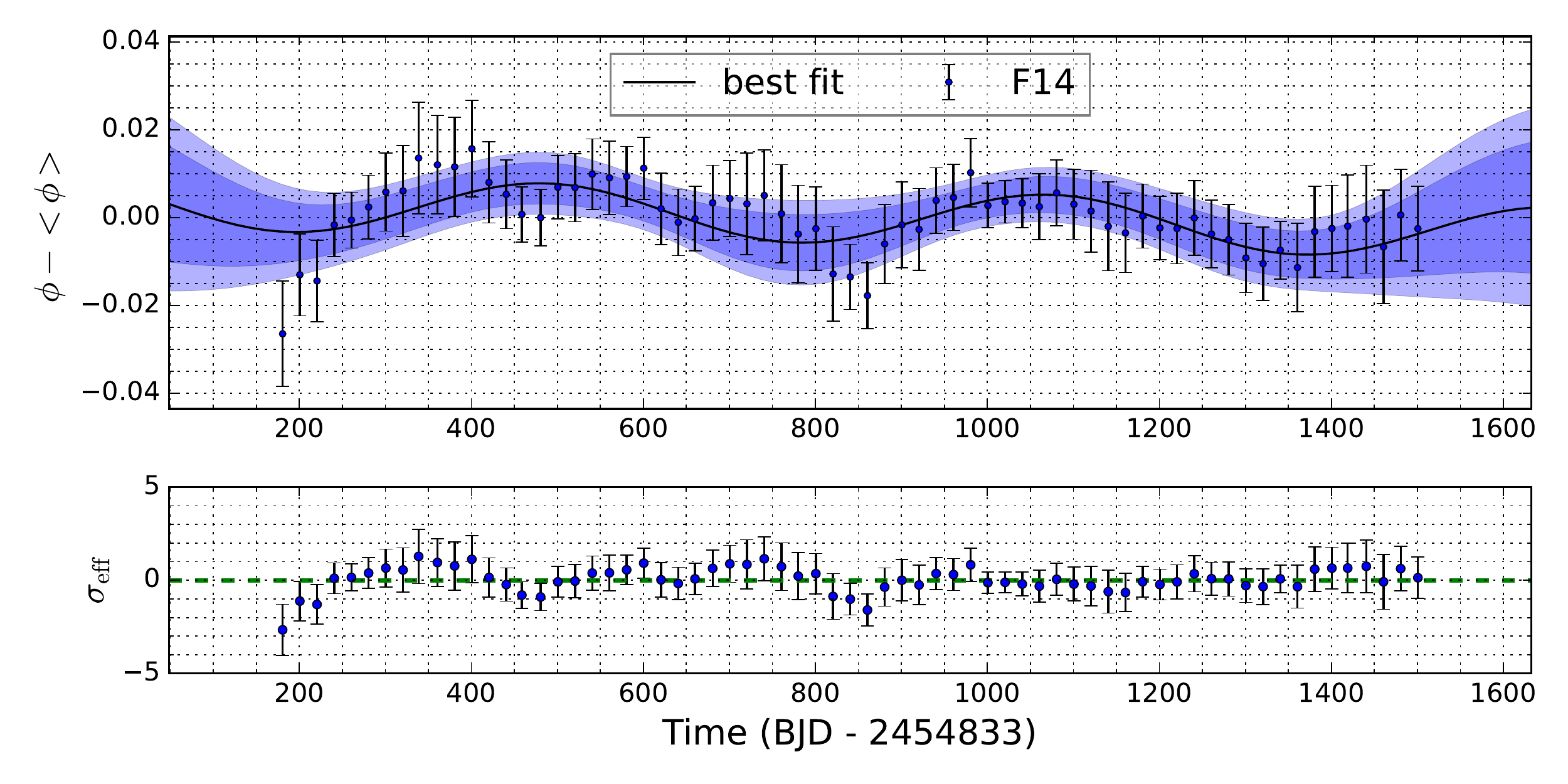}
  \includegraphics[width=0.495\textwidth]{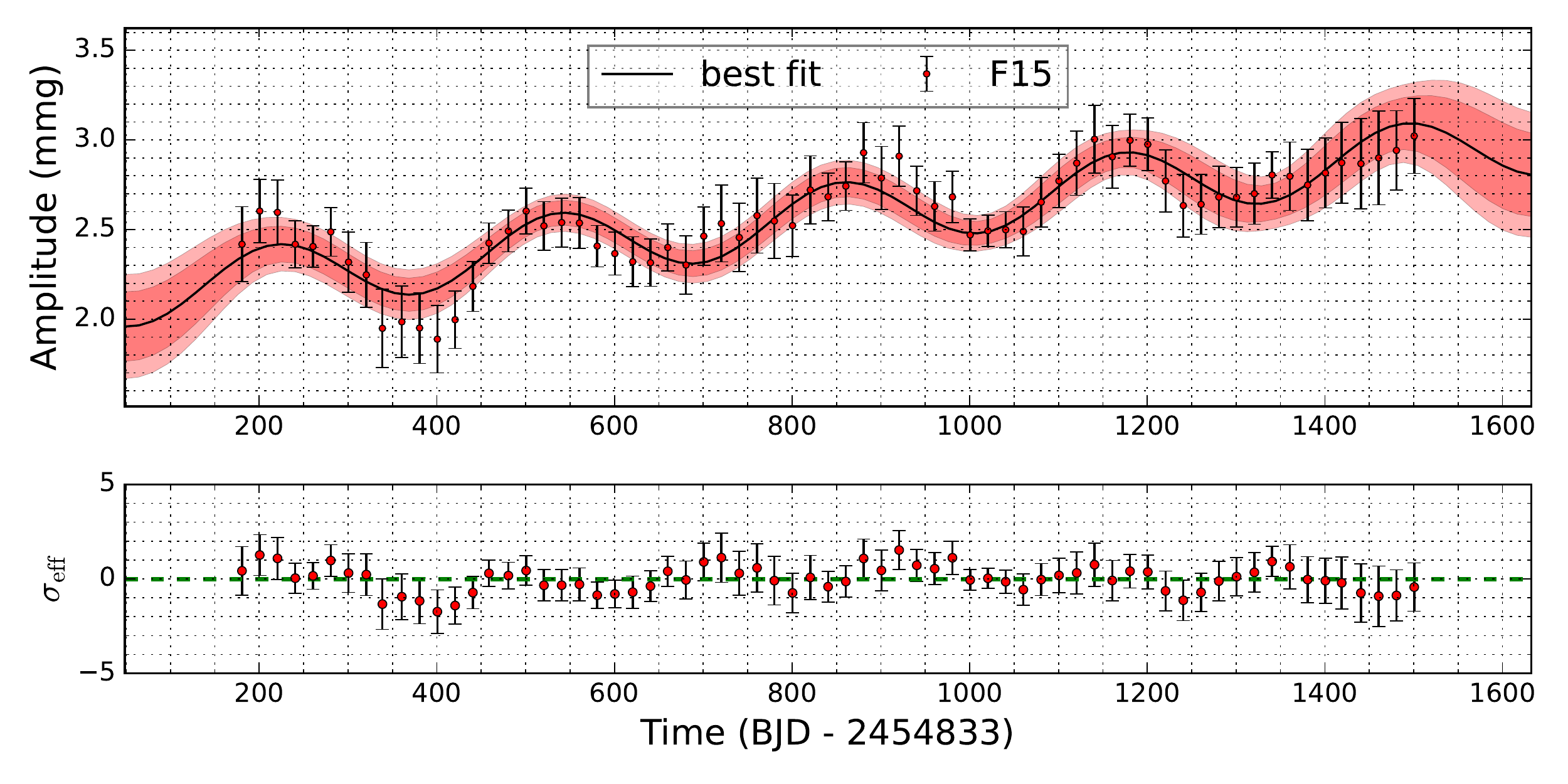}
  \includegraphics[width=0.495\textwidth]{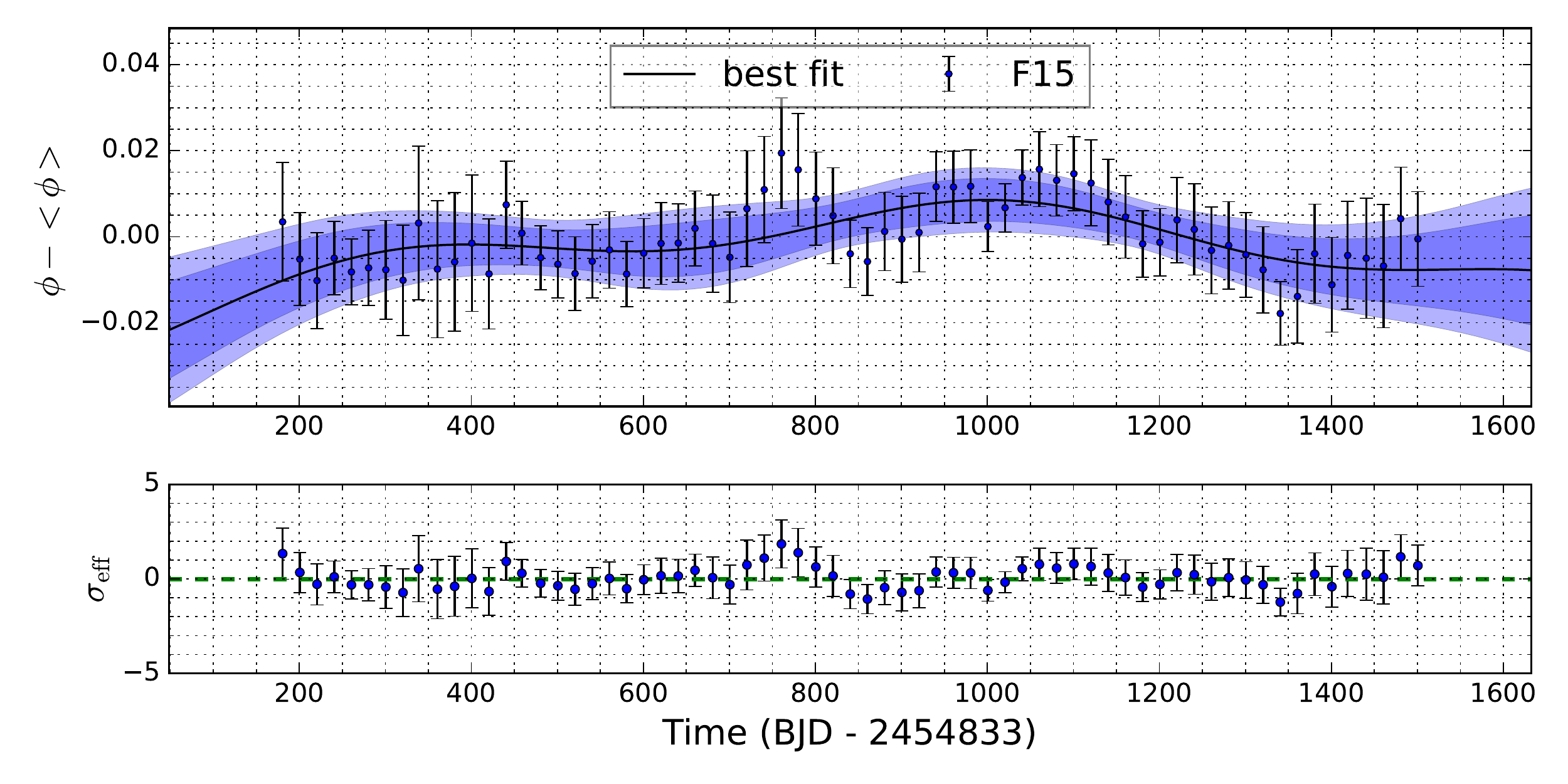}
  \includegraphics[width=0.495\textwidth]{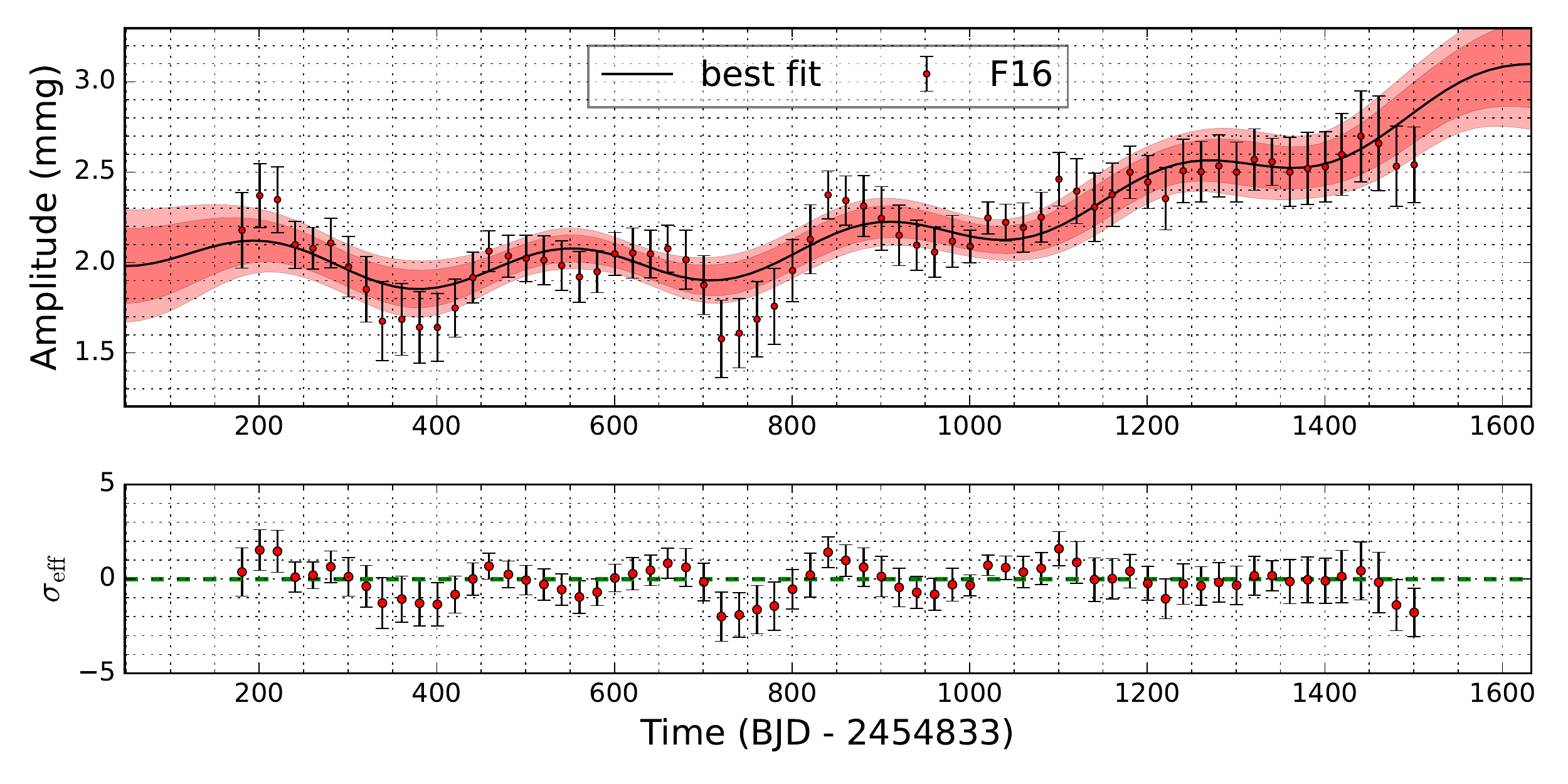}
  \includegraphics[width=0.495\textwidth]{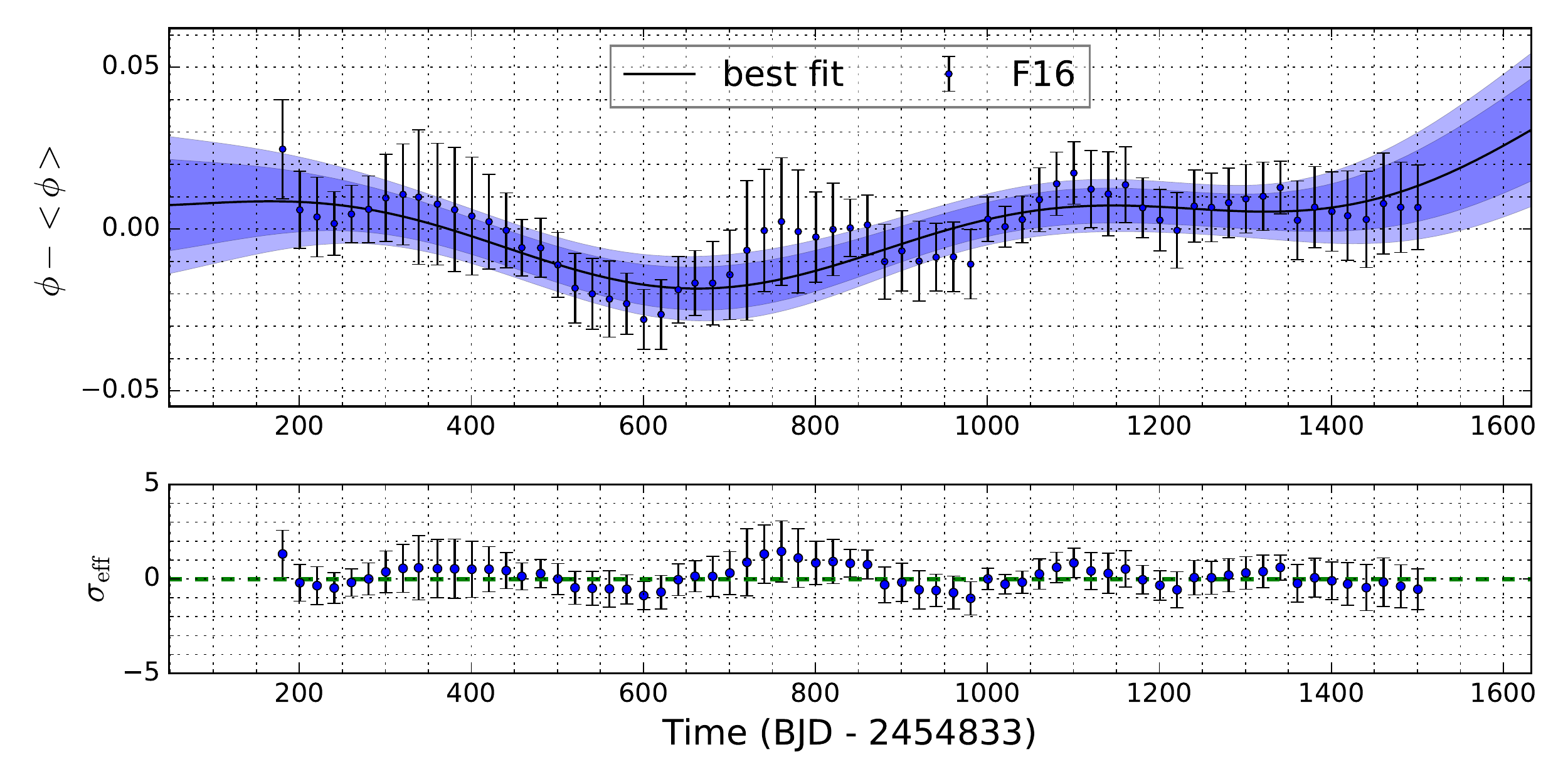}
  \includegraphics[width=0.495\textwidth]{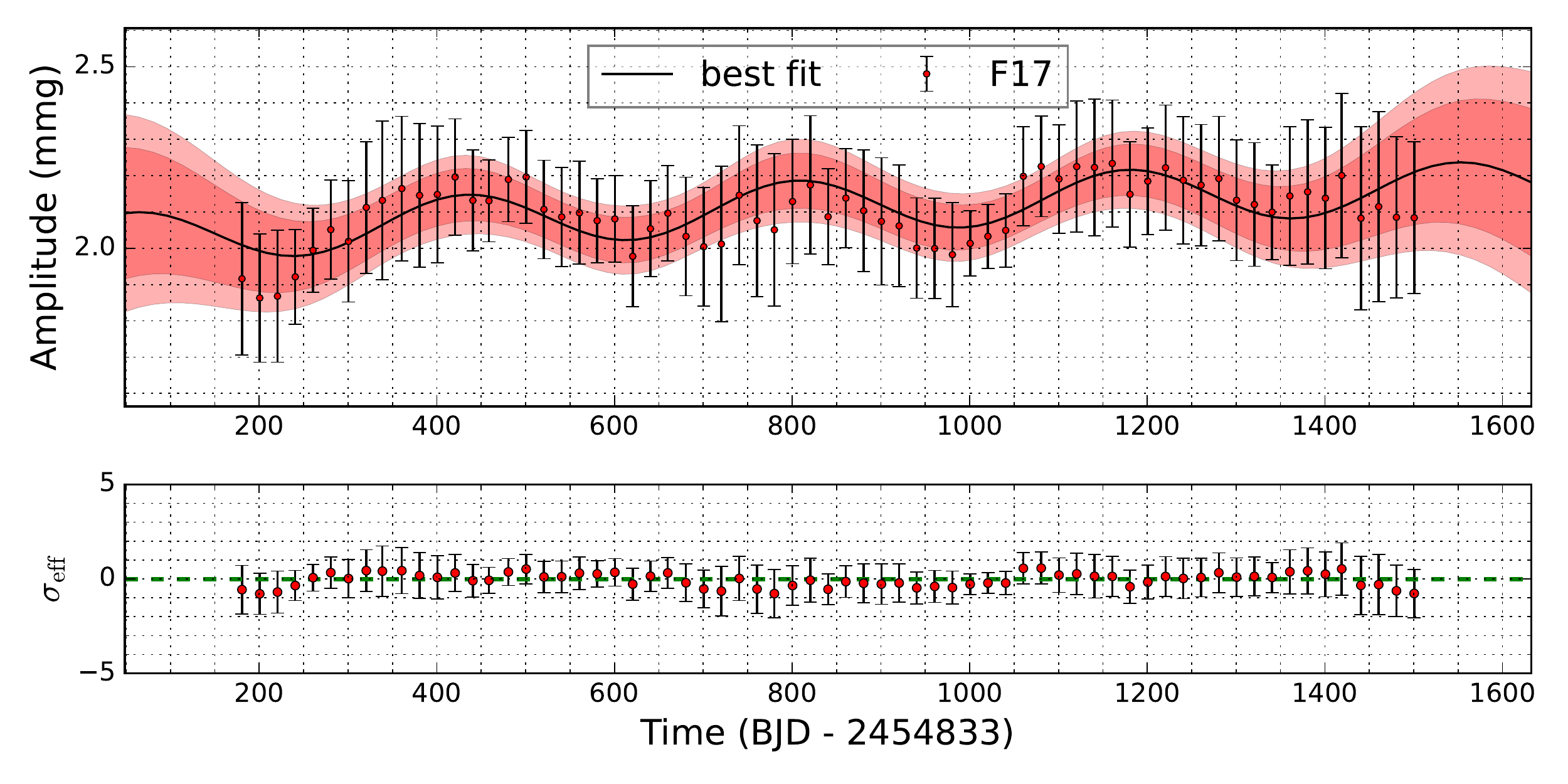}
  \includegraphics[width=0.495\textwidth]{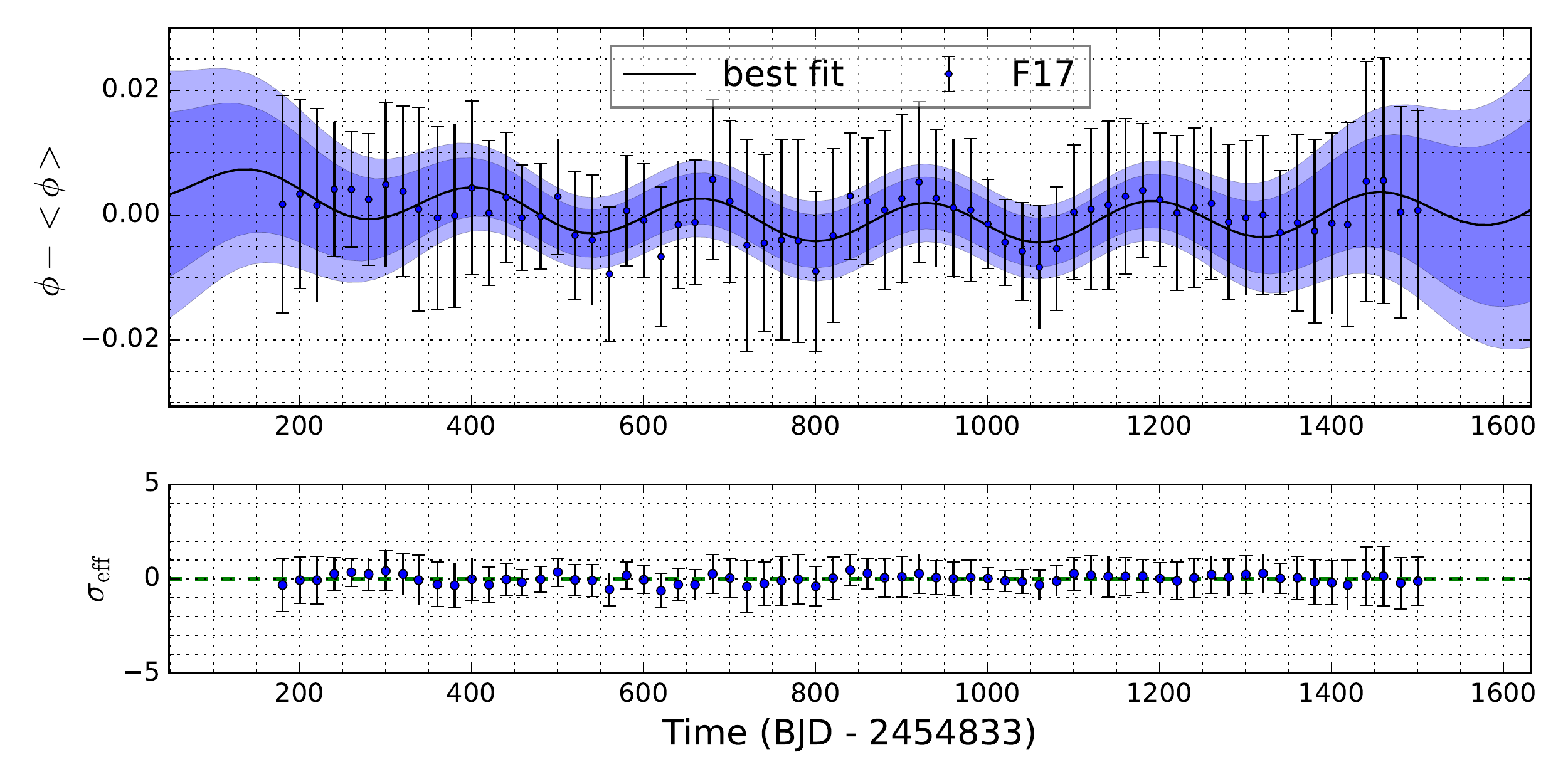}
  \includegraphics[width=0.495\textwidth]{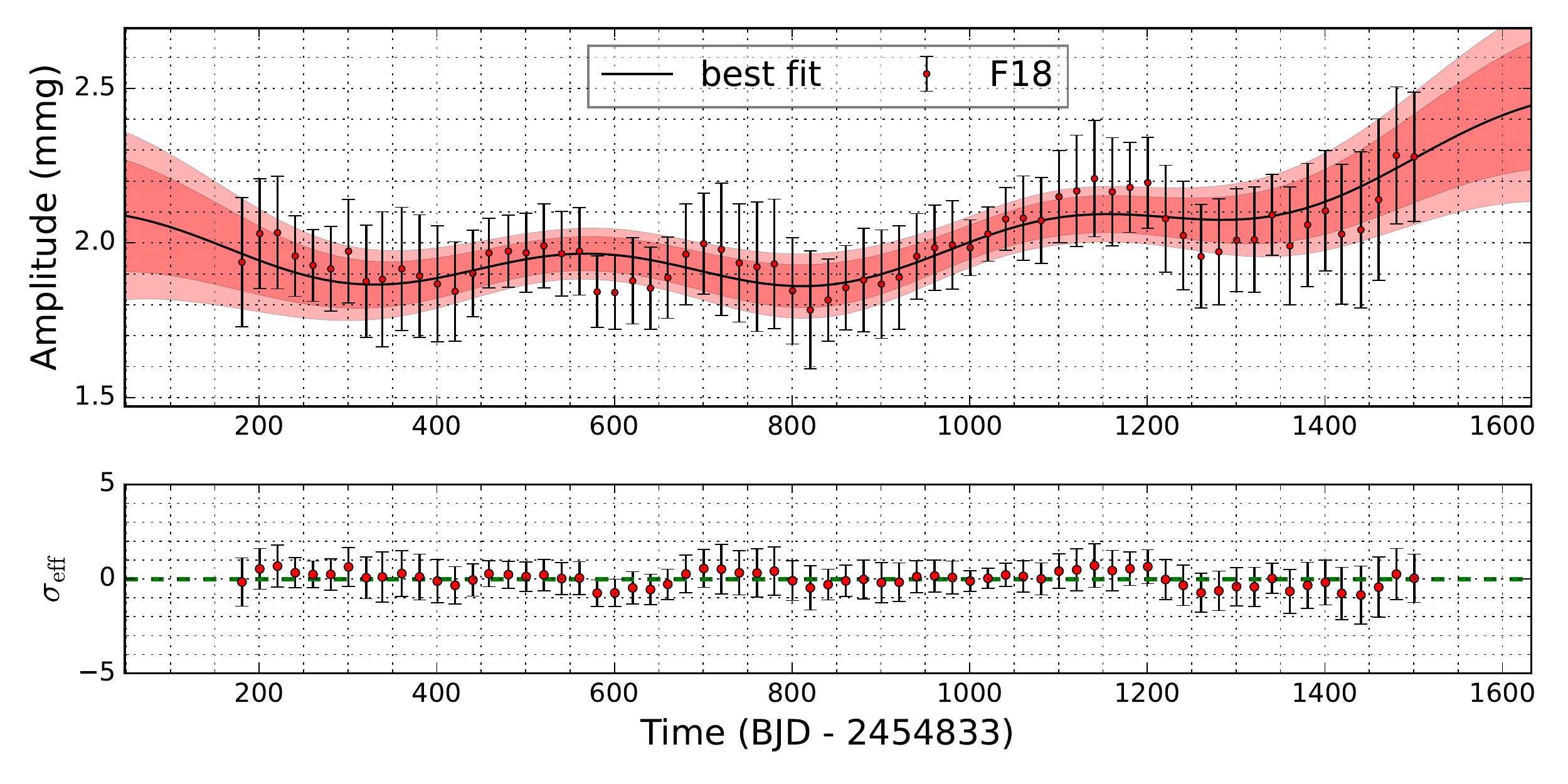}
  \includegraphics[width=0.495\textwidth]{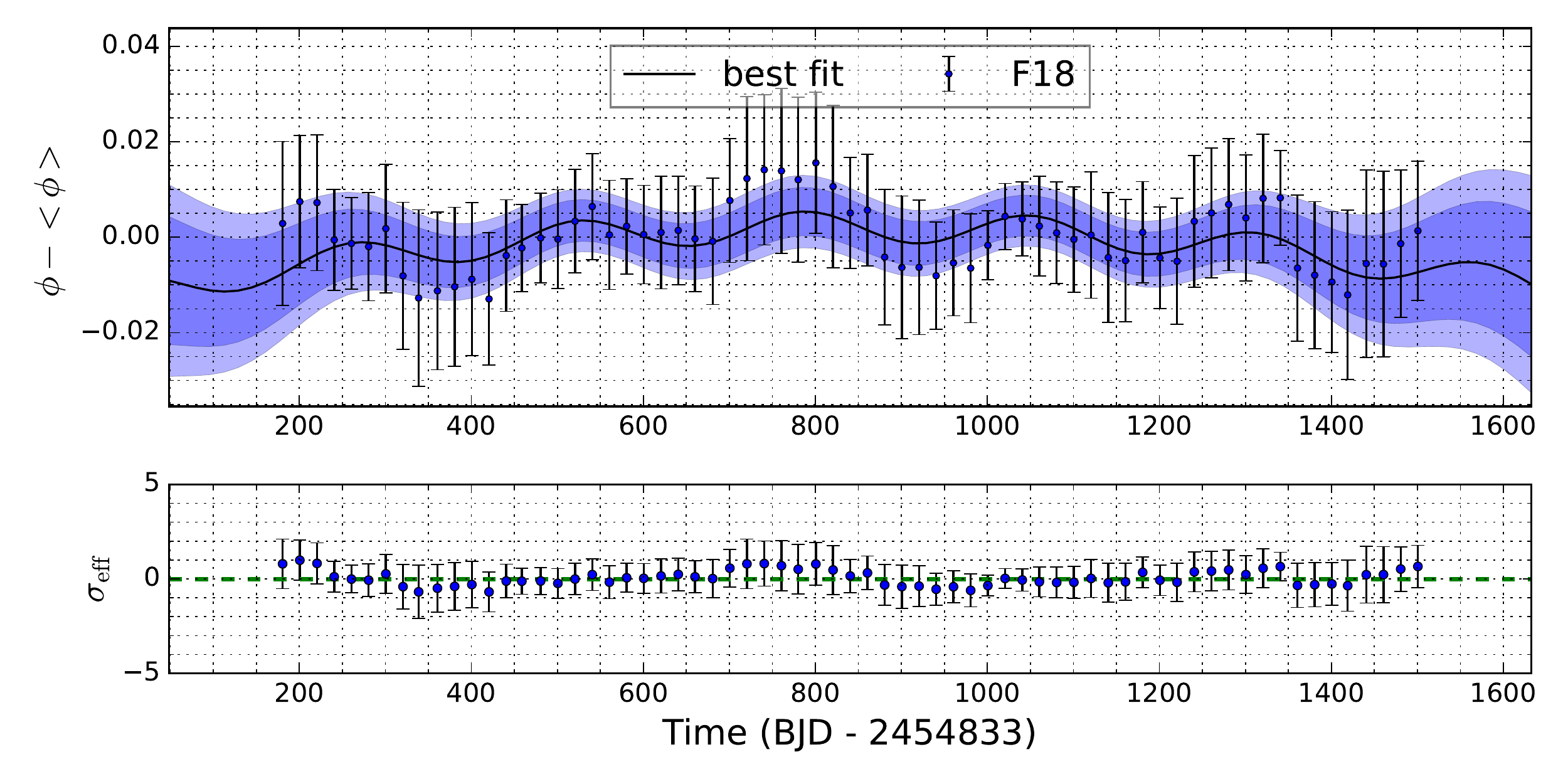}
  \caption{Variation in the amplitudes and phases of the 23 pulsation modes, Part IV.}
  \label{fig:var_amp_phase04}
\end{figure*}

\begin{figure*}[htp]
  \centering
  \includegraphics[width=0.495\textwidth]{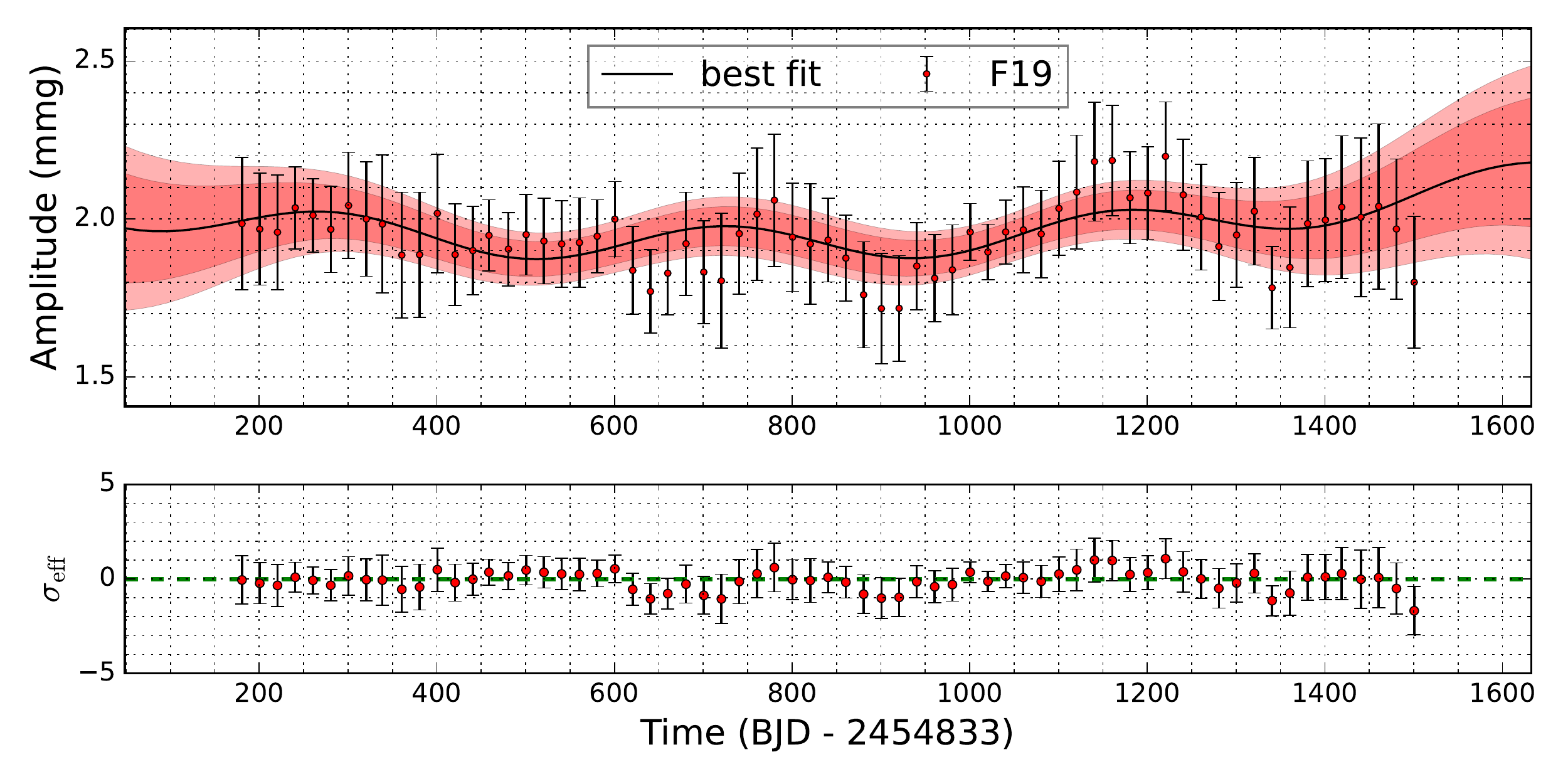}
  \includegraphics[width=0.495\textwidth]{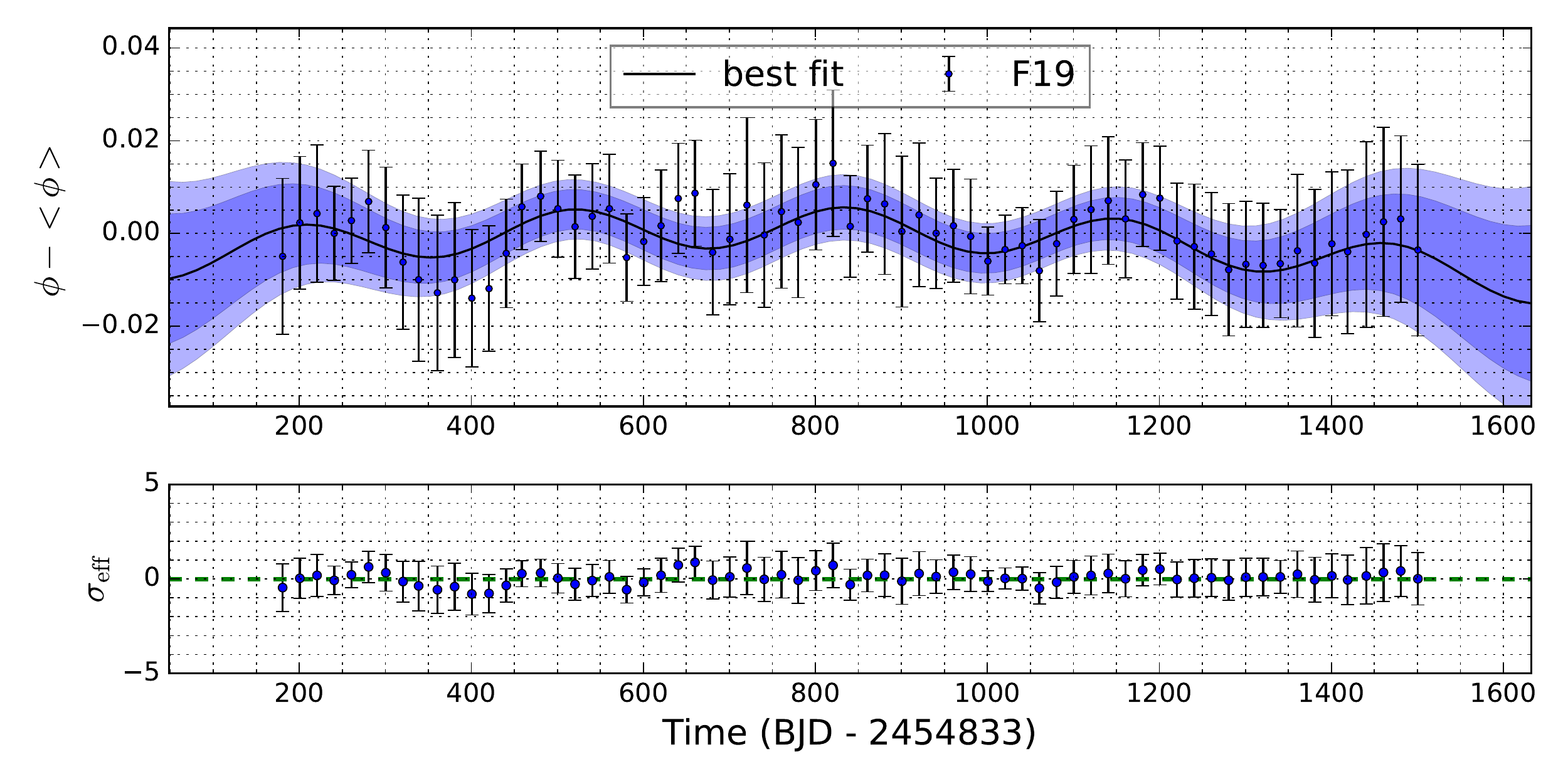}
  \includegraphics[width=0.495\textwidth]{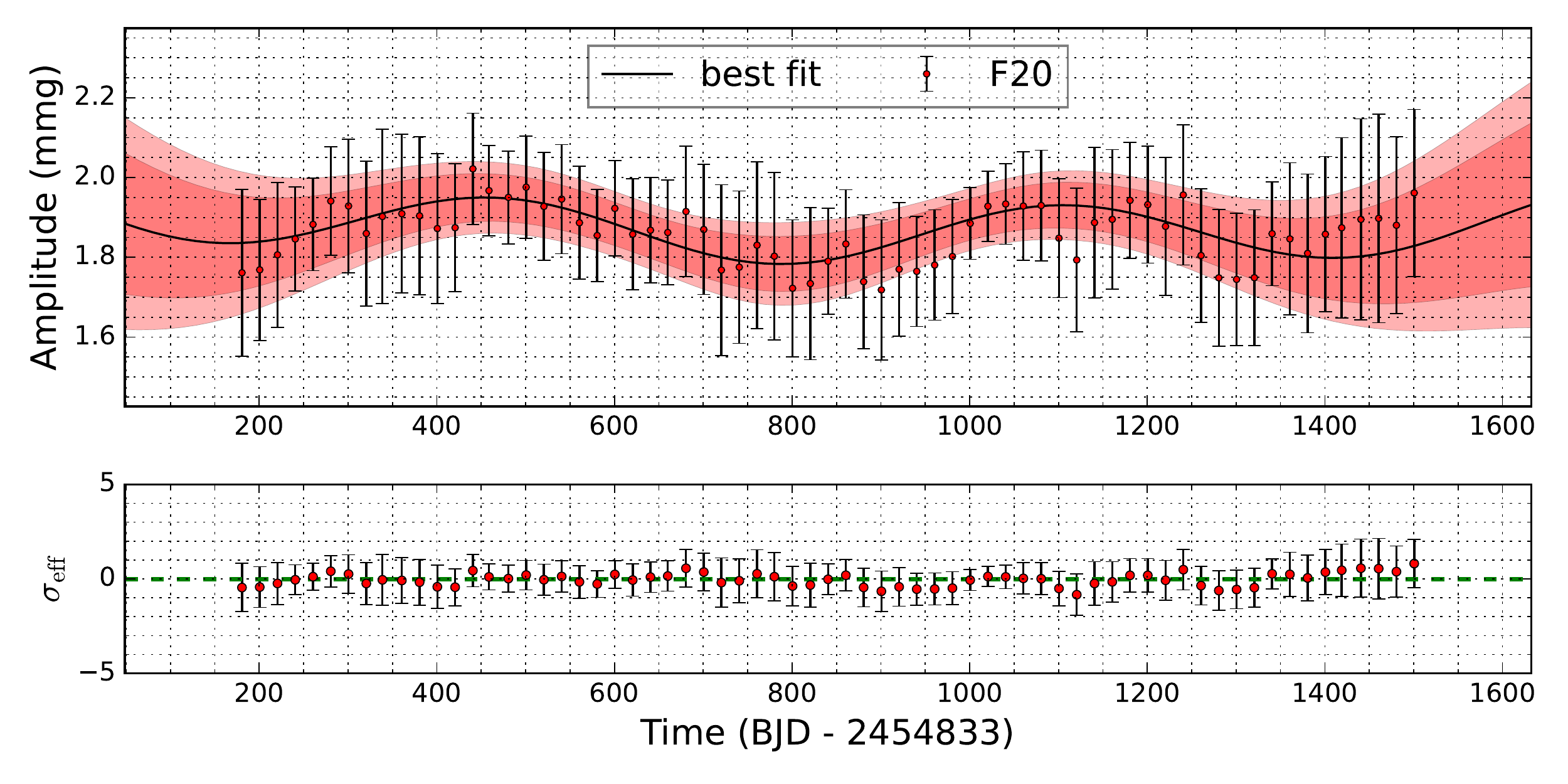}
  \includegraphics[width=0.495\textwidth]{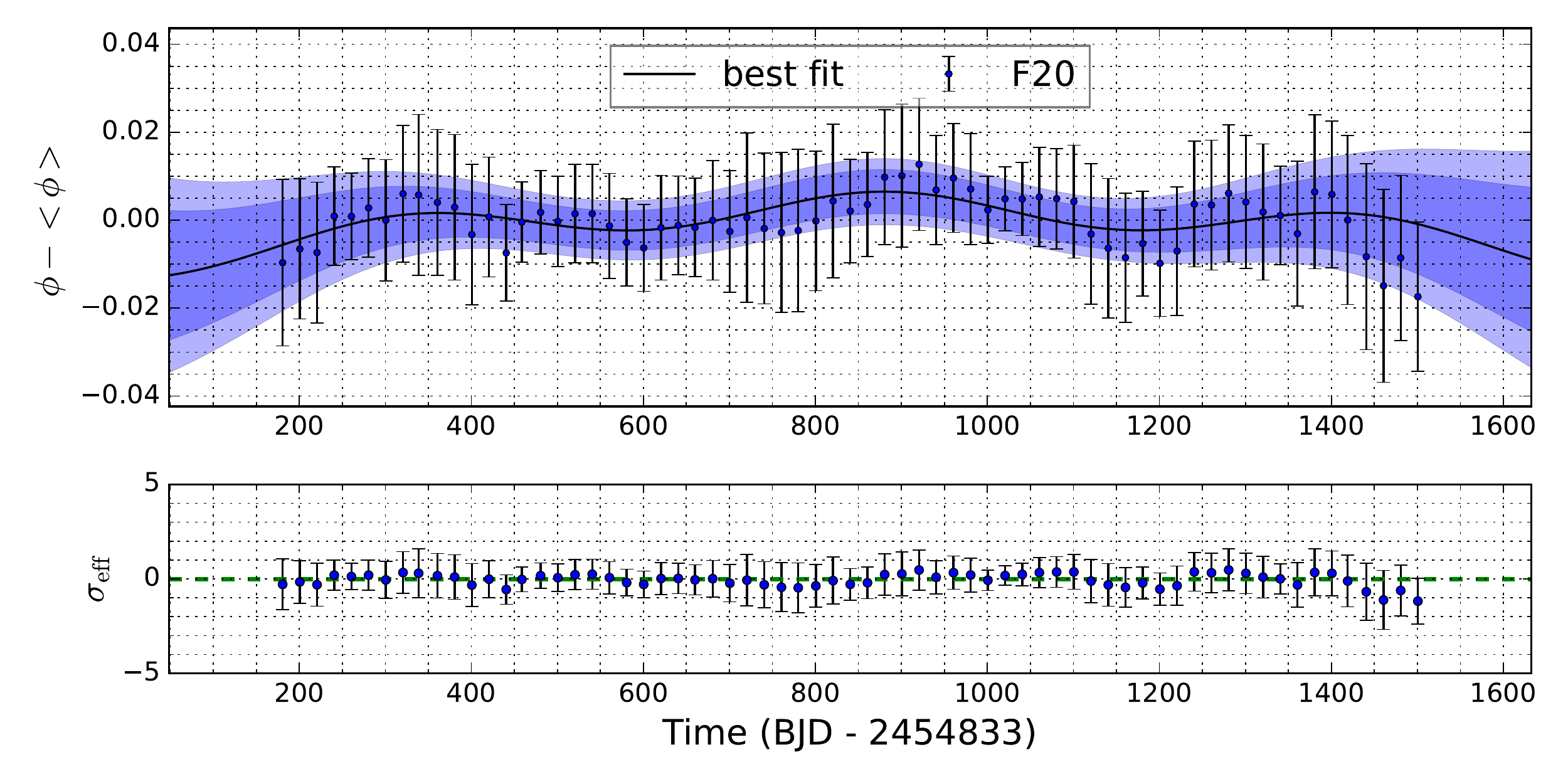}
  \includegraphics[width=0.495\textwidth]{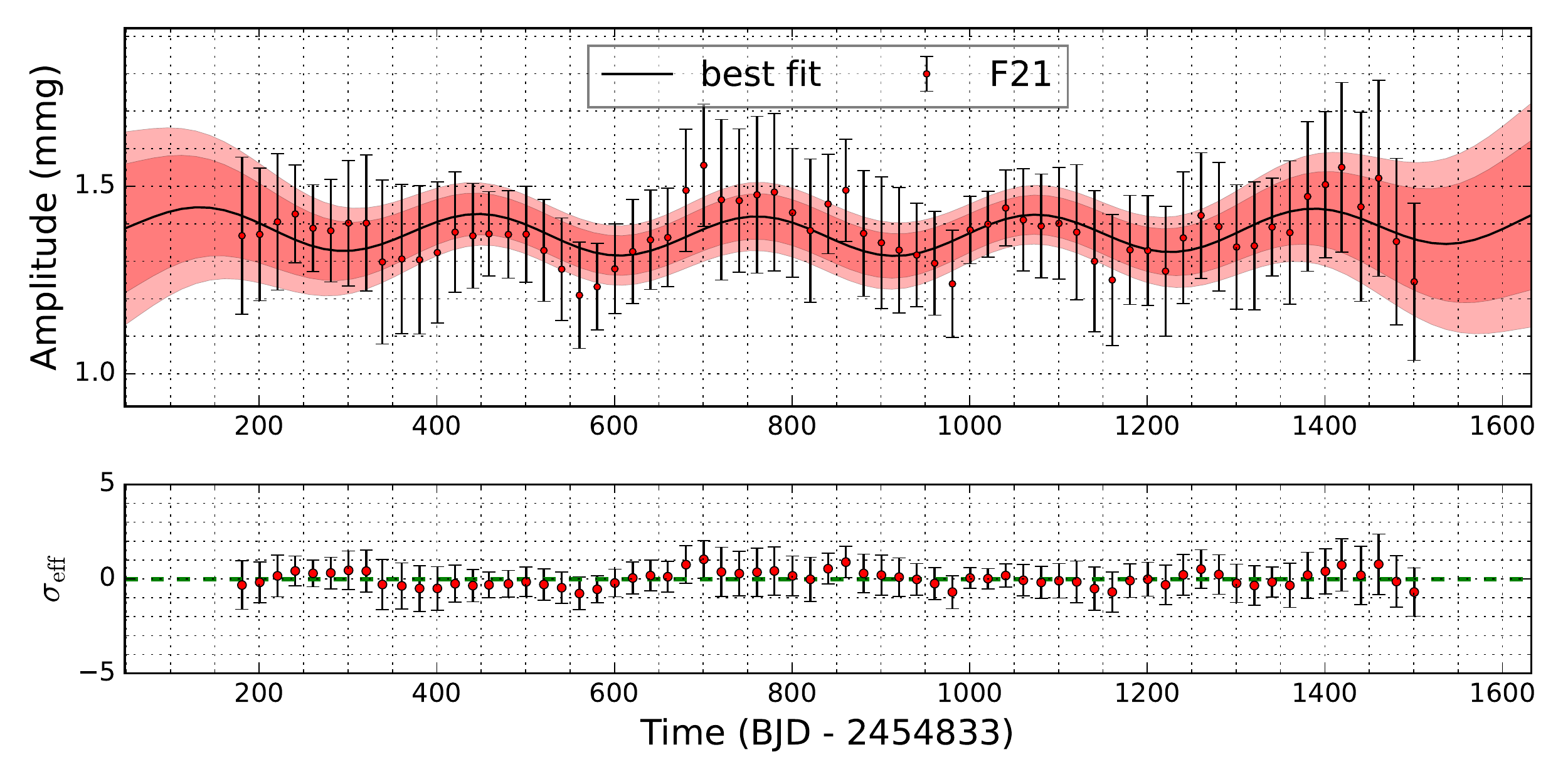}
  \includegraphics[width=0.495\textwidth]{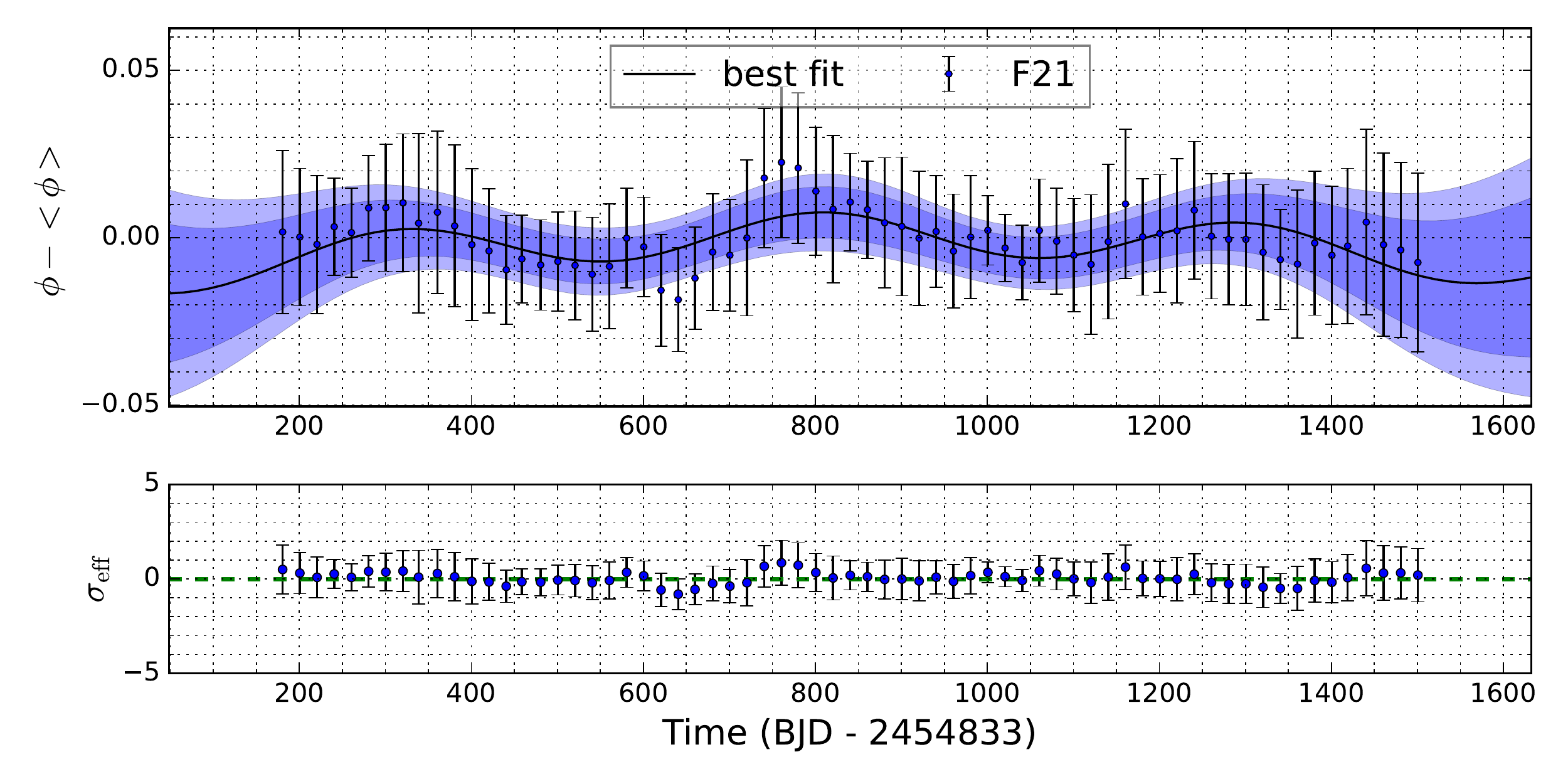}
  \includegraphics[width=0.495\textwidth]{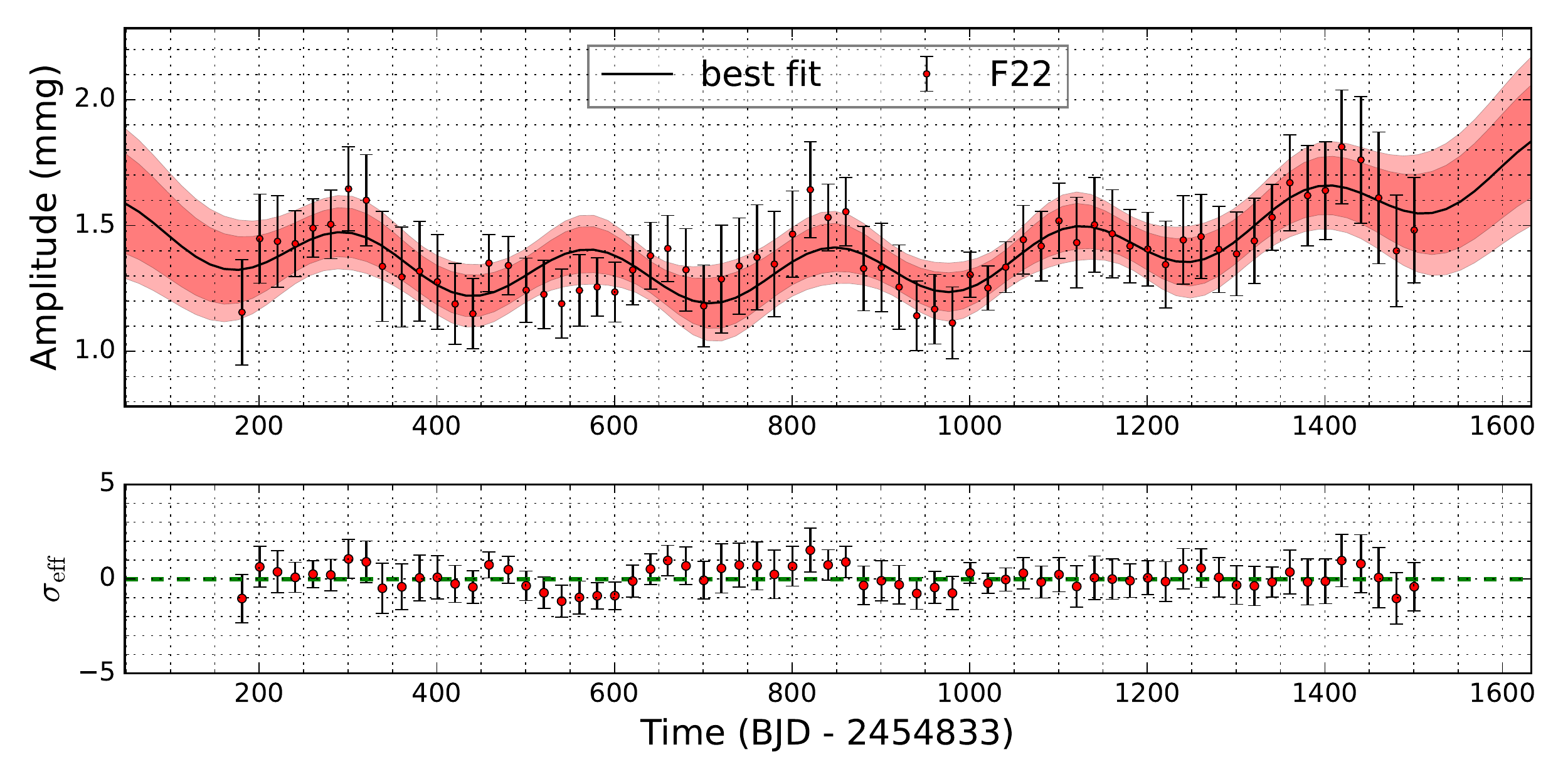}
  \includegraphics[width=0.495\textwidth]{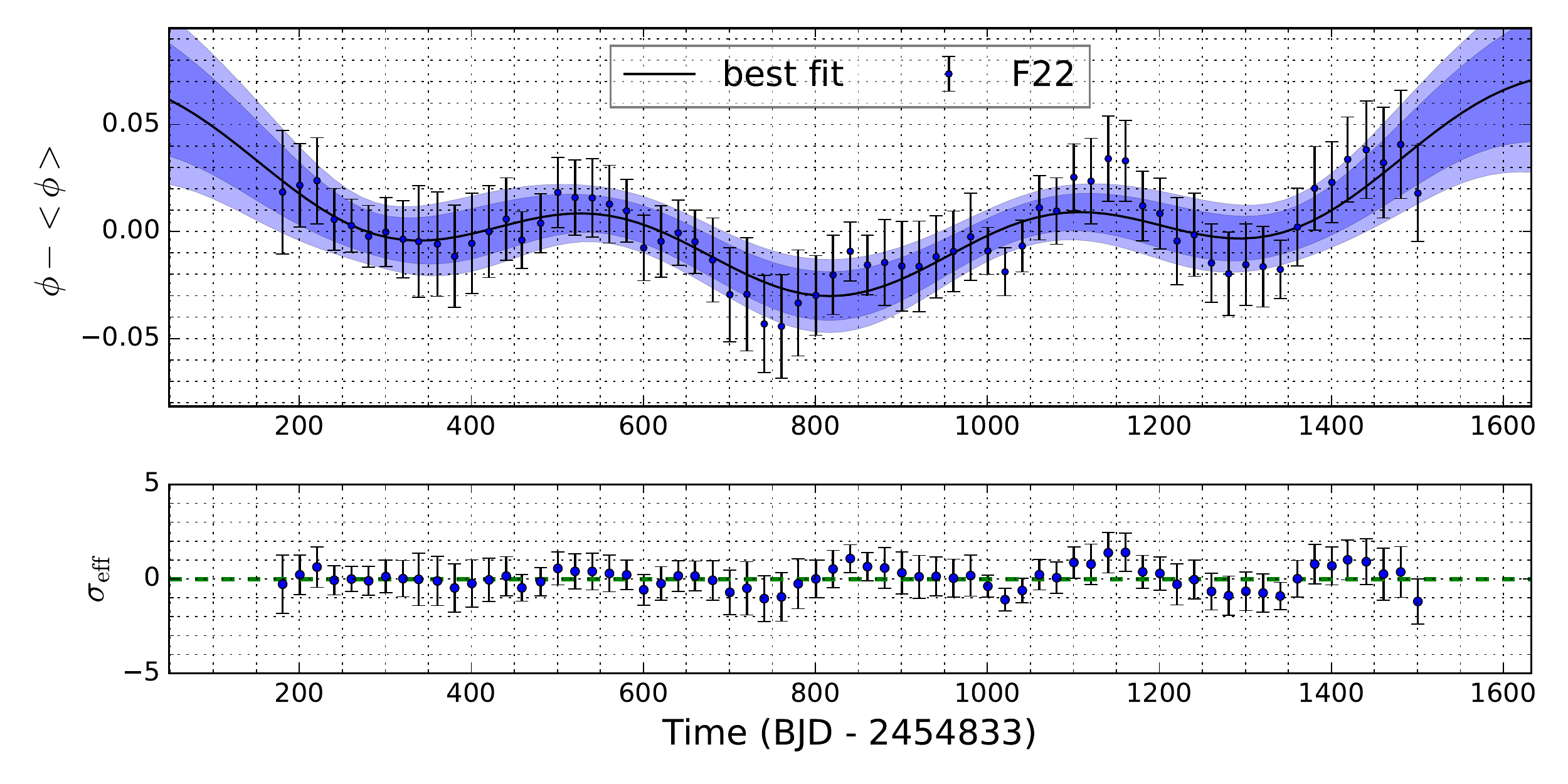}
  \caption{Variation in the amplitudes and phases of the 23 pulsation modes, Part V.}
  \label{fig:var_amp_phase05}
\end{figure*}

At last, the amplitude and phase (subtracted by its average value) for each of the 23 pulsation modes in each of the moving window were collected (with the times that were defined as the midpoints of the window). 
The variations in the amplitudes and phases of the 23 pulsation modes are presented in Figure \ref{fig:var_amp_phase01}, \ref{fig:var_amp_phase02}, \ref{fig:var_amp_phase03}, \ref{fig:var_amp_phase04}, and \ref{fig:var_amp_phase05}.

In this work, the prewhitening process was performed by the Fourier decomposition, which can be presented by the formula
\begin{equation}
  \label{eq:Fourier_de}
  m = m_{0} + \sum A_{i} \sin \left[ 2 \pi (f_{i} t + \phi_{i}) \right] ~,
\end{equation}
where $m_0$ is the shifted value, $A_{i}$ is the amplitude, $f_{i}$ is the frequency, and $\phi_{i}$ is the corresponding phase.

The uncertainties all through the work were estimated by the framework introduced in \citet{Zong2016}, in which the uncertainties of amplitudes ($\sigma_A$) are defined as the median value of the amplitudes within a Lomb-Scargle spectral window of 2 $\cd$ equally divided by the frequency peak, while the uncertainties of the frequencies ($\sigma_f$) and phases ($\sigma_{\phi}$) are estimated following the formalism proposed by \citet{Montgomery1999,Aerts2021}.

\clearpage
\section{Variations in the Amplitudes and Phases}
\label{app:2}
Considering the quasi-periodic signals in most of the amplitudes and phases, which are not the major goals in this work, we fit the variation in the amplitudes by the following formula,
\begin{equation}
\label{eq:fit_a}
a_{A} + b_{A}\cdot t + c_{A} \cdot t^2 + 10^{d_{A}} \cdot \sin[2\pi(\frac{1}{P_{A}} \cdot t + \phi_{A})].
\end{equation}

The variation in the phases are fit by a similar formula,
\begin{equation}
\label{eq:fit_p}
a_{\phi} + b_{\phi}\cdot t + c_{\phi} \cdot t^2 + 10^{d_{\phi}} \cdot \sin[2\pi(\frac{1}{P_{\phi}} \cdot t + \phi_{\phi})].
\end{equation}

In such case, the main quasi-periodic signals (from 200 to 800 days) can be fully handled, and the linear and quadratic variations in the amplitudes and phases can be obtained. 

The Markov Chain Monte Carlo (MCMC) algorithm \citep{Sharma2017} was used to determine the coefficients and their uncertainties in the above expressions, and the fitted coefficients are listed in Table \ref{tab:var_a} and \ref{tab:var_p}.

The variations in amplitudes and phases (subtracted by the average values) are presented in Figure \ref{fig:var_amp_phase01}, \ref{fig:var_amp_phase02}, \ref{fig:var_amp_phase03}, \ref{fig:var_amp_phase04}, and \ref{fig:var_amp_phase05}. In Figure \ref{fig:var_amp_phase01}, \ref{fig:var_amp_phase02}, \ref{fig:var_amp_phase03}, \ref{fig:var_amp_phase04}, and \ref{fig:var_amp_phase05}, we find that almost all these amplitudes and frequencies are modulated by quasi-periodic signals, which have the periods from about 300 to 760 days. It might come from the orbital perturbation and other annual systematic modulation from the {\it Kepler space telescope}, or the light-time perturbation from a companion in a binary system. Because we focus on the linear variation of the pulsation modes in this work, the detailed discussion about the quasi-periodic signals will be proposed in our upcoming works.

Based on the Fourier-phase diagram method \citep{Xue2018}, the linear period variation rates can be directly derived via the values of $c_{\phi}$ (which are equal to the half value of the linear frequency variation rates in $\mathrm{c\ d}^{-2}$).

Because there exist overlaps in the time domain between the data points in Figure \ref{fig:var_amp_phase01}, \ref{fig:var_amp_phase02}, \ref{fig:var_amp_phase03}, \ref{fig:var_amp_phase04} and \ref{fig:var_amp_phase05} (a time window of 120 days and a step of 20 days) and these data points are not independent, the uncertainties of the parameters in Table \ref{tab:var_a} and \ref{tab:var_p} are underestimated. Consequently, we simulated the above process using the data points without overlaps (selecting the next point 120 days away from the last one), and find that the uncertainties of the quadratic term coefficient ($c_A$ and $c_{\phi}$) are about 2-3 times larger than the original ones. For convenience, a uniform factor $\sqrt{6} \approx 2.5$ is used to rescale the uncertainties of $(1/P)(\diff P/ \diff t)$ and $\diff A/ \diff t$ in Table \ref{tab:freq_solution}. $(1/P)(\diff P/ \diff t)$ and $\diff A/ \diff t$ are also used as the criteria to mark the amplitudes and frequencies variations in Figure \ref{fig:spec01}, \ref{fig:spec02}, \ref{fig:spec03}, and \ref{fig:spec04}.

One should note that, as the criterion of the increase or decrease in the amplitudes, the $\diff A/ \diff t$ in Table \ref{tab:freq_solution} is different from the $b_A$ in Table \ref{tab:var_a}. The $\diff A/ \diff t$ (which is equivalent to the $b_{A0}$) is obtained by the following formula:
\begin{equation}
\label{eq:fit_a}
a_{A0} + b_{A0}\cdot t + 10^{d_{A0}} \cdot \sin[2\pi(\frac{1}{P_{A0}} \cdot t + \phi_{A0})].
\end{equation}

As a comparison of the results based on short-time Fourier transformation, the frequencies and amplitudes of the 23 pulsation modes in each year (2009-2013) are summarized in Table \ref{tab:freq_yr}.

\begin{deluxetable}{l|cccccc|c}[htp]
  \label{tab:var_a}
\centering
\tablecaption{Fitting results of the amplitudes of the 23 pulsation modes.}
\tabletypesize{\scriptsize}
\tablehead{
  \colhead{ID}\vline & \colhead{$a_A$}& \colhead{$b_A$}& \colhead{$c_A$} & \colhead{$d_A$} & \colhead{$P_A$} & \colhead{$\phi_A$}\vline & \colhead{$\chi^2/\mathrm{d.o.f.}$}
}
\startdata
    F0&  $79.45\pm0.10$   &  $(1.32\pm0.27)\times10^{-3}$ &$(-7.40\pm1.63)\times10^{-7}$ &  $-0.19\pm0.02$ &  $354\pm3$ &  $-0.08\pm0.02$ & 183.89/61 \\ 
    F1&  $78.05\pm0.10$ &  $(2.30\pm2.70)\times10^{-4}$ &$(-4.18\pm1.62)\times10^{-7}$ &  $-0.20\pm0.02$ &  $355\pm3$ &  $-0.07\pm0.02$ & 110.41/61 \\
    F2&  $28.49\pm0.10$ &  $(-2.04\pm2.70)\times10^{-4}$ &$(0.85\pm1.61)\times10^{-7}$ &  $-0.67\pm0.05$ &  $363\pm8$ &  $-0.03\pm0.06$ & 15.50/61 \\
    F3&  $16.19\pm0.11$  &  $(1.22\pm0.28)\times10^{-3}$ &$(-7.51\pm1.67)\times10^{-7}$ &  $-0.80\pm0.08$ &  $345\pm11$ &  $-0.18\pm0.08$ & 41.55/61 \\
    F4&  $15.99\pm0.10$ &  $(4.32\pm2.66)\times10^{-4}$ &$(-2.80\pm1.60)\times10^{-7}$ &  $-0.88\pm0.09$ &  $356\pm12$ &  $-0.06\pm0.09$ & 10.09/61 \\
    F5&  $11.87\pm0.09$ &  $(-1.55\pm0.26)\times10^{-3}$ &$(2.22\pm0.16)\times10^{-6}$ &  $-3.53\pm1.44$ &  $484\pm175$ &  $-0.11\pm0.23$ & 72.28/61 \\
    F6&  $11.74\pm0.10$ &  $(1.60\pm2.70)\times10^{-4}$ &$(-1.96\pm1.62)\times10^{-7}$ &  $-0.78\pm0.07$ &  $358\pm11$ &  $-0.06\pm0.08$ & 28.88/61 \\
    F7&  $8.35\pm0.10$ &  $(6.17\pm2.64)\times10^{-4}$ &$(-4.51\pm1.60)\times10^{-7}$ &  $-1.19\pm0.53$ &  $724\pm54$ &  $0.33\pm0.09$ & 18.59/61 \\
    F8&  $6.92\pm0.10$ &  $(-1.81\pm2.60)\times10^{-4}$ &$(1.71\pm1.58)\times10^{-7}$ &  $-3.09\pm1.56$ &  $503\pm155$ &  $-0.12\pm0.21$ & 20.90/61 \\
    F9&  $4.90\pm0.10$ &  $(9.44\pm2.62)\times10^{-4}$ &$(-5.11\pm1.57)\times10^{-7}$ &  $-0.84\pm0.09$ &  $375\pm11$ &  $0.18\pm0.07$ & 34.36/61 \\
    F10&  $4.21\pm0.10$ &  $(3.28\pm2.80)\times10^{-4}$ & $(-3.28\pm1.71)\times10^{-7}$ &  $-1.95\pm1.43$ &  $598\pm134$ &  $0.05\pm0.20$ & 28.32/61 \\
    F11&  $3.03\pm0.10$ &  $(0.16\pm2.63)\times10^{-4}$ & $(3.49\pm1.58)\times10^{-7}$ & $-2.44\pm1.61$ &  $435\pm143$ &  $-0.12\pm0.19$ & 19.65/61 \\
    F12&  $3.47\pm0.10$ &  $(-7.21\pm2.62)\times10^{-4}$ & $(4.37\pm1.58)\times10^{-7}$ & $-3.23\pm1.52$ &  $524\pm165$ &  $-0.07\pm0.23$ & 20.93/61 \\
    F13&  $3.01\pm0.10$ &  $(-3.45\pm2.63)\times10^{-4}$ & $(5.36\pm1.59)\times10^{-7}$ &  $-1.13\pm0.20$ &  $399\pm31$ &  $0.08\pm0.12$ & 11.77/61 \\
    F14&  $3.02\pm0.10$ &  $(-7.28\pm2.81)\times10^{-4}$ & $(6.62\pm1.68)\times10^{-7}$ & $-1.16\pm0.24$ &  $466\pm58$ &  $-0.36\pm0.14$ & 22.95/61 \\
    F15&  $2.12\pm0.10$ &  $(5.17\pm2.71)\times10^{-4}$ & $(0.15\pm1.63)\times10^{-7}$ &  $-0.75\pm0.07$ &  $320\pm6$ &  $-0.42\pm0.05$ & 31.77/61 \\
    F16&  $2.13\pm0.10$ &  $(-6.88\pm2.69)\times10^{-4}$ & $(7.52\pm1.61)\times10^{-7}$ &  $-1.08\pm0.23$ &  $341\pm19$ &  $-0.35\pm0.09$ & 43.41/61 \\
    F17&  $2.00\pm0.10$ &  $(1.68\pm2.59)\times10^{-4}$ & $(-0.51\pm1.56)\times10^{-7}$ &  $-3.28\pm1.53$ &  $501\pm176$ &  $-0.11\pm0.24$ & 17.21/61 \\
    F18&  $2.00\pm0.10$ &  $(-3.07\pm2.66)\times10^{-4}$ & $(2.95\pm1.60)\times10^{-7}$ &  $-3.46\pm1.45$ &  $513\pm170$ &  $-0.10\pm0.23$ & 16.29/61 \\
    F19&  $2.06\pm0.10$ &  $(-4.00\pm2.63)\times10^{-4}$ & $(2.58\pm1.59)\times10^{-7}$ &  $-3.49\pm1.43$ &  $503\pm172$ &  $-0.12\pm0.23$ & 23.28/61 \\
    F20&  $1.92\pm0.10$ &  $(-0.97\pm2.65)\times10^{-4}$ & $(0.38\pm1.60)\times10^{-7}$ & $-3.51\pm1.47$ &  $514\pm171$ &  $-0.10\pm0.24$ & 14.06/61 \\
    F21&  $1.38\pm0.10$ &  $(-0.44\pm2.60)\times10^{-4}$ & $(0.41\pm1.57)\times10^{-7}$ &  $-3.59\pm1.40$ &  $505\pm173$ &  $-0.10\pm0.23$ & 14.07/61 \\
    F22&  $1.57\pm0.10$ &  $(-7.52\pm2.74)\times10^{-4}$ & $(5.30\pm1.66)\times10^{-7}$ & $-2.70\pm1.57$ &  $467\pm169$ &  $-0.09\pm0.23$ & 38.77/61  \\
\enddata
\end{deluxetable}


\begin{deluxetable}{l|cccccc|c}
  \label{tab:var_p}
\centering
\tablecaption{Fitting results of the phases (subtracted by its average) of the 23 pulsation modes.}
\tabletypesize{\scriptsize}
\tablehead{
  \colhead{ID}\vline & \colhead{$a_{\phi}$}& \colhead{$b_{\phi}$}& \colhead{$c_{\phi}$} & \colhead{$d_{\phi}$} & \colhead{$P_{\phi}$} & \colhead{$\phi_{\phi}$}\vline & \colhead{$\chi^2/\mathrm{d.o.f.}$}
}
\startdata
    F0&  $(-1.07\pm0.20)\times10^{-3}$   &  $(3.33\pm0.54)\times10^{-6}$ &$(-2.00\pm0.32)\times10^{-9}$ &  $-3.46\pm0.07$ &  $359\pm10$ &  $-0.05\pm0.07$ & 43.53/61 \\ 
    F1&  $(-1.72\pm0.20)\times10^{-3}$ &  $(4.97\pm0.54)\times10^{-6}$ &$(-2.83\pm0.33)\times10^{-9}$ &  $-5.00\pm1.04$ &  $495\pm151$ &  $-0.08\pm0.19$ & 33.30/61 \\
    F2&  $(-2.31\pm0.54)\times10^{-3}$ &  $(6.48\pm1.47)\times10^{-6}$ &$(-3.72\pm0.89)\times10^{-9}$ &  $-3.24\pm0.14$ &  $377\pm17$ &  $0.49\pm0.10$ & 29.75/61 \\
    F3&  $(-5.83\pm0.91)\times10^{-3}$  &  $(1.64\pm0.25)\times10^{-5}$ &$(-9.34\pm1.49)\times10^{-9}$ &  $-4.88\pm1.20$ &  $506\pm172$ &  $-0.06\pm0.21$ & 37.37/61 \\
    F4&  $(-3.40\pm1.01)\times10^{-3}$ &  $(9.21\pm2.78)\times10^{-6}$ &$(-5.28\pm1.68)\times10^{-9}$ &  $-4.72\pm1.24$ &  $498\pm170$ &  $-0.08\pm0.20$ & 23.19/61 \\
    F5&  $(-13.50\pm1.36)\times10^{-3}$ &  $(4.24\pm0.36)\times10^{-5}$ &$(-2.60\pm0.21)\times10^{-8}$ &  $-2.94\pm0.18$ &  $324\pm10$ &  $-0.32\pm0.08$ & 69.78/61 \\
    F6&  $(-2.33\pm1.30)\times10^{-3}$ &  $(9.34\pm3.55)\times10^{-6}$ &$(-6.56\pm2.14)\times10^{-9}$ &  $-4.97\pm1.17$ &  $493\pm174$ &  $-0.05\pm0.20$ & 24.14/61 \\
    F7&  $(-3.82\pm1.78)\times10^{-3}$ &  $(1.18\pm0.48)\times10^{-5}$ &$(-7.25\pm2.92)\times10^{-9}$ &  $-4.91\pm1.22$ &  $500\pm171$ &  $-0.06\pm0.20$ & 21.68/61 \\
    F8&  $(-4.78\pm2.21)\times10^{-3}$ &  $(1.33\pm0.61)\times10^{-5}$ &$(-7.30\pm3.66)\times10^{-9}$ &  $-4.85\pm1.23$ &  $518\pm174$ &  $-0.06\pm0.20$ & 13.50/61 \\
    F9&  $(0.33\pm3.59)\times10^{-3}$ &  $(-4.40\pm9.36)\times10^{-6}$ &$(2.87\pm5.45)\times10^{-9}$ &  $-2.32\pm0.10$ &  $309\pm12$ &  $-0.41\pm0.15$ & 39.28/61 \\
    F10&  $(-9.83\pm3.64)\times10^{-3}$ &  $(3.15\pm0.99)\times10^{-5}$ & $(-1.93\pm0.60)\times10^{-8}$ &  $-4.01\pm1.46$ &  $564\pm160$ &  $-0.13\pm0.20$ & 19.46/61 \\
    F11&  $(-7.09\pm4.88)\times10^{-3}$ &  $(2.28\pm1.29)\times10^{-5}$ & $(-1.36\pm0.76)\times10^{-8}$ & $-4.55\pm1.38$ &  $501\pm166$ &  $-0.04\pm0.21$ & 15.55/61 \\
    F12&  $(-7.02\pm4.72)\times10^{-3}$ &  $(2.23\pm1.29)\times10^{-5}$ & $(-1.43\pm0.78)\times10^{-8}$ & $-4.58\pm1.35$ &  $516\pm170$ &  $-0.04\pm0.20$ & 13.13/61 \\
    F13&  $(-12.38\pm5.05)\times10^{-3}$ &  $(4.57\pm1.32)\times10^{-5}$ & $(-3.07\pm0.77)\times10^{-8}$ &  $-4.75\pm1.29$ &  $503\pm175$ &  $-0.05\pm0.20$ & 16.64/61 \\
    F14&  $(-0.50\pm6.09)\times10^{-3}$ &  $(0.88\pm1.69)\times10^{-5}$ & $(-0.81\pm1.01)\times10^{-8}$ & $-2.49\pm0.38$ &  $588\pm80$ &  $0.42\pm0.14$ & 30.29/61 \\
    F15&  $(-22.33\pm5.95)\times10^{-3}$ &  $(5.84\pm1.58)\times10^{-5}$ & $(-3.20\pm0.94)\times10^{-8}$ &  $-3.72\pm1.51$ &  $581\pm178$ &  $-0.09\pm0.18$ & 35.60/61 \\
    F16&  $(10.85\pm7.11)\times10^{-3}$ &  $(-6.06\pm1.88)\times10^{-5}$ & $(4.56\pm1.12)\times10^{-8}$ &  $-2.13\pm0.32$ &  $761\pm38$ &  $-0.12\pm0.06$ & 22.16/61 \\
    F17&  $(6.64\pm7.44)\times10^{-3}$ &  $(-1.83\pm2.00)\times10^{-5}$ & $(1.01\pm1.20)\times10^{-8}$ &  $-4.82\pm1.27$ &  $497\pm173$ &  $-0.05\pm0.20$ & 5.48/61 \\
    F18&  $(-4.89\pm7.59)\times10^{-3}$ &  $(1.52\pm2.06)\times10^{-5}$ & $(-0.90\pm1.23)\times10^{-8}$ &  $-4.62\pm1.36$ &  $499\pm172$ &  $-0.03\pm0.20$ & 14.21/61 \\
    F19&  $(-2.97\pm7.67)\times10^{-3}$ &  $(1.32\pm2.11)\times10^{-5}$ & $(-0.96\pm1.29)\times10^{-8}$ &  $-4.64\pm1.33$ &  $490\pm174$ &  $-0.05\pm0.20$ & 13.31/61 \\
    F20&  $(-8.05\pm8.22)\times10^{-3}$ &  $(2.48\pm2.23)\times10^{-5}$ & $(-1.46\pm1.34)\times10^{-8}$ & $-4.67\pm1.36$ &  $511\pm173$ &  $-0.06\pm0.20$ & 9.56/61 \\
    F21&  $(-1.36\pm11.10)\times10^{-3}$ &  $(0.17\pm3.03)\times10^{-5}$ & $(-0.12\pm1.83)\times10^{-8}$ &  $-4.42\pm1.46$ &  $517\pm171$ &  $-0.07\pm0.20$ & 12.13/61 \\
    F22&  $(49.17\pm12.24)\times10^{-3}$ &  $(-15.47\pm3.37)\times10^{-5}$ & $(9.56\pm2.01)\times10^{-8}$ & $-1.89\pm0.15$ &  $525\pm31$ &  $0.19\pm0.10$ & 26.45/61 \\
\enddata
\end{deluxetable}

\begin{table*}[htp]
  \centering
  \caption{Frequencies and amplitudes of the 23 pulsation modes in the year from 2009 to 2013. Frequency is in $\cd$; Amplitude is in mmag.}
  \label{tab:freq_yr}
  \scalebox{0.75}[0.75]{
\rotatebox[origin=c]{90}{
  \begin{tabular}{c|rr|rr|rr|rr|rr}
    \hline
    \hline
    &  \multicolumn{2}{c}{2009 (Q0-Q3)} & \multicolumn{2}{|c}{2010 (Q4-Q7)} & \multicolumn{2}{|c}{2011 (Q8-Q10)} & \multicolumn{2}{|c}{2012 (Q11-Q14)} & \multicolumn{2}{|c}{2013 (Q15-Q17)}\\
\hline
ID &  \multicolumn{1}{c}{Frequency} & \multicolumn{1}{c|}{Amplitude} & \multicolumn{1}{c}{Frequency} & \multicolumn{1}{c|}{Amplitude} & \multicolumn{1}{c}{Frequency} & \multicolumn{1}{c|}{Amplitude} &  \multicolumn{1}{c}{Frequency} & \multicolumn{1}{c|}{Amplitude} & \multicolumn{1}{c}{Frequency} & \multicolumn{1}{c}{Amplitude} \\
\hline
F0 & $4.90984 \pm0.00002$ &	$79.4\pm0.6	$  &  $4.909843 \pm0.000007$ & $80.0 \pm0.4 $ & $4.90984 \pm0.00001 $  & $80.0\pm0.5 $  &  $4.909844 \pm0.000008$ & $79.9 \pm0.4 $ & $4.90985 \pm0.00002$ & $80.0\pm0.6$ \\         
F1 & $6.43189 \pm0.00001$ &	$77.8\pm0.4	$  &  $6.431890 \pm0.000005$ & $78.1 \pm0.2 $ & $6.431887\pm0.000008$  & $77.8\pm0.3 $  &  $6.431884 \pm0.000005$ & $77.7 \pm0.3 $ & $6.43188 \pm0.00001$ & $77.6\pm0.4$ \\
F2 & $11.34175\pm0.00002$ &	$28.4\pm0.2	$  &  $11.341738\pm0.000008$ & $28.4 \pm0.2 $ & $11.34173\pm0.00001 $  & $28.4\pm0.2 $  &  $11.341727\pm0.000009$ & $28.4 \pm0.2 $ & $11.34173\pm0.00002$ & $28.4\pm0.3$ \\
F3 & $1.52203 \pm0.00003$ &	$16.2\pm0.2	$  &  $1.52204	\pm0.00001 $ & $16.6 \pm0.1 $ & $1.52204 \pm0.00002 $  & $16.6\pm0.2 $  &  $1.52203  \pm0.00001 $ & $16.6 \pm0.1 $ & $1.52205 \pm0.00004$ & $16.5\pm0.3$ \\
F4 & $9.81969 \pm0.00002$ &	$16.0\pm0.1	$  &  $9.819694	\pm0.000009$ & $16.15\pm0.09$ & $9.81968 \pm0.00002 $  & $16.1\pm0.1 $  &  $9.81969  \pm0.00001 $ & $16.1 \pm0.1 $ & $9.81969 \pm0.00003$ & $16.0\pm0.2$ \\
F5 & $8.03546 \pm0.00003$ &	$11.5\pm0.1 $  &  $8.03545	\pm0.00001 $ & $11.68\pm0.07$ & $8.03540 \pm0.00002 $  & $12.2\pm0.1 $  &  $8.03540  \pm0.00001 $ & $13.24\pm0.09$ & $8.03547 \pm0.00003$ & $14.5\pm0.2$ \\
F6 & $12.86381\pm0.00002$ &	$11.8\pm0.1	$  &  $12.86379	\pm0.00001 $ & $11.77\pm0.08$ & $12.86377\pm0.00002 $  & $11.8\pm0.1 $  &  $12.86378 \pm0.00001 $ & $11.61\pm0.08$ & $12.86383\pm0.00003$ & $11.5\pm0.2$ \\
F7 & $16.25157\pm0.00003$ &	$8.45\pm0.09$  &  $16.25158	\pm0.00001 $ & $8.57 \pm0.08$ & $16.25158\pm0.00002 $  & $8.57\pm0.09$  &  $16.25158 \pm0.00001 $ & $8.43 \pm0.08$ & $16.25166\pm0.00004$ & $ 8.6\pm0.1$ \\
F8 & $7.95391 \pm0.00004$ &	$7.0 \pm0.1	$  &  $7.95394	\pm0.00001 $ & $6.84 \pm0.07$ & $7.95394 \pm0.00003 $  & $6.9 \pm0.1 $  &  $7.95392  \pm0.00002 $ & $7.01 \pm0.07$ & $7.95396 \pm0.00005$ & $ 7.0\pm0.2$ \\
F9 & $3.38771 \pm0.00008$ &	$5.0 \pm0.2	$  &  $3.38779	\pm0.00003 $ & $5.23 \pm0.09$ & $3.38777 \pm0.00006 $  & $5.4 \pm0.1 $  &  $3.38780  \pm0.00003 $ & $5.3  \pm0.1 $ & $3.3878  \pm0.0001 $ & $ 5.4\pm0.2$ \\
F10& $17.77368\pm0.00004$ &	$4.24\pm0.07$  &  $17.77358	\pm0.00003 $ & $4.29 \pm0.07$ & $17.77363\pm0.00003 $  & $4.29\pm0.07$  &  $17.77360 \pm0.00003 $ & $4.12 \pm0.07$ & $17.77367\pm0.00008$ & $ 4.2\pm0.1$ \\
F11& $12.94528\pm0.00006$ &	$3.06\pm0.08$  &  $12.94523	\pm0.00003 $ & $3.15 \pm0.07$ & $12.94531\pm0.00005 $  & $3.30\pm0.08$  &  $12.94523 \pm0.00003 $ & $3.56 \pm0.06$ & $12.94522\pm0.00009$ & $ 3.9\pm0.1$ \\
F12& $22.68354\pm0.00005$ &	$3.32\pm0.06$  &  $22.68347	\pm0.00003 $ & $3.24 \pm0.07$ & $22.68341\pm0.00005 $  & $3.15\pm0.07$  &  $22.68347 \pm0.00003 $ & $3.22 \pm0.06$ & $22.68343\pm0.00009$ & $ 3.3\pm0.1$ \\
F13& $14.46749\pm0.00006$ &	$2.88\pm0.07$  &  $14.46735	\pm0.00003 $ & $3.01 \pm0.07$ & $14.46727\pm0.00005 $  & $3.14\pm0.07$  &  $14.46730 \pm0.00003 $ & $3.33 \pm0.06$ & $14.46745\pm0.00008$ & $ 3.7\pm0.1$ \\
F14& $1.6038  \pm0.0001	$ & $2.9 \pm0.2	$  &  $1.60347	\pm0.00005 $ & $2.76 \pm0.09$ & $1.6036  \pm0.0001  $  & $2.9 \pm0.1 $  &  $1.60349  \pm0.00005 $ & $3.1  \pm0.1 $ & $1.6035  \pm0.0002 $ & $ 3.2\pm0.2$ \\
F15& $3.1255  \pm0.0002	$ & $2.4 \pm0.2	$  &  $3.12562	\pm0.00006 $ & $2.28 \pm0.09$ & $3.1256  \pm0.0001  $  & $2.7 \pm0.1 $  &  $3.12550  \pm0.00005 $ & $2.7  \pm0.1 $ & $3.1256  \pm0.0002 $ & $ 2.9\pm0.2$ \\
F16& $3.3060  \pm0.0002	$ & $2.2 \pm0.2	$  &  $3.30620	\pm0.00007 $ & $1.90 \pm0.09$ & $3.3063  \pm0.0001  $  & $2.0 \pm0.1 $  &  $3.30632  \pm0.00006 $ & $2.4  \pm0.1 $ & $3.3063  \pm0.0002 $ & $ 2.6\pm0.2$ \\
F17& $14.72953\pm0.00008$ &	$1.98\pm0.06$  &  $14.72952	\pm0.00005 $ & $2.11 \pm0.07$ & $14.72961\pm0.00007 $  & $2.05\pm0.07$  &  $14.72956 \pm0.00004 $ & $2.16 \pm0.06$ & $14.7295 \pm0.0001 $ & $ 2.2\pm0.1$ \\
F18& $9.5574  \pm0.0001	$ & $1.94\pm0.08$  &  $9.55751	\pm0.00004 $ & $1.91 \pm0.05$ & $9.55733 \pm0.00009 $  & $1.91\pm0.08$  &  $9.55746  \pm0.00004 $ & $2.09 \pm0.06$ & $9.5576  \pm0.0002 $ & $ 2.1\pm0.1$ \\
F19& $19.29575\pm0.00008$ &	$1.97\pm0.06$  &  $19.29574	\pm0.00006 $ & $1.91 \pm0.07$ & $19.29569\pm0.00008 $  & $1.90\pm0.07$  &  $19.29567 \pm0.00005 $ & $1.99 \pm0.06$ & $19.2957 \pm0.0002 $ & $ 1.9\pm0.1$ \\
F20& $24.20562\pm0.00007$ &	$1.84\pm0.06$  &  $24.20550	\pm0.00005 $ & $1.89 \pm0.07$ & $24.20554\pm0.00008 $  & $1.76\pm0.07$  &  $24.20549 \pm0.00005 $ & $1.87 \pm0.06$ & $24.2055 \pm0.0001 $ & $ 2.0\pm0.1$ \\
F21& $21.1616 \pm0.0001	$ & $1.37\pm0.06$  &  $21.16138	\pm0.00008 $ & $1.32 \pm0.07$ & $21.1613 \pm0.0001  $  & $1.40\pm0.07$  &  $21.16142 \pm0.00006 $ & $1.36 \pm0.06$ & $21.1616 \pm0.0002 $ & $ 1.4\pm0.1$ \\
F22& $3.0438  \pm0.0003	$ & $1.3 \pm0.2 $  &  $3.0441	\pm0.0001  $ & $1.31 \pm0.09$ & $3.0443  \pm0.0002  $  & $1.3 \pm0.1 $  &  $3.0440   \pm0.0001  $ & $1.43 \pm0.09$ & $3.0436  \pm0.0003 $ & $ 1.6\pm0.2$ \\
    \hline
  \end{tabular}
  }
  }
\end{table*}

\clearpage
\section{interaction diagram}
\label{app:3}
The interaction diagram is a useful tool to show the global interaction structures between the pulsation modes based on their amplitude, frequency, and phase variations, and clusters the modes with similar behaviors together when they interact with all the other modes.

The interaction strength of a characteristic quantity (amplitude/frequency/phase) between two pulsation modes is measured by the correlation coefficient of their variation over time. In this work, we use the Spearman correlation coefficient (which is a measure of how well the relationship between two variables can be described by a monotonic function) rather than the Pearson correlation coefficient because of (i) the dataset is nonnormal; (ii) it is more robust for the outliers; (iii) it is more suitable for the measurement of the nonlinear relations, which focus on the relationships of the trends rather than the magnitudes of the variations.

The correlation coefficients of the $\text{F}i\ (0 \le i \le n-1, i \in \mathbb{Z})$ mode with all the $n$ pulsation modes (including itself) form a vector ($\mathbf{k}^{i} = (\rho^{i}_{0},\rho^{i}_{1},\rho^{i}_{2},...,\rho^{i}_{n-1}),\ (0 \le i \le n-1, i \in \mathbb{Z})$) in the interaction space, which is a $n$-dimensional Euclid space spanned by $n$ unit vectors $\mathbf{e}_{i} = (0,...,1^{(i\text{-th})},...,0),\ (0 \le i \le n-1, i \in \mathbb{Z})$.\footnote{These $\mathbf{e}_{i}$ present the $n$ imaginary pulsation modes independent with each other.}
Based on the above structures, one can use the Euclidean distance between the $\mathbf{k}^{i}$s to measure the similarities of each pulsation modes in the interaction space.
Moreover, the agglomerative hierarchical clustering (AHC) process \citep{AHC} can be performed to cluster the pulsation modes who have similar behaviors in the interaction space together (i.e., group a set of modes in such a way that modes in the same group (called a cluster) are more similar (in some sense) to each other than to those in other groups (clusters)).

The color of a small square represents the correlation coefficient of the characteristic quantity between the two  labeled pulsation modes whose column and row intersect at the square. The dendrograms on the upper and left represent the AHC process in the interaction space, which reorders the pulsation modes in order to cluster those who have similar behaviors together.

\begin{figure*}[htp]
  \centering
  \includegraphics[width=0.99\textwidth]{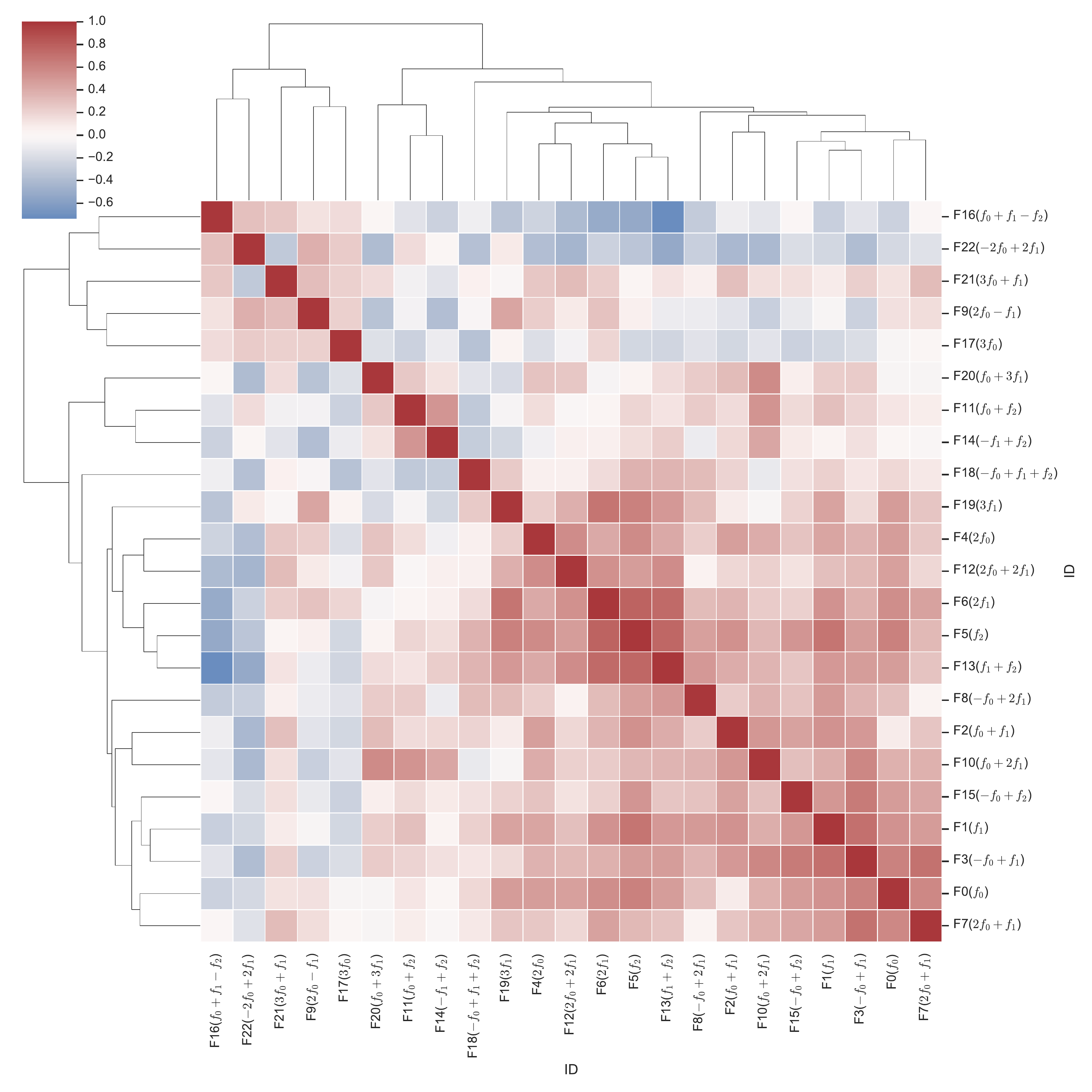}
  \caption{Interaction diagram of phases of the 23 pulsation modes.}
  \label{fig:corr_phase}
\end{figure*}

All the relationships and structures shown in the interaction diagram of amplitudes indicate that there exist interactions or energy transfer between the independent pulsation modes and their harmonics/combinations. Moreover, the interaction diagram of amplitudes help us find the particularity of $f_{2}$ related modes, discover their rotationally locked modes, and identify $f_{2}$ as a nonradial $l=1, m=1$ mode, which indicates its potential in future research.
On the other hand, comparing with the complicated interactions shown in Figure \ref{fig:corr_amp}, the Interactions Diagram of the phases (see Figure \ref{fig:corr_phase}) shows that almost all the pulsation modes have the same variation trend except for the F16, F22, and F17 modes, which are caused by the decreasing trend of their periods. Excluding the period variation rate of F17 with large uncertainty, that of F16 can be interpreted by its $-f_2$ component and that of F22 could be interpreted by its interaction with F15.

\clearpage
\section{Theoretical Model Calculation}
\label{app:4}
In order to determine the stellar mass and evolutionary stage based on the single-star evolutionary models, the open-source 1D stellar evolution code Modules for Experiments in Stellar Astrophysics (MESA, r15140, \citet{Paxton2011,Paxton2013,Paxton2015,Paxton2018,Paxton2019}) was used to construct the structural and evolutionary models. At each step along with the evolutionary tracks, the pulsation frequencies of the specific structure were calculated by the stellar oscillation code GYRE \citep{Townsend2013,Townsend2018,Goldstein2020}.

The initial parameters used to construct pre-main sequence evolutionary models of KIC 6382916 were configured as follows. Different metallicity [Fe/H] with values of $0.040$ \citep{Mathur2017}, $0.111$ \citep{Xiang2019}, and  $0.266$ \citep{Luo2019} dex were considered as the initial metallicity of the evolutionary models.
The following formulas were used to calculate the initial heavy element abundance $Z$ and initial hydrogen abundance $X$:
\begin{equation}
  \mathrm{[Fe/H]} = \log \frac{Z}{X} - \log \frac{Z_\odot}{X_\odot} ,
\end{equation}
\begin{equation}
  \label{equ:Y(Z)}
  Y=0.24 + 3Z ,
\end{equation}
\begin{equation}
  X+Y+Z=1 ,
\end{equation}
where $X_\odot =0.7381$ and $Z_\odot =0.0134$ \citep{Asplund2009}. Eq.(\ref{equ:Y(Z)}) was provided by \citet{Mowlavi1998}. Based on the given values of $\mathrm{[Fe/H]}$, we got ($X=0.704$, $Z=0.014$), ($X=0.695$, $Z=0.016$), and ($X=0.672$, $Z=0.022$), as the initial inputs of the evolutionary models.
At the same time, the values of the option {\tt Zbase} (which provides the reference metallicity necessary to calculate element variations) were set to be the initial metallicity of the star (0.014, 0.016, and 0.022, respectively), and {\tt initial\_zfracs = 'AGSS09\_zfracs'} was selected. For the opacity, {\tt kap\_file\_prefix = 'a09'}, {\tt kap\_CO\_prefix = 'a09\_co'} and {\tt kap\_lowT\_prefix = 'lowT\_fa05\_a09p'} were selected.

The initial mass of the models was set in the interval from 1.5 $\mathrm{M_\odot}$ to 2.5 $\mathrm{M_\odot}$ with a step of $0.01\ \mathrm{M_{\odot}}$, covering the typical mass range of HADS.  The rotation of the star was not considered in the model calculation because of the relative slow rotation of the star ($\Delta \omega = 0.0815$). 
The mixing-length parameter was set to be a value of $\alpha_{\rm{MLT}} = 1.89$ \citep{Niu2017}. 
The exponential scheme proposed in \citet{Herwig2000} was adopted to account for the overshooting, and the overshooting parameter was set to the fixed value $f_{\mathrm{ov}} = 0.015$ \citep{Magic2010}.

All the evolutionary tracks were calculated from the pre-MS to the red giant branch.

\begin{figure*}[htp]
  \centering
  \includegraphics[width=0.495\textwidth]{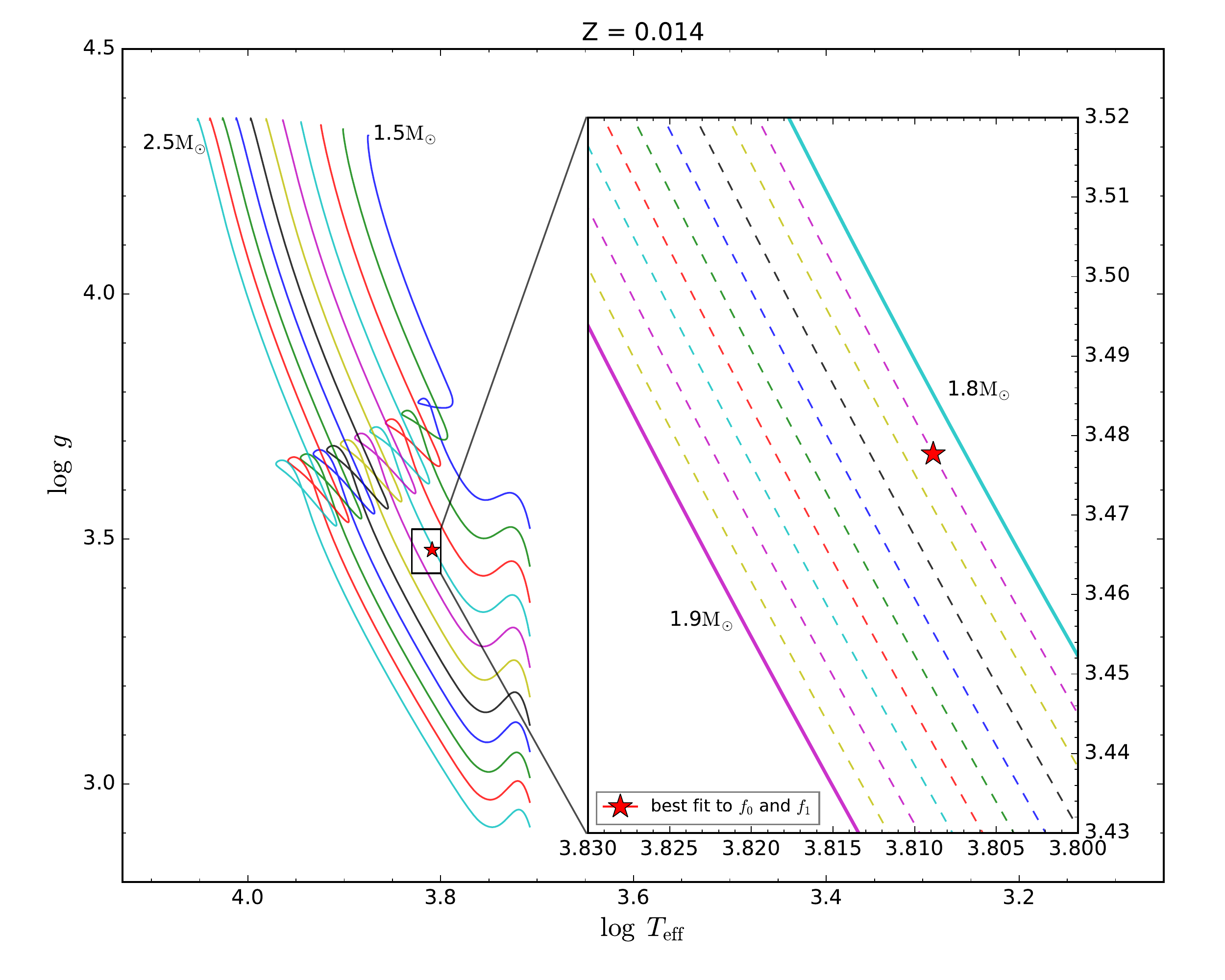}
  \includegraphics[width=0.495\textwidth]{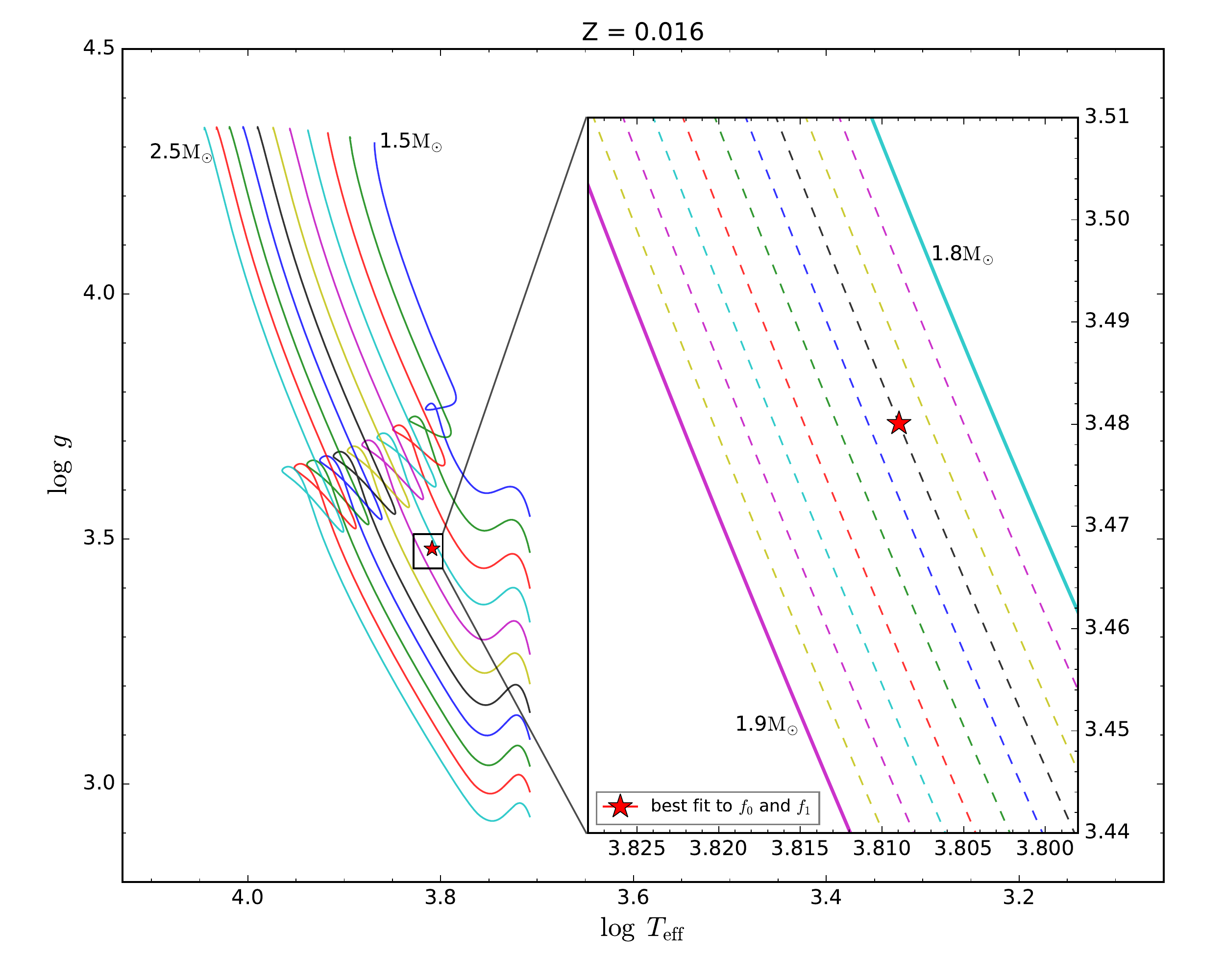}
  \includegraphics[width=0.495\textwidth]{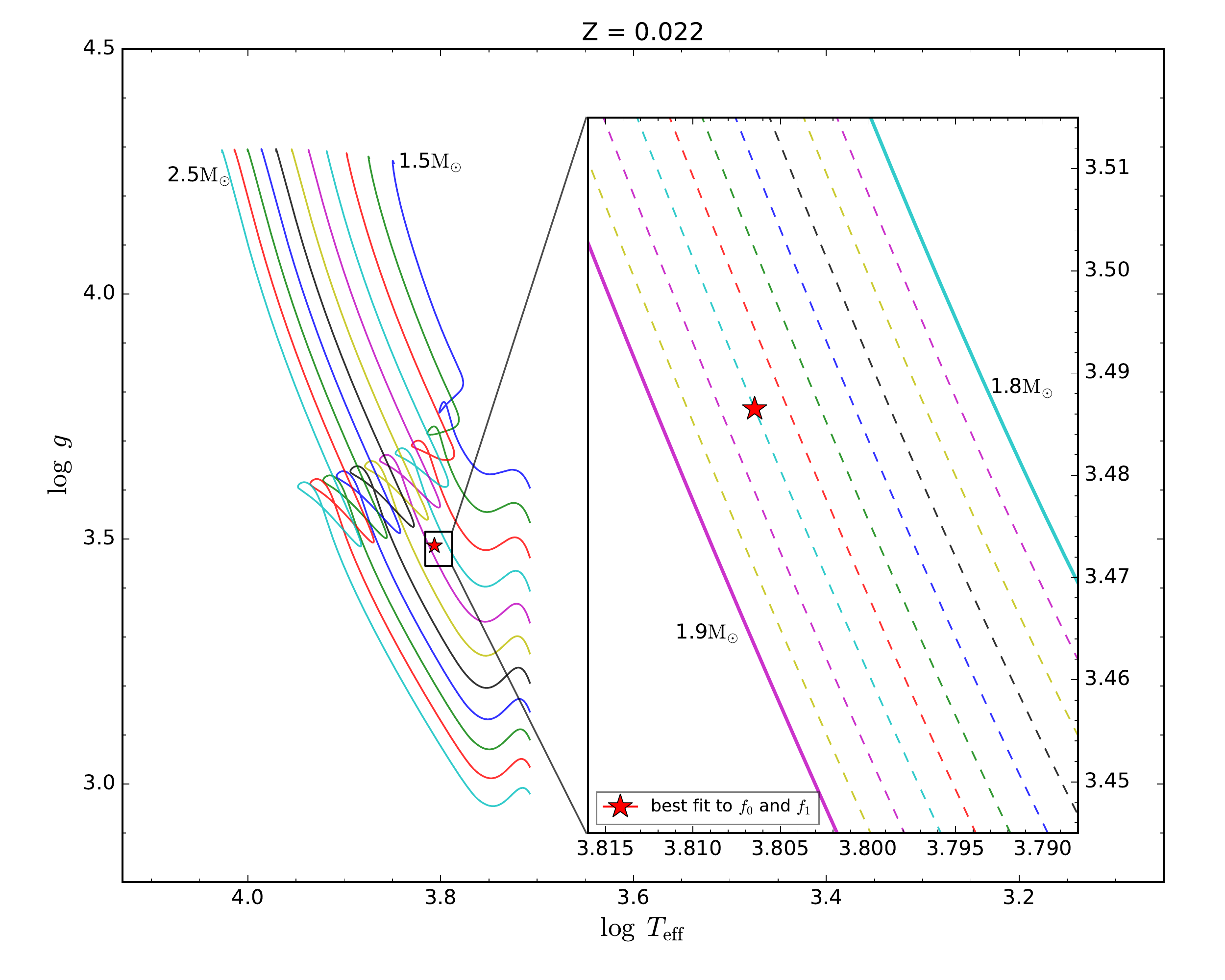}
  \caption{$\log \Teff$ vs. $\log g$ diagram with all the evolutionary tracks and the best-fit seismic models of KIC 6382916 with different $Z$ values. The colored solid evolutionary tracks show from the zero-age MS to the post-MS evolutionary stage, with initial masses from 1.50 $M_{\odot}$ to 2.50 $M_{\odot}$. The regions surrounded by the black rectangular boxes are selected to zoom in the best-fit seismic models, which are represented by the red stars. The colored dash lines present the evolutionary tracks with initial mass steps of 0.01 $M_{\odot}$.}
  \label{fig:best_model_logg}
\end{figure*}

\begin{figure*}[htp]
  \centering
  \includegraphics[width=0.495\textwidth]{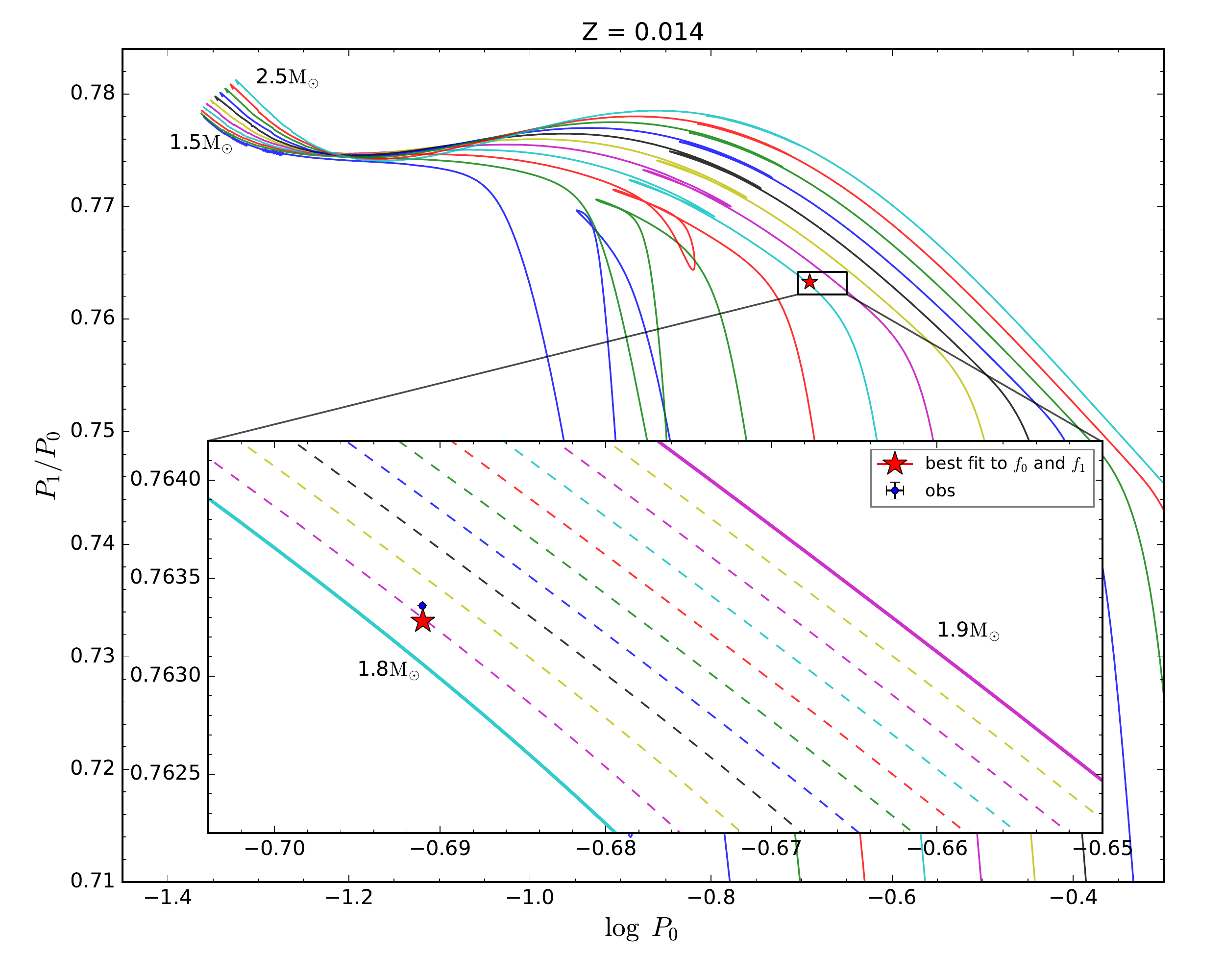}
  \includegraphics[width=0.495\textwidth]{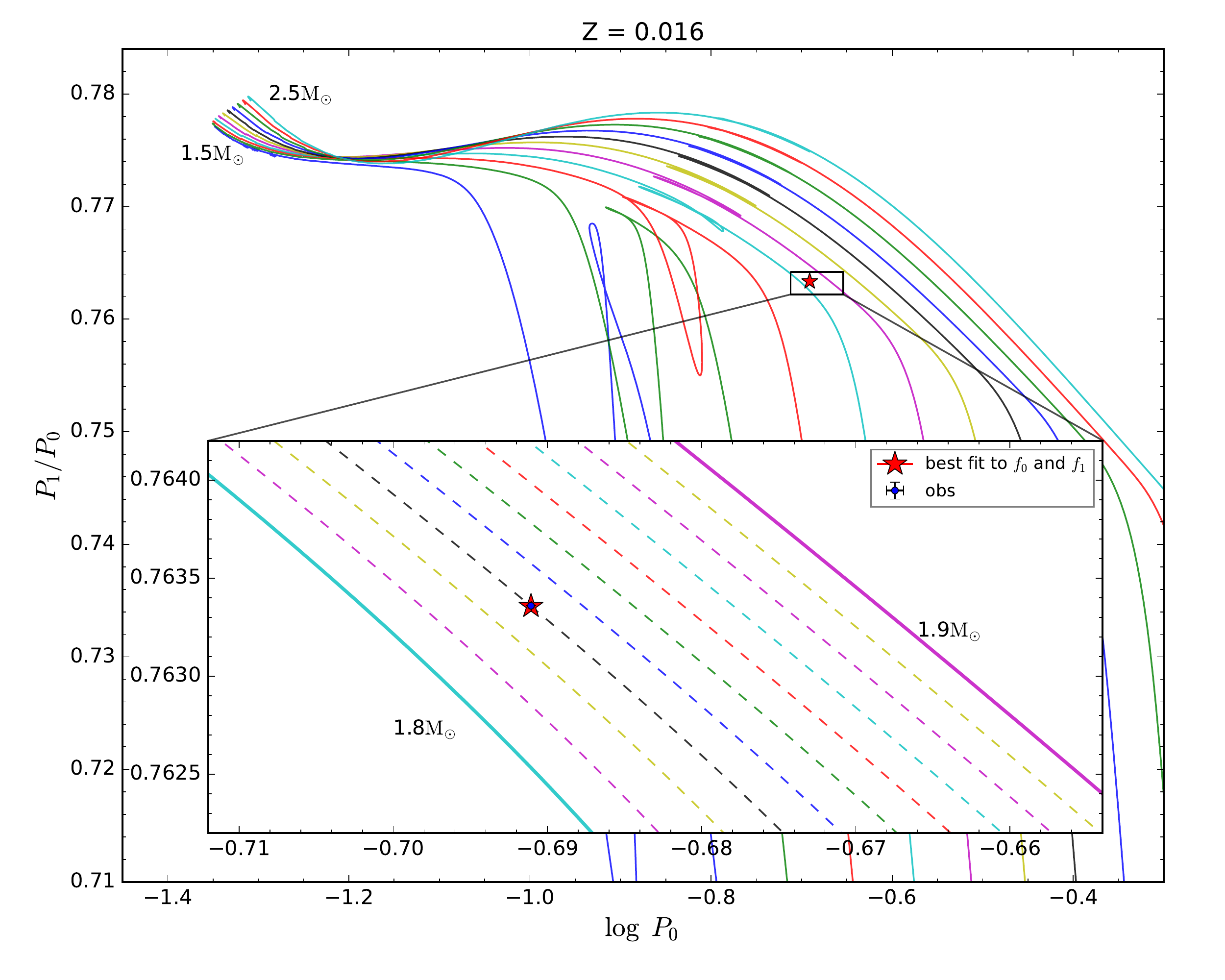}
  \includegraphics[width=0.495\textwidth]{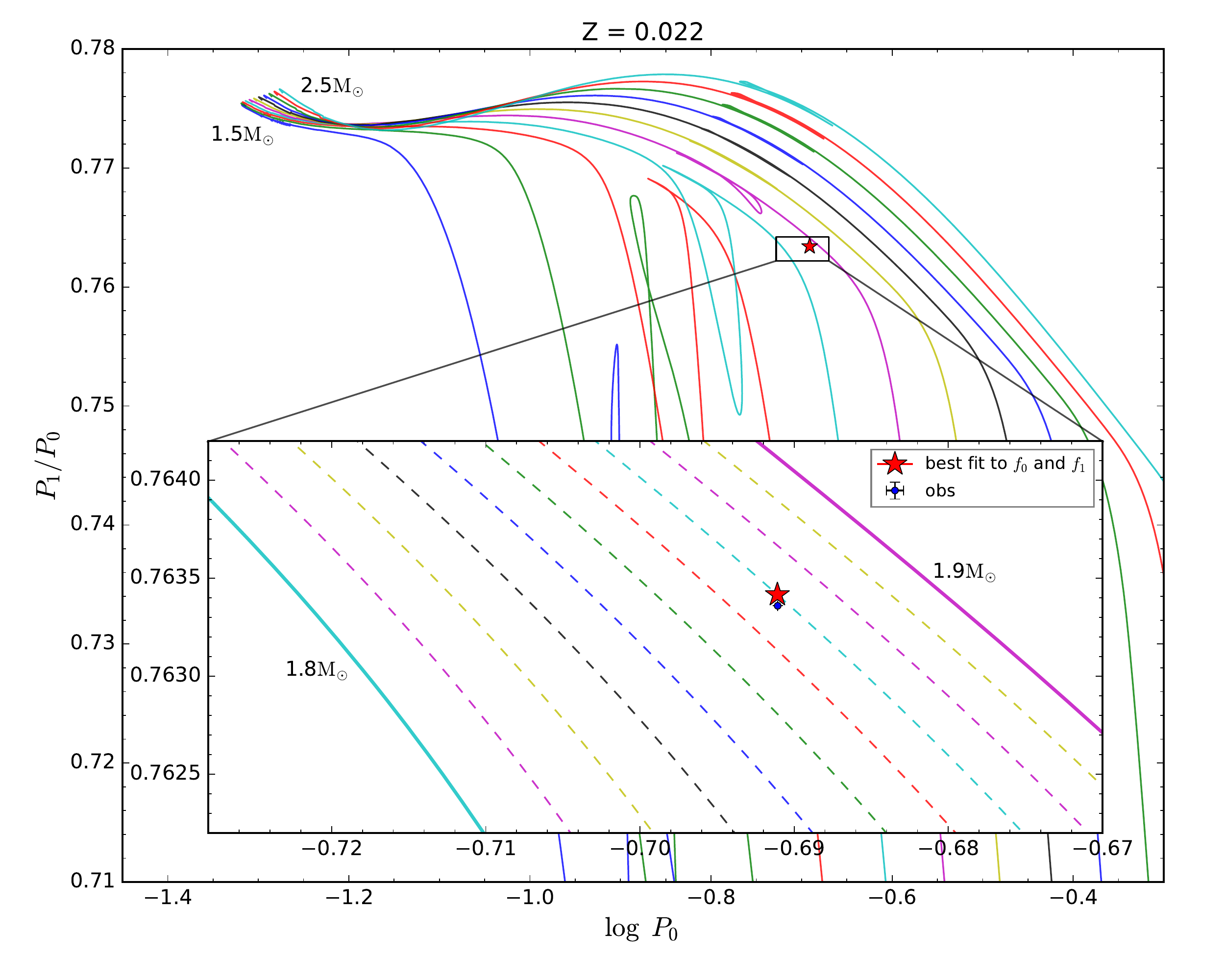}
  \caption{Petersen diagram with all the evolutionary tracks and the best-fit seismic models of KIC 6382916 with different $Z$ values. The colored solid evolutionary tracks show the stages from the zero-age MS to the post-MS evolutionary stage, with initial masses from 1.50 $M_{\odot}$ to 2.50 $M_{\odot}$. The regions surrounded by the black rectangular boxes are selected to zoom in the best-fit seismic models, which are represented by the red stars. The colored dashed lines present the evolutionary tracks with initial mass steps of 0.01 $M_{\odot}$.}
  \label{fig:best_model_petersen}
\end{figure*}

Based on the pulsation frequencies calculated in every step along with the evolutionary tracks, we got the best-fit seismic models (which have the smallest $\chi^2$ with respect to the observed values of $f_0$ and $f_1$) with the different $Z$ values (see in Table \ref{tab:best-fit} for detailed information). The H-R diagram, $\log \Teff$ vs $\log g$ diagram, and Petersen diagram are shown in Figure \ref{fig:best_model}, Figure \ref{fig:best_model_logg} and Figure \ref{fig:best_model_petersen}, respectively.

\clearpage
\section{Variation of $\Delta \omega$}
\label{app:5}

Based on the results in Table \ref{tab:var_p}, five pairs (F3-F14, F5-F8, F6-F11, F9-F16, and F15-F22) can be used to calculate the variation rate of the rotation frequency ($\diff \Delta \omega/\diff t$), which can be obtained by the differences in the frequency variation rates ($2 \cdot c_{\phi}$) in each of the pairs.

At the same time, all the above pairs follow the frequency relation
\begin{equation}
\Delta \omega = f_0 - 2f_1 + f_2,
\end{equation}
which can be differentiated on both sides with respect to time:
\begin{equation}
\frac{\diff \Delta \omega}{\diff t} \equiv \dot{\Delta \omega} = \dot{f}_0 - 2\dot{f}_1 + \dot{f}_2.
\end{equation}
$\dot{f}_0$, $\dot{f}_1$, and $\dot{f}_2$ can also be found in Table \ref{tab:var_p}.

About the uncertainty estimation, all the related frequency variation rates were taken as independent variables, and a uniform factor $\sqrt{6} \approx 2.5$ was used to rescale the uncertainties as that in Appendix \ref{app:2}. The numerical results were collected in Table \ref{tab:omega}.

For the best-fit seismic models of different values of $Z$, the variation rates of the rotation frequency were calculated based on an assumption of the conservation of angular momentum in this evolutionary phase and a uniform rotation of the star, whose numerical results were also collected in Table \ref{tab:omega}.

\begin{table*}[htp]
  \centering
  \caption{Variation rate of $\Delta \omega$.}
  \label{tab:omega}
  \begin{tabular}{c|c}
    \hline
    \hline
Origins     &  $\diff \Delta \omega/\diff t\ (\mathrm{c\ d^{-2}})$ \\
\hline
F3-F14 & $-(3 \pm 50)\times 10^{-9}$\\
F5-F8 & $-(37 \pm 20)\times 10^{-9}$\\
F6-F11 & $-(14 \pm 40)\times 10^{-9}$\\
F9-F16 & $-(85 \pm 63)\times 10^{-9}$\\
F15-F22 & $-(255 \pm 110)\times 10^{-9}$\\
\hline
F0-F1-F5 & $-(45 \pm 10)\times 10^{-9}$\\
\hline
Model (Z=0.014) & $ -9\times 10^{-10}$\\
Model (Z=0.016) & $ -1.1\times 10^{-9}$\\
Model (Z=0.022) & $ -1.4\times 10^{-9}$\\
\hline
\end{tabular}
\end{table*}

\clearpage



\label{lastpage}

\end{CJK*}
\end{document}